\documentclass[preprint,12pt]{article}
\pdfoutput=1
\usepackage{jheppub}
\usepackage{relsize}
\usepackage{units}
\usepackage{xspace} 

\usepackage{array}
\newcolumntype{C}[1]{>{\centering\let\newline\\\arraybackslash\hspace{0pt}}m{#1}}

\usepackage{amssymb}
\usepackage{amsthm}

\usepackage{lineno}

\usepackage[section]{placeins}
\usepackage{rotating}
\usepackage{relsize}

\usepackage{color}
\usepackage{graphicx}
\usepackage{subfigure}
\usepackage{amsmath}
\usepackage{ifthen} 
\newboolean{uprightparticles}
\setboolean{uprightparticles}{false} 
\usepackage{multirow}
\usepackage{tablefootnote}

\newcommand{\eqnref}[1]{Eq.~\ref{#1}}

\newcommand{\eqnsref}[2]{Eqs.~\ref{#1}~-~\ref{#2}}

\newcommand{\Tabref}[1]{Table~\ref{#1}}
\newcommand{\tabref}[1]{Tab.~\ref{#1}}

\newcommand{\appref}[1]{Appendix~\ref{#1}}

\newcommand{\secref}[1]{Sec.~\ref{#1}}

\newcommand{\secsref}[2]{Sections~\ref{#1}~-~\ref{#2}}

\newcommand{\Figref}[1]{Figure~\ref{#1}}
\newcommand{\figref}[1]{Fig.~\ref{#1}}

\def\lhcb {LHCb\xspace}

\def\ux85 {UX85\xspace}

\def\lhc {LHC\xspace}

\def\babar  {BABAR\xspace}
\def\belle  {BELLE\xspace}

\def\cleo   {CLEO\xspace}

\ifthenelse{\boolean{uprightparticles}}%
{

 \def\Ppi         {\ensuremath{\uppi}\xspace}

 \def\Ppsi        {\ensuremath{\uppsi}\xspace}

 \def\PDelta      {\ensuremath{\Delta}\xspace}                 
 \def\PXi      {\ensuremath{\Xi}\xspace}                 
 \def\PLambda      {\ensuremath{\Lambda}\xspace}                 
 \def\PSigma      {\ensuremath{\Sigma}\xspace}                 
 \def\POmega      {\ensuremath{\Omega}\xspace}                 
 \def\PUpsilon      {\ensuremath{\Upsilon}\xspace}                 
 
 \def\PB      {\ensuremath{\mathrm{B}}\xspace}                 
                  
 \def\PD      {\ensuremath{\mathrm{D}}\xspace}

 \def\PK      {\ensuremath{\mathrm{K}}\xspace}

 \def\PW      {\ensuremath{\mathrm{W}}\xspace}

 \def\Pb      {\ensuremath{\mathrm{b}}\xspace}                 
 \def\Pc      {\ensuremath{\mathrm{c}}\xspace}                 
                  
 \def\Pe      {\ensuremath{\mathrm{e}}\xspace}

 \def\Pi      {\ensuremath{\mathrm{i}}\xspace}

 \def\Ps      {\ensuremath{\mathrm{s}}\xspace}                 
                  
 \def\Pu      {\ensuremath{\mathrm{u}}\xspace}

}
{

 \def\Ppi         {\ensuremath{\pi}\xspace}

 \def\Ppsi        {\ensuremath{\psi}\xspace}                 
                  
 \mathchardef\PDelta="7101
 \mathchardef\PXi="7104
 \mathchardef\PLambda="7103
 \mathchardef\PSigma="7106
 \mathchardef\POmega="710A
 \mathchardef\PUpsilon="7107
                  
 \def\PB      {\ensuremath{B}\xspace}                 
                  
 \def\PD      {\ensuremath{D}\xspace}

 \def\PK      {\ensuremath{K}\xspace}

 \def\PW      {\ensuremath{W}\xspace}

 \def\Pb      {\ensuremath{b}\xspace}                 
 \def\Pc      {\ensuremath{c}\xspace}                 
                  
 \def\Pe      {\ensuremath{e}\xspace}

 \def\Pi      {\ensuremath{i}\xspace}

 \def\Ps      {\ensuremath{s}\xspace}                 
                  
 \def\Pu      {\ensuremath{u}\xspace}

}

\def\ep         {\ensuremath{\Pe^+}\xspace}
\def\epm        {\ensuremath{\Pe^\pm}\xspace}

\def\Wm     {\ensuremath{\PW^-}\xspace}

\def\u     {\ensuremath{\Pu}\xspace}
\def\ubar  {\ensuremath{\overline \u}\xspace}

\def\s     {\ensuremath{\Ps}\xspace}

\def\c     {\ensuremath{\Pc}\xspace}
\def\cbar  {\ensuremath{\overline \c}\xspace}

\def\b     {\ensuremath{\Pb}\xspace}

\def\pion  {\ensuremath{\Ppi}\xspace}
\def\piz   {\ensuremath{\pion^0}\xspace}

\def\pip   {\ensuremath{\pion^+}\xspace}
\def\pim   {\ensuremath{\pion^-}\xspace}
\def\pipm  {\ensuremath{\pion^\pm}\xspace}
\def\pimp  {\ensuremath{\pion^\mp}\xspace}

\def\kaon  {\ensuremath{\PK}\xspace}
  \def\Kbar  {\kern 0.2em\overline{\kern -0.2em \PK}{}\xspace}

\def\Kz    {\ensuremath{\kaon^0}\xspace}
\def\Kzb   {\ensuremath{\Kbar^0}\xspace}
\def\KzKzb {\ensuremath{\Kz \kern -0.16em \Kzb}\xspace}
\def\Kp    {\ensuremath{\kaon^+}\xspace}
\def\Km    {\ensuremath{\kaon^-}\xspace}
\def\Kpm   {\ensuremath{\kaon^\pm}\xspace}
\def\Kmp   {\ensuremath{\kaon^\mp}\xspace}
\def\KpKm  {\ensuremath{\Kp \kern -0.16em \Km}\xspace}
\def\KS    {\ensuremath{\kaon^0_{\rm\scriptscriptstyle S}}\xspace} 
\def\KL    {\ensuremath{\kaon^0_{\rm\scriptscriptstyle L}}\xspace}

\def\Kstarm  {\ensuremath{\kaon^{*-}}\xspace}

  \def\Dbar    {\kern 0.2em\overline{\kern -0.2em \PD}{}\xspace}
\def\D       {\ensuremath{\PD}\xspace}

\def\Dz      {\ensuremath{\D^0}\xspace}
\def\Dzb     {\ensuremath{\Dbar^0}\xspace}
\def\DzDzb   {\ensuremath{\Dz {\kern -0.16em \Dzb}}\xspace}
\def\Dp      {\ensuremath{\D^+}\xspace}
\def\Dm      {\ensuremath{\D^-}\xspace}

\def\DpDm    {\ensuremath{\Dp {\kern -0.16em \Dm}}\xspace}

\def\Dstarp  {\ensuremath{\D^{*+}}\xspace}

\def\B       {\ensuremath{\PB}\xspace}
  \def\Bbar    {\kern 0.18em\overline{\kern -0.18em \PB}{}\xspace}

\def\Bu      {\ensuremath{\B^+}\xspace}
\def\Bub     {\ensuremath{\B^-}\xspace}
\def\Bp      {\ensuremath{\Bu}\xspace}
\def\Bm      {\ensuremath{\Bub}\xspace}
\def\Bpm     {\ensuremath{\B^\pm}\xspace}

\def\psiprpr  {\ensuremath{\Ppsi(3770)}\xspace}

  \def\Y#1S{\ensuremath{\PUpsilon{(#1S)}}\xspace}

\newcommand{\decay}[2]{\ensuremath{#1\!\to #2}\xspace}         

\def\to                 {\ensuremath{\rightarrow}\xspace}

\def\CP                {\ensuremath{C\!P}\xspace}

\def\AT#1     {\ensuremath{A_T^{#1}}\xspace}           

\def\C#1      {\ensuremath{\mathcal{C}_{#1}}\xspace}                       
\def\Cp#1     {\ensuremath{\mathcal{C}_{#1}^{'}}\xspace}                    
\def\Ceff#1   {\ensuremath{\mathcal{C}_{#1}^{\mathrm{(eff)}}}\xspace}        
\def\Cpeff#1  {\ensuremath{\mathcal{C}_{#1}^{'\mathrm{(eff)}}}\xspace}       
\def\Ope#1    {\ensuremath{\mathcal{O}_{#1}}\xspace}                       
\def\Opep#1   {\ensuremath{\mathcal{O}_{#1}^{'}}\xspace}                    

\newcommand{\tev}{\ensuremath{\mathrm{\,Te\kern -0.1em V}}\xspace}
\newcommand{\gev}{\ensuremath{\mathrm{\,Ge\kern -0.1em V}}\xspace}
\newcommand{\mev}{\ensuremath{\mathrm{\,Me\kern -0.1em V}}\xspace}
\newcommand{\kev}{\ensuremath{\mathrm{\,ke\kern -0.1em V}}\xspace}
\newcommand{\ev}{\ensuremath{\mathrm{\,e\kern -0.1em V}}\xspace}

\def\invpb {\ensuremath{\mbox{\,pb}^{-1}}\xspace}

\def\invfb   {\ensuremath{\mbox{\,fb}^{-1}}\xspace}

\def\gsim{{~\raise.15em\hbox{$>$}\kern-.85em
          \lower.35em\hbox{$\sim$}~}\xspace}
\def\lsim{{~\raise.15em\hbox{$<$}\kern-.85em
          \lower.35em\hbox{$\sim$}~}\xspace}

\def\rad{\ensuremath{\rm \,rad}\xspace}

\def\root       {\mbox{\textsc{Root}}\xspace}

\def\cpp        {\mbox{\textsc{C\raisebox{0.1em}{{\footnotesize{++}}}}}\xspace}

\def\tell1  {TELL1\xspace}
\def\ukl1   {UKL1\xspace}

\newcommand{\eg}{\mbox{\itshape e.g.}\xspace}
\newcommand{\ie}{\mbox{\itshape i.e.}\xspace}

\newcommand{\etc}{\mbox{\itshape etc.}\xspace}

\newcommand{\fourpi}{\ensuremath{4\pipm}\xspace}

\def\BtoDK  {\decay{\Bpm}{\D\Kpm}\xspace}

\def\BptoDKp  {\decay{\Bp}{\D\Kp}\xspace}
\def\BmtoDzKm  {\decay{\Bm}{\Dz\Km}\xspace}
\def\BmtoDzbKm  {\decay{\Bm}{\Dzb\Km}\xspace}
\def\BmtoDKm  {\decay{\Bm}{\D\Km}\xspace}

\newcommand{\pspointf}{\ensuremath{\mathbf{p}}\xspace}
\newcommand{\pspointg}{\ensuremath{\mathbf{q}}\xspace}
\newcommand{\pspointbarf}{\ensuremath{\mathbf{\overline{p}}}\xspace}

\newcommand{\DDb}{\ensuremath{\D\Dbar}\xspace}
\newcommand{\si}{\ensuremath{s_{i}^{f}}\xspace}
\newcommand{\sj}{\ensuremath{s_{j}^{g}}\xspace}
\newcommand{\ci}{\ensuremath{c_{i}^{f}}\xspace}
\newcommand{\cj}{\ensuremath{c_{j}^{g}}\xspace}
\newcommand{\ki}{\ensuremath{K_{i}^{f}}\xspace}
\newcommand{\kbi}{\ensuremath{\bar{K}_{i}^{f}}\xspace}
\newcommand{\kj}{\ensuremath{K_{j}^{g}}\xspace}
\newcommand{\kbj}{\ensuremath{\bar{K}_{j}^{g}}\xspace}

\newcommand{\Fri}{\ensuremath{T_{i}^{f}}\xspace}
\newcommand{\Frbi}{\ensuremath{\bar{T}_{i}^{f}}\xspace}

\newcommand{\Frifourpi}{\ensuremath{T_{i}^{4\pi}}\xspace}
\newcommand{\Frbifourpi}{\ensuremath{\bar{T}_{i}^{4\pi}}\xspace}

\newcommand{\smi}{\ensuremath{s_{-i}^{f}}\xspace}
\newcommand{\cmi}{\ensuremath{c_{-i}^{f}}\xspace}
\newcommand{\kmi}{\ensuremath{K_{-i}^{f}}\xspace}
\newcommand{\kbmi}{\ensuremath{\bar{K}_{-i}^{f}}\xspace}

\newcommand{\sifourpi}{\ensuremath{s_{i}^{4\pi}}\xspace}
\newcommand{\cifourpi}{\ensuremath{c_{i}^{4\pi}}\xspace}
\newcommand{\kifourpi}{\ensuremath{K_{i}^{4\pi}}\xspace}
\newcommand{\kbifourpi}{\ensuremath{\bar{K}_{i}^{4\pi}}\xspace}

\newcommand{\sikspipi}{\ensuremath{s_{i}^{\KS\pi\pi}}\xspace}
\newcommand{\cikspipi}{\ensuremath{c_{i}^{\KS\pi\pi}}\xspace}

\newcommand{\Frikspipi}{\ensuremath{F_{i}^{\KS\pi\pi}}\xspace}
\newcommand{\Frbikspipi}{\ensuremath{\bar{F}_{i}^{\KS\pi\pi}}\xspace}

\newcommand{\siklpipi}{\ensuremath{s_{i}^{\KL\pi\pi}}\xspace}
\newcommand{\ciklpipi}{\ensuremath{c_{i}^{\KL\pi\pi}}\xspace}
\newcommand{\kiklpipi}{\ensuremath{K_{i}^{\KL\pi\pi}}\xspace}
\newcommand{\kbiklpipi}{\ensuremath{\bar{K}_{i}^{\KL\pi\pi}}\xspace}
\newcommand{\Friklpipi}{\ensuremath{F_{i}^{\KL\pi\pi}}\xspace}
\newcommand{\Frbiklpipi}{\ensuremath{\bar{F}_{i}^{\KL\pi\pi}}\xspace}

\newcommand{\kjall}{\ensuremath{K^{g}}\xspace}
\newcommand{\kbjall}{\ensuremath{\bar{K}^{g}}\xspace}
\newcommand{\sjall}{\ensuremath{s^{g}}\xspace}
\newcommand{\cjall}{\ensuremath{c^{g}}\xspace}

\newcommand{\fsi}{\ensuremath{f_{i}}\xspace}
\newcommand{\fsj}{\ensuremath{g_{j}}\xspace}

\newcommand{\fsp}{\ensuremath{f_{\pspointf}}\xspace}
\newcommand{\fsq}{\ensuremath{g_{\pspointg}}\xspace}

\newcommand{\half}{\ensuremath{\frac{1}{2}}\xspace}

\newcommand{\Af}{\ensuremath{A^f_\pspointf}\xspace}
\newcommand{\Abf}{\ensuremath{\bar{A}^f_\pspointf}\xspace}
\newcommand{\Afb}{\ensuremath{A^f_\pspointbarf}\xspace}

\newcommand{\Afourpi}{\ensuremath{A^{4\pi}_\pspointf}\xspace}
\newcommand{\Abfourpi}{\ensuremath{\bar{A}^{4\pi}_\pspointf}\xspace}

\newcommand{\Deldelf}{\ensuremath{\Delta\delta^f_\pspointf}\xspace}

\newcommand{\Deldelfourpi}{\ensuremath{\Delta\delta^{4\pi}_\pspointf}\xspace}
\newcommand{\Deldelbfourpi}{\ensuremath{\Delta\delta^{4\pi}_\pspointbarf}\xspace}
\newcommand{\Deldelkspipi}{\ensuremath{\Delta\delta^{\KS\pi\pi}_\pspointf}\xspace}

\newcommand{\Deldelklpipi}{\ensuremath{\Delta\delta^{\KL\pi\pi}_\pspointf}\xspace}

\newcommand{\denStatesf}{\ensuremath{ \phi(\pspointf) }\xspace}

\newcommand{\CPp}{\ensuremath{ \CP\raisebox{.3\height}{\scalebox{.7}{+}} }\xspace}
\newcommand{\CPm}{\ensuremath{ \CP\raisebox{.1\height}{\scalebox{.9}{-}} }\xspace}

\newcommand{\xmix}{\ensuremath{ x_D }\xspace}
\newcommand{\ymix}{\ensuremath{ y_D }\xspace}

\newcommand{\Kenu}{\ensuremath{\Kmp\epm\nu}\xspace}
\newcommand{\Kpi}{\ensuremath{\Kmp\pipm}\xspace}
\newcommand{\Kpipiz}{\ensuremath{\Kmp\pipm\piz}\xspace}
\newcommand{\Kpipipi}{\ensuremath{\Kmp\pipm\pimp\pipm}\xspace}

\newcommand{\Kmpi}{\ensuremath{\Km\pip}\xspace}
\newcommand{\Kmpipiz}{\ensuremath{\Km\pip\piz}\xspace}

\newcommand{\KK}{\ensuremath{\Kp\Km}\xspace}
\newcommand{\pipi}{\ensuremath{\pip\pim}\xspace}
\newcommand{\KLpiz}{\ensuremath{\KL\piz}\xspace}
\newcommand{\KLomega}{\ensuremath{\KL\omega}\xspace}
\newcommand{\KSpizpiz}{\ensuremath{\KS\piz\piz}\xspace}

\newcommand{\KSpiz}{\ensuremath{\KS\piz}\xspace}
\newcommand{\KSomega}{\ensuremath{\KS\omega}\xspace}
\newcommand{\KSeta}{\ensuremath{\KS\eta}\xspace}
\newcommand{\KSetap}{\ensuremath{\KS\eta'}\xspace}

\newcommand{\KSpipi}{\ensuremath{\KS\pip\pim}\xspace}
\newcommand{\KLpipi}{\ensuremath{\KL\pip\pim}\xspace}

\newcommand{\pipipiz}{\ensuremath{\pip\pim\piz}\xspace}

\newcommand{\mbc}{\ensuremath{m_{\mathrm{bc}}}\xspace}
\newcommand{\avembc}{\ensuremath{m^{\mathrm{ave}}_{\mathrm{bc}}}\xspace}
\newcommand{\missm}{\ensuremath{m^2_{\mathrm{miss}}}}
\newcommand{\umiss}{\ensuremath{U_{\mathrm{miss}}}}

\numberwithin{equation}{section}

\begin{document}

\title{ Model-independent determination of the strong phase difference between \Dz and \decay{\Dzb}{\pip\pim\pip\pim} amplitudes. 
}


\author[a]{Samuel Harnew}
\author[a]{Paras Naik}
\author[a]{Claire Prouve}
\author[a]{Jonas Rademacker}
\author[b]{David Asner}

\affiliation[a]{University of Bristol, H H Wills Physics Laboratory, UK}
\affiliation[b]{Pacific Northwest National Laboratory, Richland, WA 99354, USA}
\emailAdd{sam.harnew@bristol.ac.uk}
\emailAdd{jonas.rademacker@bristol.ac.uk}

\abstract{ 
For the first time, the strong phase difference between \Dz and \decay{\Dzb}{\pip\pim\pip\pim} amplitudes is determined in bins of the decay phase space.
The measurement uses 818\invpb of $e^+e^-$ collision data that is taken at the \psiprpr resonance and collected by the CLEO-c experiment.
The measurement is important for the determination of the \CP-violating phase $\gamma$ in \BtoDK (and similar) decays , where the \D meson (which represents a superposition of \Dz and \Dzb) subsequently decays to $\pip\pim\pip\pim$.
To obtain optimal sensitivity to $\gamma$, the phase space of the $\decay{\D}{\pip\pim\pip\pim}$ decay is divided into bins based on a recent amplitude model of the decay.
Although an amplitude model is used to define the bins, the measurements obtained are model-independent. 
The \CP-even fraction of the $\decay{\D}{\pip\pim\pip\pim}$ decay is determined to be $F_{+}^{4\pi} = 0.769 \pm 0.021 \pm 0.010$, where the uncertainties are statistical and systematic, respectively. 
Using simulated $\BtoDK, \decay{\D}{\pip\pim\pip\pim}$ decays, it is estimated that by the end of the current \lhc run, the \lhcb experiment could determine $\gamma$ from this decay mode with an uncertainty of $(\pm10\pm7)^\circ$, where the first uncertainty is statistical based on estimated \lhcb event yields, and the second is due to the uncertainties on the parameters determined in this paper. 
}

\maketitle


\section{Introduction}
\label{sec:introduction}
A primary goal in modern flavour physics is to constrain the unitarity triangle (UT); an abstract representation of the famous Cabibbo-Kobayashi-Maskawa matrix that describes transitions between different quark flavours~\cite{CKMCabibbo, CKM}. 
Key to determining the UT is are better experimental constraints on the angle $\gamma$ (or $\phi_{3}$), which is related to the phase difference between $\b\to\u\ \Wm$ and $\b\to\c\ \Wm$ quark transitions. 
Currently, $\gamma$ is the least-well constrained angle of the UT, and can be determined, for example, using \BmtoDKm decays\footnote{Charge-conjugate decays are implied throughout this paper.}, where \D represents a superposition of \Dz and \Dzb states~\cite{GLW1,GLW2, ADS, DalitzGamma1, DalitzGamma2, Rademacker:2006zx}.
The amplitudes \BmtoDzbKm and \BmtoDzKm are overwhelmingly dominated by the tree-level transitions $\b \to \u \, \cbar \s$ and $\b \to \c \, \ubar \s$, respectively, and therefore offer an extremely clean method to measure $\gamma$. 
In order to obtain the necessary interference between \BmtoDzKm and \BmtoDzbKm amplitudes, a final state $f$ must be chosen that is accessible from both \Dz and \Dzb, such as $\pip\pim\pip\pim$ (\fourpi).

To determine $\gamma$ in \BmtoDKm decays, one must know the relative magnitude and phase of \decay{\Dz}{f} and \decay{\Dzb}{f} amplitudes, collectively known as the \decay{\D}{f} hadronic parameters.
The relative magnitudes can be determined by measuring \decay{\Dstarp}{\Dz\pip} decays that are subsequently followed by a \decay{\Dz}{f} decay; this is possible at a large variety of collider experiments, such as \lhcb and the \B-factories.
Measuring the relative phase, however, is more challenging. 
One method is to infer the relative phase through use of an amplitude model; in principle this is the best way to exploit the available statistics, but theoretical uncertainties in determining the model can lead to large systematic uncertainties on $\gamma$.
The relative phase can also be determined model-independently by using samples of \decay{\D}{f} decays, where the \D meson is in a known superposition of \Dz and \Dzb states.
Previously, such data samples have been obtained from two sources: correlated \DDb pairs from the decay of a \psiprpr meson~\cite{gammaADS, ModelIndepGammaTheory,
coherenceCLEO, Libby:2010nu, Briere:2009aa, CLEO:DeltaKpi,
Insler:2012pm, Libby:2014rea} (the first charmonia resonance above the charm threshold);
and the decay \decay{\Dstarp}{\D\pip}, where the superposition of \Dz and \Dzb states depends on the \D meson decay-time ~\cite{selfcite,selfcite2,K3piLHCb}. 
In this paper we determine the relative magnitude and phase of \decay{\Dz}{\fourpi} and \decay{\Dzb}{\fourpi} amplitudes using \psiprpr decays collected by the \cleo-c experiment.

In multi-body \D decays, such as \decay{\D}{\fourpi}, there are infinitely many configurations of the final state momenta, each with a different amplitude.  
The parameter space that describes these final state configurations is known as the phase space of the decay. 
For the \fourpi final state, a phase space-integrated measurement was performed in Ref.~\cite{FPlusFourPi} to determine the \CP-content of the inclusive decay, and then applied in a \BtoDK study at LHCb~\cite{cpObsTwoAndFourBody}.
However, to better exploit the information available in multi-body \D decays, the phase space can be divided into bins such that regions of constructive and destructive interference do not dilute each other. 
Such a method has already been applied to the \KSpipi final state~\cite{KzPiPiCiSi} which gives the best single measurement of $\gamma$ to date~\cite{LHCbKSPiPi}; here an amplitude-model was used to group regions of phase space that have a similar phase difference between \decay{\Dz}{\KSpipi} and \decay{\Dzb}{\KSpipi} amplitudes~\cite{Bondar:2008hh}. 
Recently an amplitude model for \decay{\D}{\fourpi} has become available~\cite{fourpimodel}, so in this paper a similar technique is applied to the \fourpi final state.
It is important to note that although the binning scheme is defined by an amplitude model, this will not result in any model-dependent bias. 
If the model is incorrect, this will just result in an increased \emph{statistical} uncertainty. 

This paper is organised as follows: \secref{sec:formalism} gives an overview of the formalism for correlated $\psiprpr\to\DDb$ decays; 
\secref{sec:model} introduces the \decay{\Dz}{\fourpi} amplitude model that is used in \secref{sec:binning} to inspire the phase space binning schemes; 
\secref{sec:selection} discusses the dataset used in the analysis and the selection criteria applied; 
\secref{sec:fitter} describes the fit used to obtain constraints on the \decay{\D}{\fourpi} hadronic parameters; \secref{sec:systematics} discusses the systematic uncertainties associated to the results in \secref{sec:results}; \secref{sec:gamma} uses the measured hadronic parameters to estimate the $\gamma$ constraints that are possible with current and future \lhcb datasets; 
finally a summary is given in \secref{sec:summary}.

\section{Formalism}
\label{sec:formalism}
The mass eigenstates of the \D meson, $| \D_{1,2} \rangle$, can be written in terms of the flavour eigenstates,
\begin{align}
   | \D_{1,2} \rangle = | \Dz \rangle \pm | \Dzb \rangle,
\end{align}
where the convention $\CP | \Dz \rangle = +| \Dzb \rangle$ is followed such that $| \D_{1} \rangle$ and $| \D_{2} \rangle$ are the \CPp and \CPm eigenstates, respectively. 
Throughout this paper \CP violation in the \D meson system is neglected, which is a good assumption given current experimental limits~\cite{HFAGCKM2016}. 
The masses and widths of $\D_{1,2}$ are given by $m_{1,2}$ and $\Gamma_{1,2}$ respectively, which allows the average width, $\Gamma_D = \half(\Gamma_{1} + \Gamma_{2})$, and the charm mixing parameters, $\xmix = (m_1 - m_2)/\Gamma_D$ and $\ymix = (\Gamma_{1} - \Gamma_{2})/2\Gamma_D$, to be defined. 
Due to the effects of \D-mixing, a \D meson produced in a $| \Dz \rangle$ eigenstate at $t_0=0$ evolves to an admixture of $| \Dz \rangle$ and $| \Dzb \rangle$ states, denoted $| \Dz(t) \rangle$, after time $t$. 
Similarly, the $| \Dzb \rangle$ eigenstate evolves to $| \Dzb(t) \rangle$.

The \Dz and \Dzb decay amplitudes for a final state $f$ are defined $\Af = \langle \fsp | H | \Dz \rangle$ and $\Abf = \langle \fsp | H | \Dzb \rangle$, where $H$ is the relevant Hamiltonian.
The parameter \pspointf describes a point in the phase space of the $\D\to f$ decay, and has a dimensionality that depends on the number of final state particles and their spin.
For two-, three- and four-body pseudo-scalar final states the phase space dimensionality is 0, 2 and 5, respectively.

In this paper, the measured observables will always be integrated over bins of phase space. 
For the final states $f$ and $g$, these regions are labeled by $i$ and $j$, respectively\footnote{Having labels for two final states will later be important for describing correlated \D decays.}. 
The branching fraction for $\Dz\to\fsi$ and $\Dzb\to\fsi$ decays are defined,
\begin{align}
\ki = \int_i |\Af |^2 \denStatesf \mathrm{d} \pspointf   \ \ \ \ \ \ \   \kbi = \int_i |\Abf |^2 \denStatesf \mathrm{d} \pspointf,\label{eqn:kkbar}
\end{align}
where \denStatesf gives the density of states at \pspointf.
From these follow the quantities $\Fri = \ki / \sum \ki$ and $\Frbi = \kbi/ \sum \kbi$, which give the fraction of \decay{\Dz}{f} and \decay{\Dzb}{f} decays that populate phase space bin $i$, respectively\footnote{This is the fraction with respect to all phase space bins considered in an analysis, which is not necessarily the entire phase space.}.
To describe the interference of $\Dz\to f$ and $\Dzb\to f$ amplitudes integrated over the region $i$, the bin-averaged sine and cosine are defined,
\begin{align}
\ci = \frac{1}{\sqrt{\ki\kbi}} \int_i  |\Af | |\Abf | \cos \left ( \Deldelf \right ) \denStatesf \mathrm{d} \pspointf , \label{eqn:ci}\\
\si = \frac{1}{\sqrt{\ki\kbi}} \int_i  |\Af | |\Abf | \sin \left ( \Deldelf \right ) \denStatesf \mathrm{d} \pspointf  , \label{eqn:si}
\end{align}
where $\Deldelf = \arg ( \Af ) - \arg( \Abf )$. 
Collectively, the parameters \ci, \si, \ki and \kbi are referred to as the hadronic parameters of the \decay{\D}{f} decay.

Using the formalism above, the decay $\psiprpr \to \DDb \to \fsi\fsj$ is now considered. 
The strong decay \decay{\psiprpr}{\DDb} results in a correlated \DDb pair in a $C=-1$ state. Therefore,
\begin{align}
   | \psiprpr \rangle \to | \Dz \Dzb \rangle - | \Dzb \Dz \rangle.
\end{align}
Since the two \D mesons evolve coherently, \D-mixing has no observable consequences until one meson decays.
Therefore, when studying such decays, what is important is the time difference, $\delta t$, between the $\D\to f$ and $\D\to g$ decays.
The decay amplitude for $\psiprpr \to \DDb \to \fsp\fsq$ is given by~\cite{Rama:mixingTheory},
\begin{align}
  A(  \psiprpr& \to \DDb \to \fsp\fsq ) \propto \notag \\
  &\langle \fsp | H | \Dz \rangle \langle \fsq | H | \Dzb(\delta t) \rangle - \langle \fsp | H | \Dzb \rangle \langle \fsq | H | \Dz(\delta t) \rangle.
\end{align}
To obtain the decay rate, the magnitude of this amplitude is squared and integrated over the phase space regions $i$ and $j$, and all decay-times. 
Expanding to second order in the small parameters \xmix and \ymix gives,
\begin{align}
 \Gamma[\psiprpr &\to \DDb \to \fsi\fsj] \propto \notag \\
   &\left (  1 + \frac{\ymix^2 - \xmix^2}{2} \right ) \left [ \ki \kbj +  \kbi \kj  
 - 2 \sqrt{  \ki \kbj \kbi \kj } \left ( \ci \cj + \si \sj \right ) \right ] \notag \\
 + &\left ( \phantom{ 1 + \ } \frac{\ymix^2 + \xmix^2}{2} \right ) \left [ \ki \kj +  \kbi \kbj  
 - 2 \sqrt{  \ki \kbj \kbi \kj } \left ( \ci \cj - \si \sj \right ) \right ]. \label{eqn:masterFormula} 
\end{align}
This single formula is used to describe all decays studied in this paper. 
Note that \eqnref{eqn:masterFormula} can be significantly simplified for some final states; for example \CP eigenstates such as $\Kp\Km$ (\CPp) and $\KS\piz$ (\CPm) have $\kj \equiv \kbj$, $\sj \equiv 0$ and $\cj = \eta_{\CP}$, where $\eta_{\CP} = \pm 1$ for \CPp and \CPm eigenstates, respectively\footnote{This follows from the convention $\CP | \Dz \rangle = +| \Dzb \rangle$ that was chosen earlier.}.

When only one of the \D meson final states is reconstructed it is known as a single-tag.
In this case, the final state $g$ represents \emph{all} possible \D meson final states and $\kjall \equiv \kbjall \equiv 1$, $\sjall \equiv 0$ and $\cjall = \ymix$, leading to,
\begin{align}
 \Gamma[\psiprpr \to \DDb \to \fsi X ] \propto  \left (  1 + \ymix^2  \right ) \left [ \ki +  \kbi 
 - 2 \sqrt{  \ki \kbi  } \ci y \right ]. \label{eqn:masterSTFormula} 
\end{align}
The $\sjall \equiv 0$ can be understood by realising that for every final state $g$, there is a charge-conjugate final state $\overline{g}$ that has $s^{\overline{g}} = -s^g$. 
The $\cjall = \ymix$ can be understood by rewriting \eqnref{eqn:ci} as,
\begin{align}
\ci &= \frac{1}{\sqrt{\ki\kbi}} \int \frac{1}{2}\left [ \left | \frac{\Af  + \Abf}{\sqrt{2}} \right |^2  -  \left | \frac{\Af  - \Abf}{\sqrt{2}} \right |^2 \right ]  \denStatesf \mathrm{d} \pspointf , \\
 &= \frac{\Gamma[ \D^{\CPp} \to f ] - \Gamma[ \D^{\CPm} \to f ]}{ 2 \sqrt{ \Gamma[ \Dz \to f ] \Gamma[ \Dzb \to f ] } }. \label{eqn:citoy}
\end{align}
Therefore if $g$ represents all final states, $\cj = (\Gamma_{1} - \Gamma_{2})/2\Gamma_D = \ymix$.

Although the decay $\BtoDK, \decay{\D}{f}$ is not measured in this paper, it is important to consider its decay rate so that the \decay{\D}{\fourpi} binning schemes defined in \secref{sec:binning} give optimum sensitivity to $\gamma$ in a future measurement of \BtoDK decays. 
The ratio of $\BmtoDzbKm$ to $\BmtoDzKm$ amplitudes is given by $A(\BmtoDzbKm)/A(\BmtoDzKm) = r_B e^{i(  \delta_B - \gamma)}$, where $r_B$ is that ratio of their magnitudes, and $\delta_B$ is the strong phase difference.
The $\BtoDK, \D\to \fsp$ decay rates are then given by,
\begin{align}
\Gamma [  \BmtoDKm, \D\to\fsp ] &\propto \left | \Abf \right |^2  r_B^2 + \left | \Af \right |^2  + 2 \left | \Af \Abf \right | \left [ x_- \cos( \Deldelf )  + y_- \sin( \Deldelf ) \right ], \label{eqn:bmtodkunbinned} \\
\Gamma [  \BptoDKp, \D\to\fsp ] &\propto \left | \Af \right |^2  r_B^2 + \left | \Abf \right |^2  + 2 \left | \Af \Abf \right | \left [ x_+ \cos( \Deldelf )  + y_+ \sin( \Deldelf ) \right ], \label{eqn:bptodkunbinned}
\end{align}
where $x_\pm = r_B \cos (\delta_B \pm \gamma)$ and  $y_\pm = r_B \sin (\delta_B \pm \gamma)$.
Integrating this expression over a phase space bin $i$ then gives,
\begin{align}
\Gamma [  \BmtoDKm, \D\to\fsi ] &\propto \kbi r_B^2 + \ki + 2\sqrt{ \ki \kbi } (\ci x_- + \si y_-), \label{eqn:bmtodkbinned} \\
\Gamma [  \BptoDKp, \D\to\fsi ] &\propto \ki r_B^2 + \kbi + 2\sqrt{ \ki \kbi } (\ci x_+ - \si y_+). \label{eqn:bptodkbinned}
\end{align}

\section{Amplitude model for \decay{\Dz}{\fourpi} decays }
\label{sec:model}
An amplitude model is used to define how the five-dimensional phase space is divided into bins.
Such a model has recently become available~\cite{fourpimodel}, which was determined from a fit to flavour tagged \decay{\Dz}{\fourpi} decays collected by the \cleo-c experiment. 
To construct the total amplitude, the isobar approach was used, which assumes the decay can be factorised into consecutive two-body decay amplitudes. 
The dominant contributions to the model are \decay{\Dz}{a_1(1260)^+ \pim}, \decay{\Dz}{\sigma f_0(1370)} and \decay{\Dz}{\rho\rho}. 
In addition to the main (`nominal') model, Ref.~\cite{fourpimodel} also includes a further 8 alternative models which use a different set of amplitude components - these are used for systematic studies.

Since \CP conservation in \decay{\D}{\fourpi} decays is assumed, the \decay{\Dz}{\fourpi} model implies the \decay{\Dzb}{\fourpi} model, since $\Abf \equiv \Afb$. 
Here \pspointbarf is the \CP conjugate point of \pspointf, which has all charges reversed ($C$) and three-momenta flipped ($P$). 
The assumption of \CP conservation in \decay{\D}{\fourpi} decays is explicitly tested in Ref.~\cite{fourpimodel} by determining \Af and \Abf independently from samples of \Dz and \Dzb tagged decays, respectively.
The results are consistent with the \CP conservation hypothesis.

\section{Binning}
\label{sec:binning}

The definition of the \fourpi phase space bins strongly influences sensitivity to $\gamma$ in $\BtoDK, \decay{\D}{\fourpi}$ decays.
To best exploit the symmetries of the self-conjugate \fourpi final state, phase space bins are defined in pairs that map to each other under the \CP operation. 
The bins are labeled such that bin $+i$ is paired with bin $-i$, therefore, for any point \pspointf in $+i$, the \CP conjugate point $\pspointbarf$ will fall into bin $-i$.
This choice of binning means that the following relations exist between the hadronic parameters of $+i$ and $-i$ bins:
$\kmi \equiv \kbi$, $\kbmi \equiv \ki$, $\cmi \equiv \ci$ and $\smi \equiv -\si$.

Since the relative magnitude and phase of \Afourpi and \Abfourpi varies over the $\decay{\D}{\fourpi}$ phase space, so will the relative size of the interference term in $\BtoDK, \decay{\D}{\fourpi}$ decays.  
If a single bin contains regions of phase space with differing levels of interference (for example, constructive and destructive interference) the overall interference is diluted, and the sensitivity to $\gamma$ is reduced. 
It is therefore preferable for both $r^{4\pi}_{\mathbf{p}} = |\Afourpi / \Abfourpi|$ and $\Deldelfourpi= \arg (\Afourpi / \Abfourpi)$ to be approximately constant within each bin.
This is possible by using an amplitude model to assign each point in phase space a value of $r^{4\pi}_{\mathbf{p}}$ and \Deldelfourpi, which are used to determine the bin number.
Although a model is used to determine the bin number, this will not introduce any model-dependent systematic uncertainties, since the hadronic parameters will still be determined model-independently. 
An incorrect model will only lead to a non-optimal binning, and an increased \emph{statistical} uncertainty. 

\begin{figure}[h]
\centering
      \includegraphics[width=0.40\textwidth]{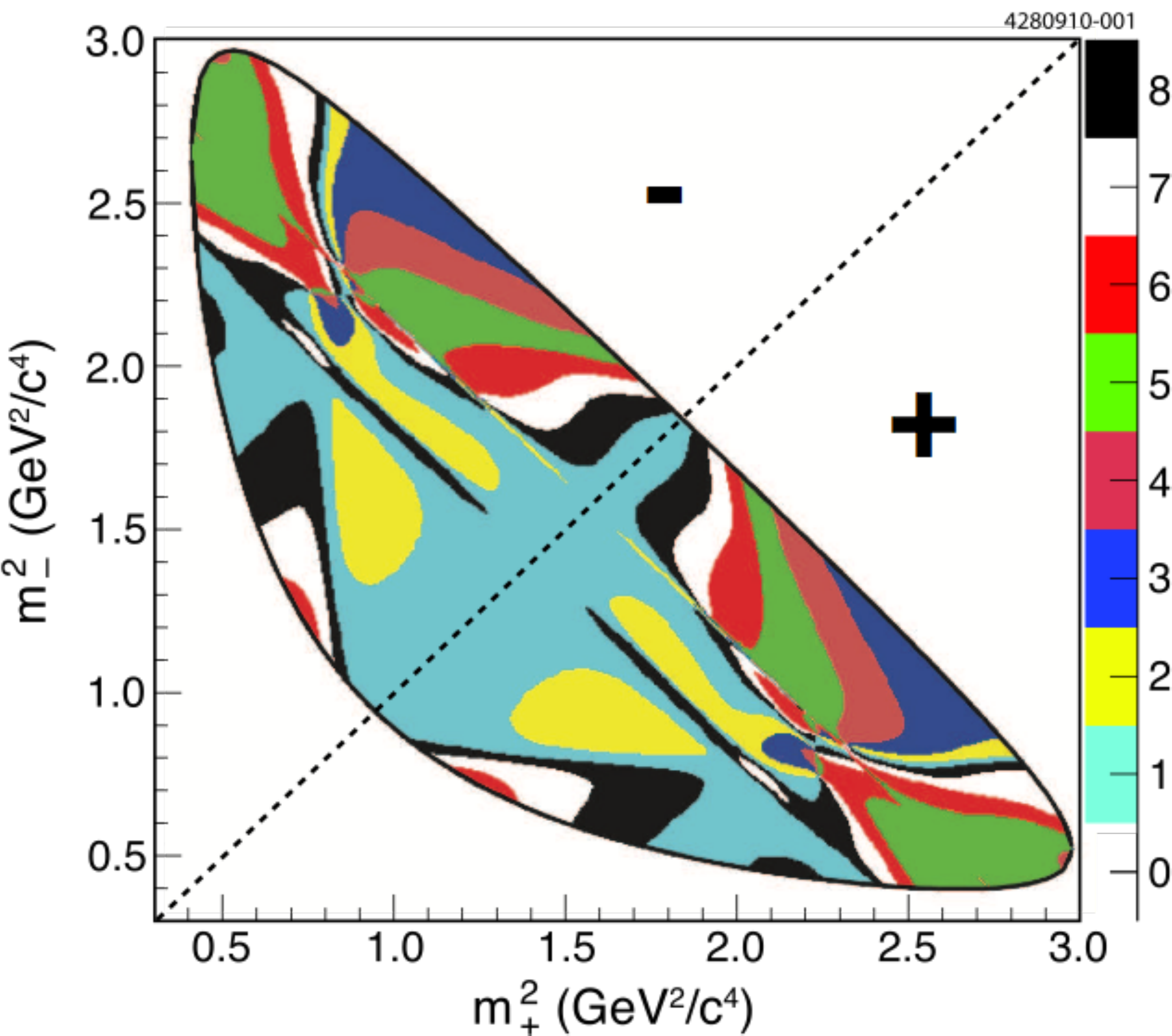}  
\caption{ The equal $\Delta\delta_D$ binning from Ref.~\cite{KzPiPiCiSi}. The absolute bin number, $|i|$, is indicated by the colouring. The positive bins are defined in the region $m^2_+ > m^2_-$, and the negative bins in the region $m^2_+ < m^2_-$. \label{fig:kspipiBinning} }
\end{figure}

Before discussing the \decay{\D}{\fourpi} binning scheme used in this paper, it is informative to review previous work on the final state \KSpipi in Ref.~\cite{KzPiPiCiSi}.
This decay has a two-dimensional phase space (the Dalitz plot) which can be parameterised by the variables $m^2_+=m^2(\KS\pip)$ and $m^2_-=m^2(\KS\pim)$.
The region $m^2_+ > m^2_-$ is divided into $\mathcal{N}$ bins, labelled $+1$ to $+\mathcal{N}$,
which are reflected over the line $m^2_+ = m^2_-$ to obtain the $-\mathcal{N}$ to $-1$ bins (a reflection over this line is equivalent to \CP).
Using the line $m^2_+ = m^2_-$ to divide the Dalitz plot 
is a good choice since most Cabibbo favoured (CF) amplitudes, such as \decay{\Dz}{\Kstarm\pip}, fall in the region $m^2_+ > m^2_-$, whereas most Doubly Cabibbo suppressed (DCS) amplitudes fall into the region $m^2_- > m^2_+$.
This is beneficial since it makes $r^{\KS\pi\pi}_{\mathbf{p}}$ consistently large (small) over the $+i$ ($-i$) bins.
To determine the absolute bin numbers, the model prediction for \Deldelkspipi is divided into 8 equal regions.
The \KSpipi binning scheme for $\mathcal{N}=8$ is shown in \figref{fig:kspipiBinning}.
The authors of Ref.~\cite{KzPiPiCiSi} also provide a fine granularity lookup table that describes the binning shown in \figref{fig:kspipiBinning}; this is very useful because the amplitude model is not necessary to reproduce the binning scheme.  
A similar idea will be used for the \fourpi binning schemes.

\subsection{\KS veto bin}
\label{sec:ksveto}

A large peaking background to \decay{\D}{\fourpi} decays is \decay{\D}{\KS\pip\pim} where $\decay{\KS}{\pip\pim}$.
In order to remove the majority of this background, a \KS-veto bin is included in all \decay{\D}{\fourpi} binning schemes that are later described in \secsref{sec:eqDeldelBinning}{sec:optimalBinning}. 
The region of phase space that contains any $\pip\pim$ pair satisfying $480\mev < m(\pip\pim) < 505\mev$ is designated as the \KS-veto bin.
Using the nominal \decay{\Dz}{\fourpi} amplitude model, the \KS-veto was found to remove approximately 10\% of signal.


\subsection{Equal / variable \Deldelfourpi binning}
\label{sec:eqDeldelBinning}

When comparing the \KSpipi to the \fourpi final state, one clear difference is the decay amplitudes that contribute. 
As discussed, \KSpipi has contributions from both CF and DCS amplitudes, whereas \fourpi only has contributions from singly Cabibbo suppressed (SCS) amplitudes. 
This means that there is no clear way to divide the phase space, like the line $m^2_+ = m^2_-$ in the \KSpipi Dalitz plot.
A different approach is therefore followed.
The baseline amplitude model from Ref.~\cite{fourpimodel} is used to assign each point \pspointf a value of \Deldelfourpi, then a bin number is assigned using,
\begin{align}
 +i &:=  \forall \mathbf{p} : \phantom{-}\delta_{i-1} <  \Deldelfourpi < \phantom{-} \delta_i\notag \\
 -i &:=  \forall \mathbf{p} : -\delta_{i-1} > \Deldelfourpi > -\delta_i \label{eqn:equalDelBins}
\end{align}
where $\delta_0 \equiv 0$, $\delta_{\mathcal{N}} \equiv \pi$ and $\delta_i < \delta_{i + 1}$.
This automatically fulfils the requirement that bin $+i$ maps to bin $-i$ under \CP, since $\Deldelfourpi \equiv -\Deldelbfourpi$.
The values of $\delta_i$ are chosen using two methods: 
the equal \Deldelfourpi binning, for which $\delta_i = i\pi/\mathcal{N}$; 
and the variable \Deldelfourpi binning, for which the values of $\delta_i$ are chosen such that $\kifourpi + \kbifourpi$ is approximately the same in each bin. 

Since amplitude models are difficult to reproduce, it is desirable to have a model-implementation-independent
binning scheme.
This is possible by splitting the five dimensional phase space into many small hypervolumes, each of which is assigned a bin number. 
The overall bin is then formed from the combination of all hypervolumes with that bin number.
To create a model-implementation-independent binning scheme, referred to as a hyper-binning, a set of variables must be defined that parameterises the five-dimensional phase space of \D\to\fourpi decays. 
The variables $\{ m_+, m_-, \cos \theta_+, \cos \theta_-, \phi \}$ are chosen, where $m_+$ ($m_-$) is the invariant mass of the $\pip\pip$ ($\pim\pim$) pair; $\theta_+$ ($\theta_-$) is the helicity angle of the $\pip\pip$ ($\pim\pim$) pair; and $\phi$ is the angle between the $\pip\pip$ and $\pim\pim$ decay planes (a full definition of these variables can be found in \appref{app:helicityVars}).
Since the hyper-binning is most easily implemented with square phase space boundaries, the following transformation is made,
\begin{align}
m_{\pm}' = m_\pm + \delta \ \ \ \mathrm{where} \ \ \ \delta = \mathrm{min}\{ m_{+}, m_{-} \} - m_{\mathrm{min}}, \label{eqn:mprime}
\end{align}
where $m_{\mathrm{min}}$ is the minimum value kinematically possible for $m_{+}$ (or $m_{-}$).
When using the variables $\{ m_+', m_-', \cos \theta_+, \cos \theta_-, \phi \}$, the kinematically allowed region of phase space is a hypervolume defined by the corners $\{ m_{\mathrm{min}}$, $m_{\mathrm{min}}$, $-1$, $-1$, $-\pi \}$ and $\{ m_{\mathrm{max}}$, $m_{\mathrm{max}}$, $1$, $1$, $\pi \}$.
This set of variables has been chosen to exploit the symmetries of the system, these being \CP-conjugation and identical particle interchange:
\begin{align}
\mathcal{CP} \{  m_{+}', m_{-}', \cos\theta_{+}, \cos\theta_{-}, \phi   \} &\to \{  m_{-}', m_{+}', \cos\theta_{-}, \cos\theta_{+}, -\phi   \} \label{eqn:CPtrans} \\
[ \pi^{+}_{1} \leftrightarrow \pi^{+}_{2} ]  \{  m_{+}', m_{-}', \cos\theta_{+}, \cos\theta_{-}, \phi   \} &\to \{  m_{+}', m_{-}', -\cos\theta_{+}, \cos\theta_{-}, \phi - \pi   \} \\
[ \pi^{-}_{1} \leftrightarrow \pi^{-}_{2} ]  \{  m_{+}', m_{-}', \cos\theta_{+}, \cos\theta_{-}, \phi   \} &\to \{  m_{+}', m_{-}', \cos\theta_{+}, -\cos\theta_{-}, \phi - \pi   \} 
\end{align}
The symmetries for identical particle exchange allow the phase space to be `folded' twice along the lines $\cos\theta_{+} = 0$ and $\cos\theta_{-} = 0$, reducing the phase space volume by a factor of four. 
A further folding is also possible by considering the \CP operation; for a point \pspointf with bin number $i$, it follows that point \pspointbarf has bin number $-i$. 

\begin{figure}[h]
     \centering

      \hspace{-14mm}
      \includegraphics[width=0.39\textwidth]{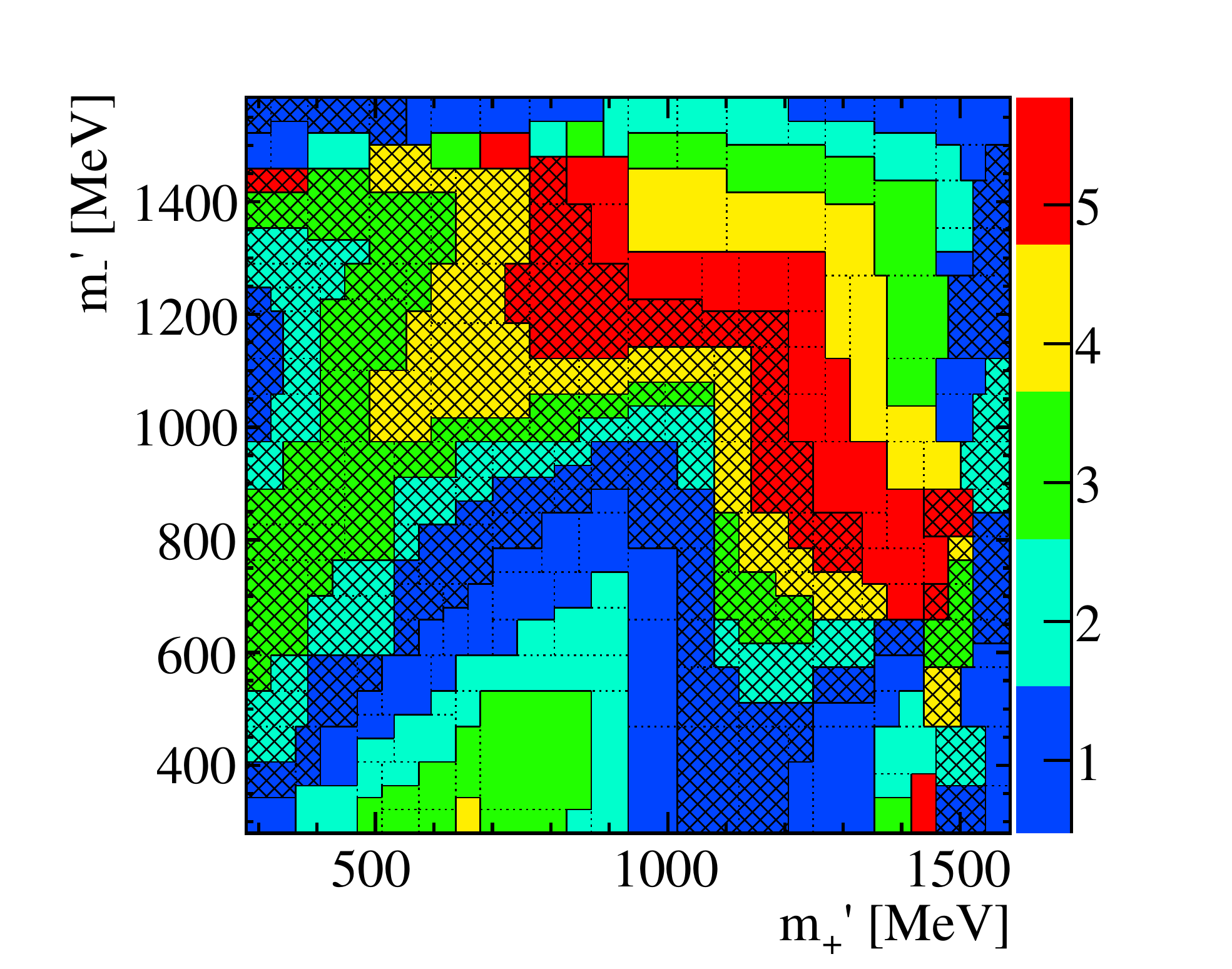}\hspace{-8mm}
      \includegraphics[width=0.39\textwidth]{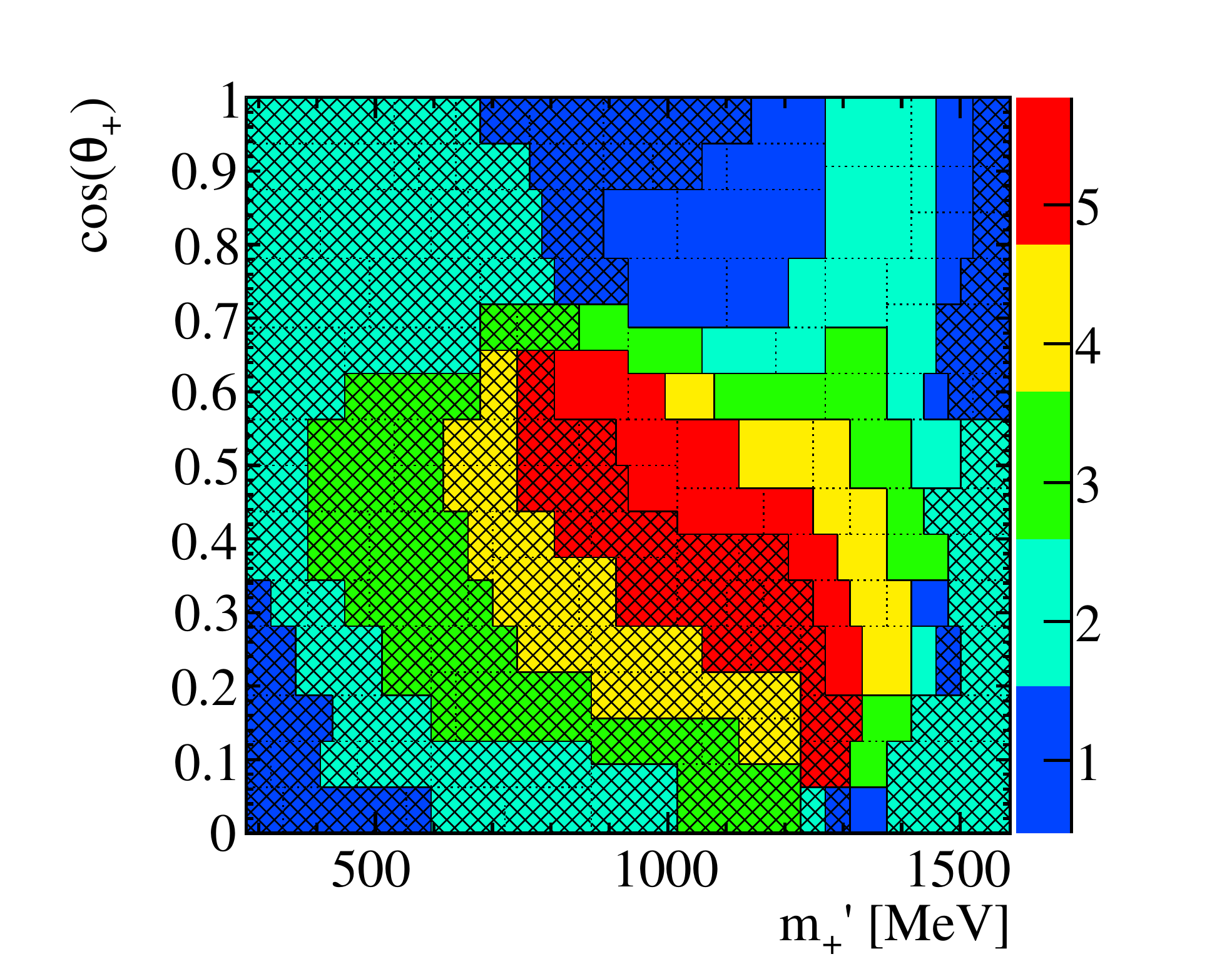}\hspace{-8mm}  
      \includegraphics[width=0.39\textwidth]{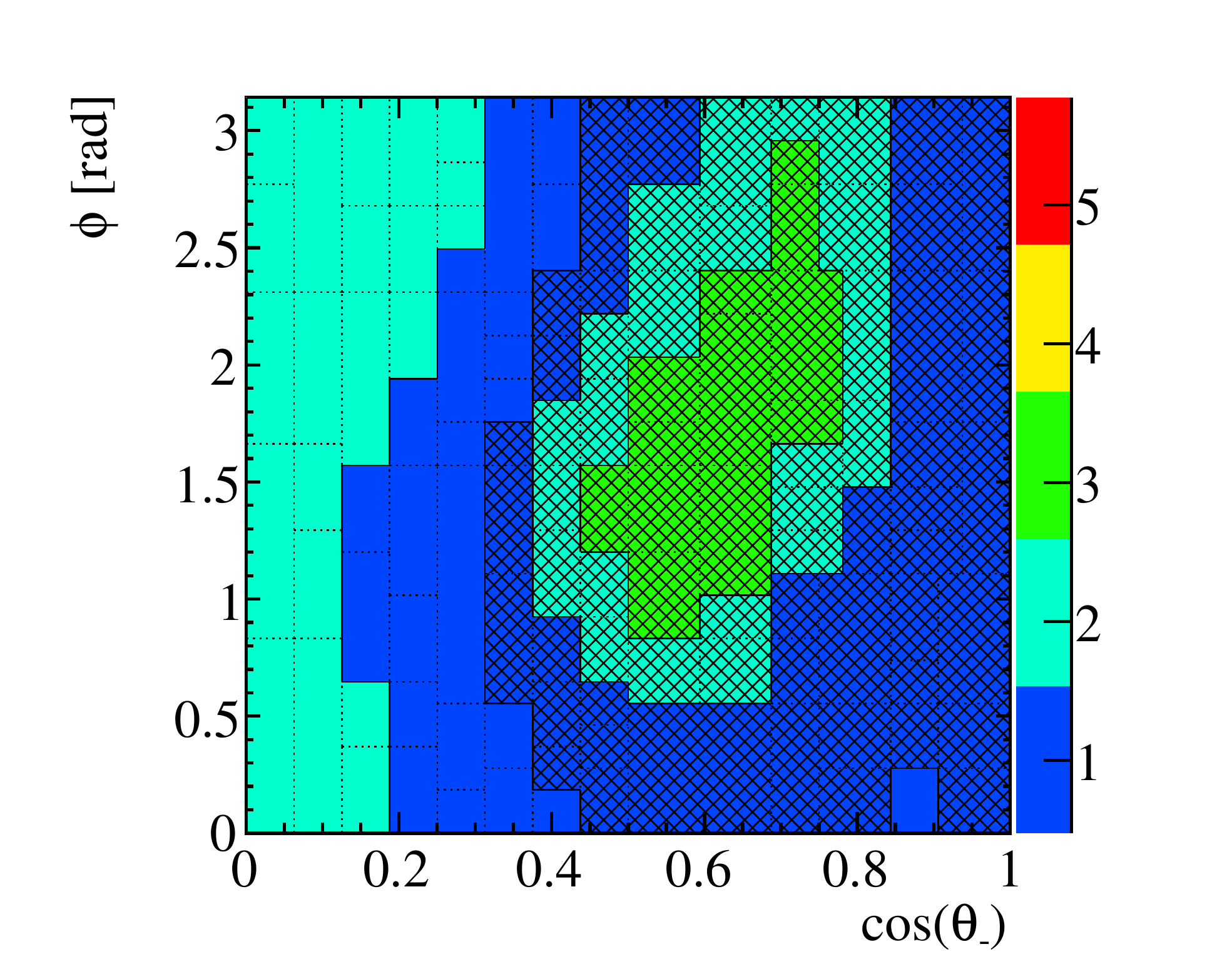}\hspace{-8mm}  
       
\caption{  Two-dimensional slices of the \decay{\D}{\fourpi} phase space showing the equal $\Deldelfourpi$ binning with $\mathcal{N}=5$. The colour denotes the absolute value of the bin number, and the cross hatching denotes a negative bin number.   \label{fig:hyperBinningSlices} }
\end{figure}

An adaptive binning algorithm is used to create a hyper-binning scheme. 
At the beginning of the algorithm one hypervolume is defined with corners $\{ m_{\mathrm{min}}$, $m_{\mathrm{min}}$, $0$, $0$, $0 \}$ and $\{ m_{\mathrm{max}}$, $m_{\mathrm{max}}$, $1$, $1$, $\pi \}$.
At each iteration of the algorithm, the hypervolumes from the previous iteration are split in two, choosing to split in the dimension that has the fastest varying \Deldelfourpi, and picking a split point that is as close as possible to one of the bin boundaries defined in \eqnref{eqn:equalDelBins}. 
The algorithm runs until either: splitting a hypervolume will always result in two hypervolumes with the same bin number; splitting a hypervolume will always result in a hypervolume that has an edge length narrower than the minimum allowed.
Several minimum edge lengths were tested and the values $\{ 39\mev, 39\mev, 0.06, 0.06, 0.19\rad \}$ were chosen since this results in a reasonable number of volumes ($\sim 250,000$) while reproducing the parameters $c_i$ and $s_i$ to within $2\%$ compared to a binning scheme that uses the model directly.
It is possible to visualise the hyper-binning by taking two-dimensional slices of the five-dimensional phase space.
Some examples are shown for the equal $\Deldelfourpi$ binning with $\mathcal{N}=5$ in \figref{fig:hyperBinningSlices}.
The full binning schemes used in this paper are provided in both ASCII and \root format as supplementary material. 

\subsection{Model predictions of the hadronic parameters}
\label{sec:modPred}

\begin{figure}[h]
     \centering
      \hspace{-12mm}
      \includegraphics[width=0.39\textwidth]{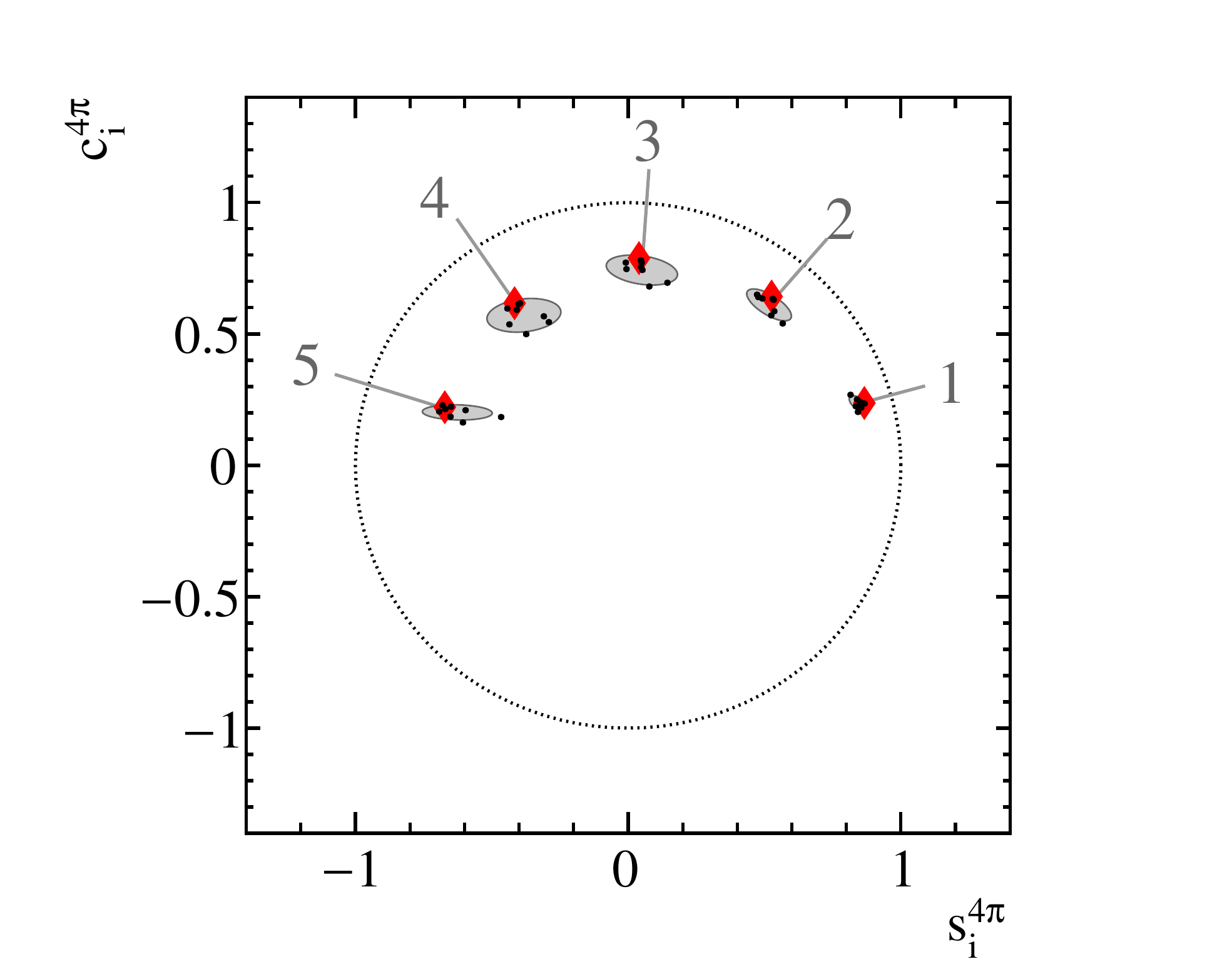} \hspace{-12mm}
      \includegraphics[width=0.39\textwidth]{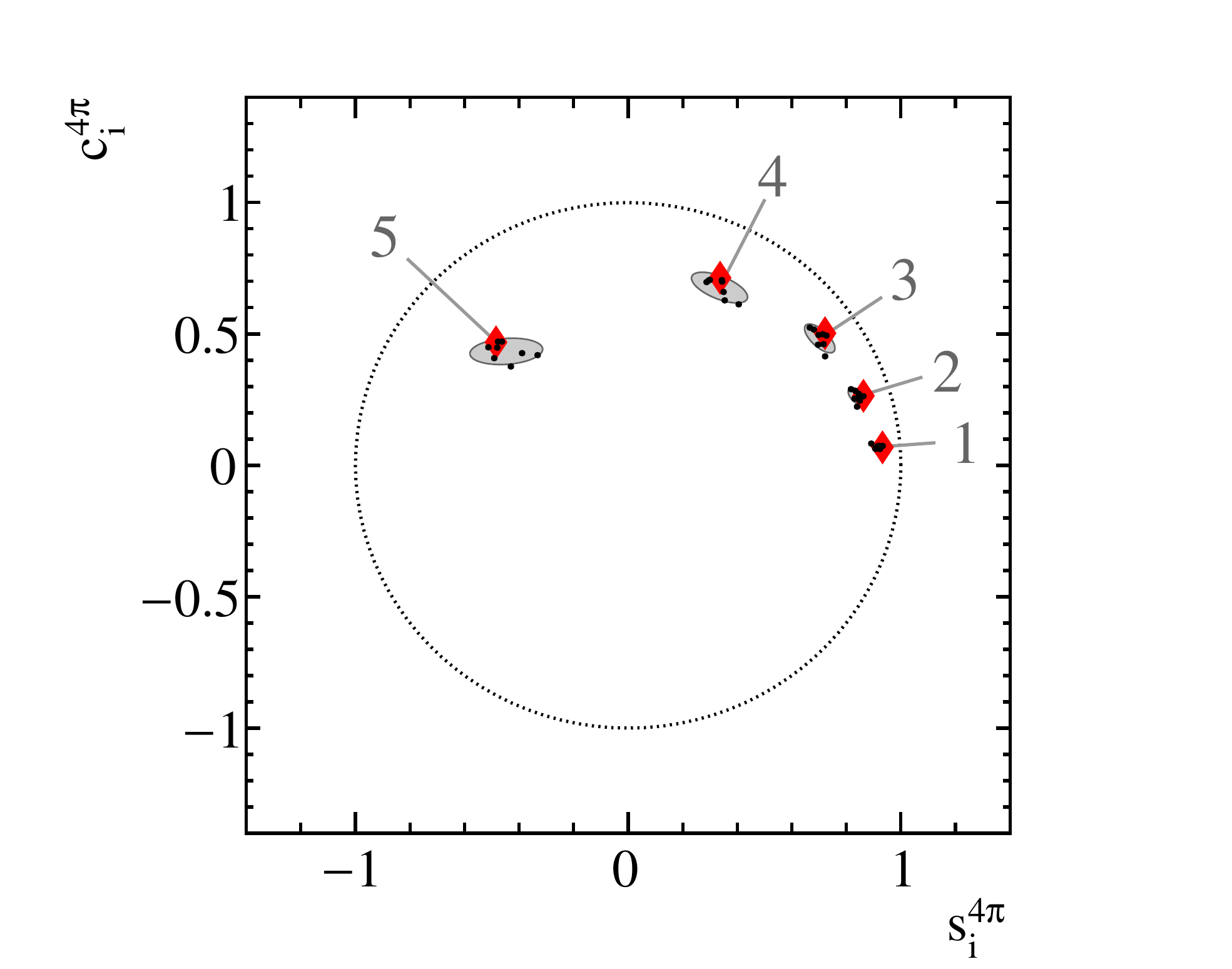} \hspace{-12mm}  
      \includegraphics[width=0.39\textwidth]{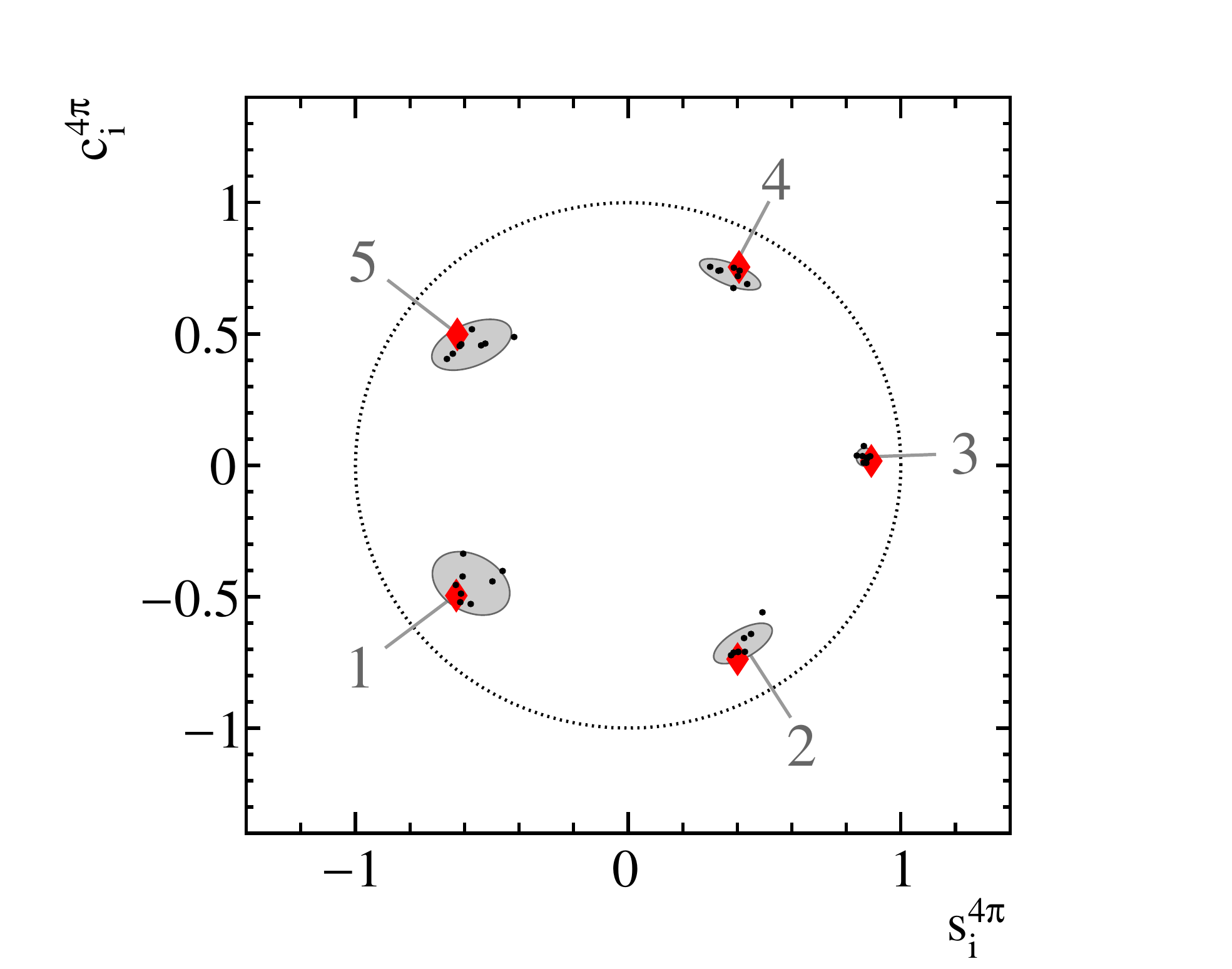} \hspace{-12mm}
      
      \includegraphics[width=0.32\textwidth]{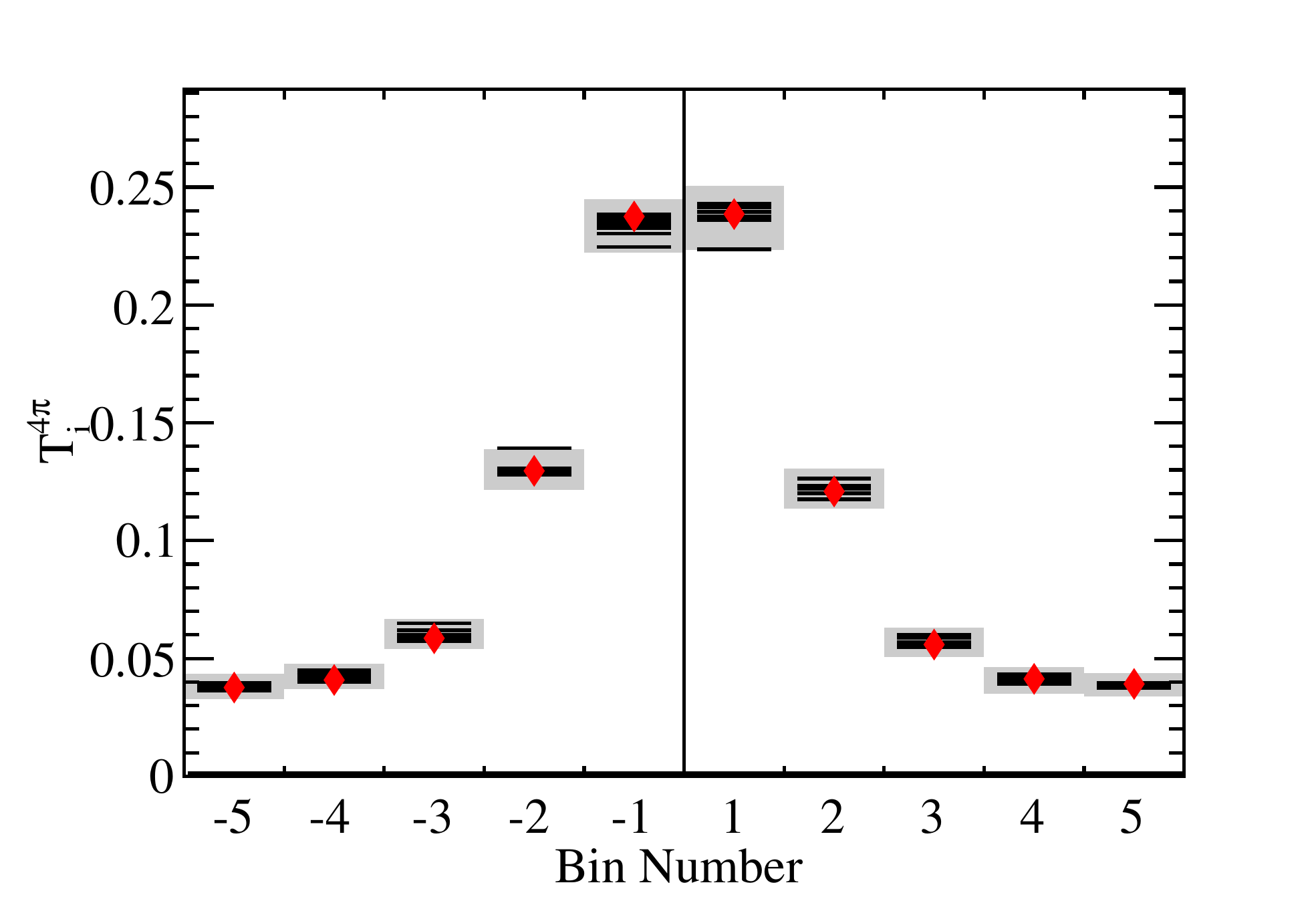}
      \includegraphics[width=0.32\textwidth]{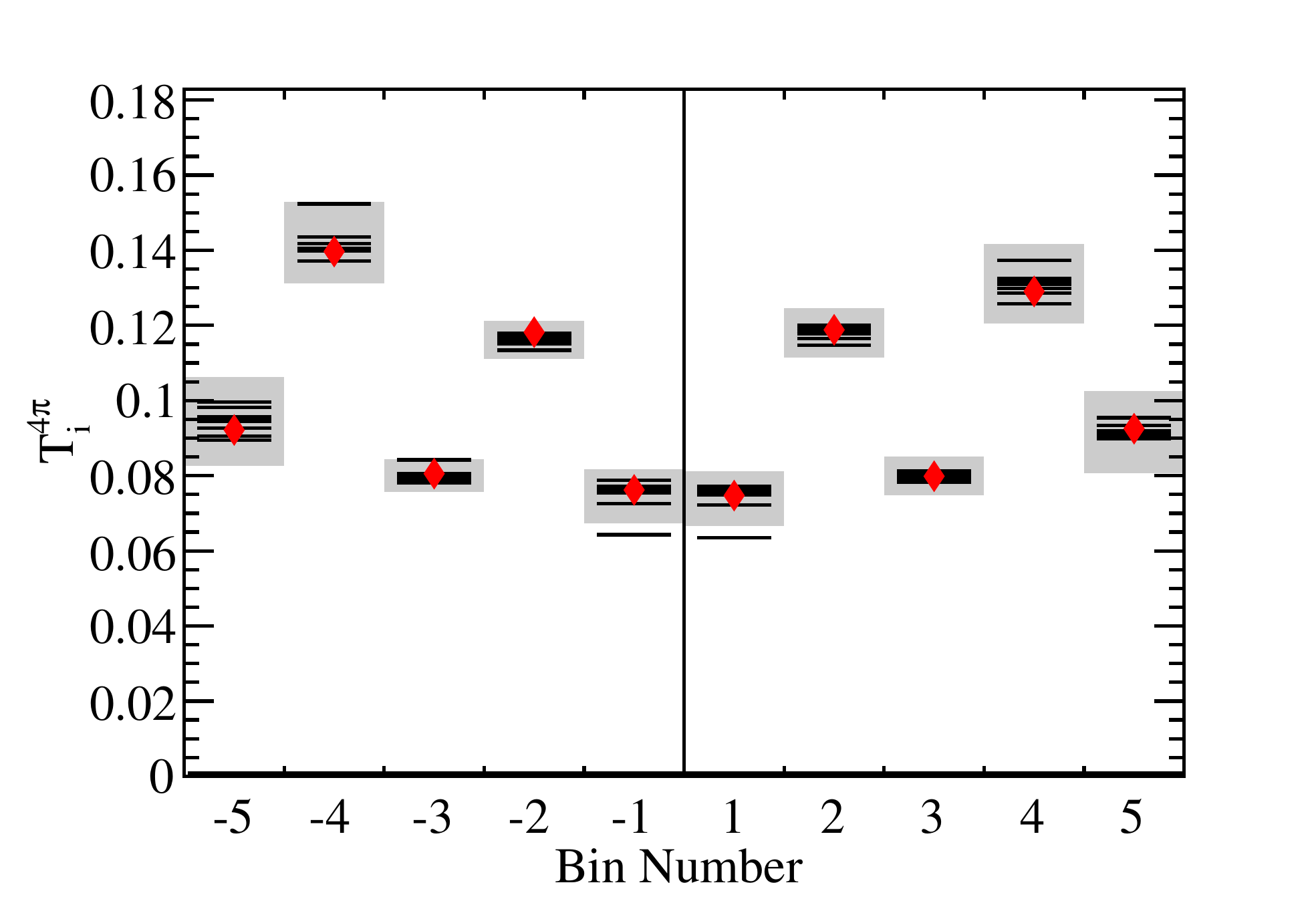}        
      \includegraphics[width=0.32\textwidth]{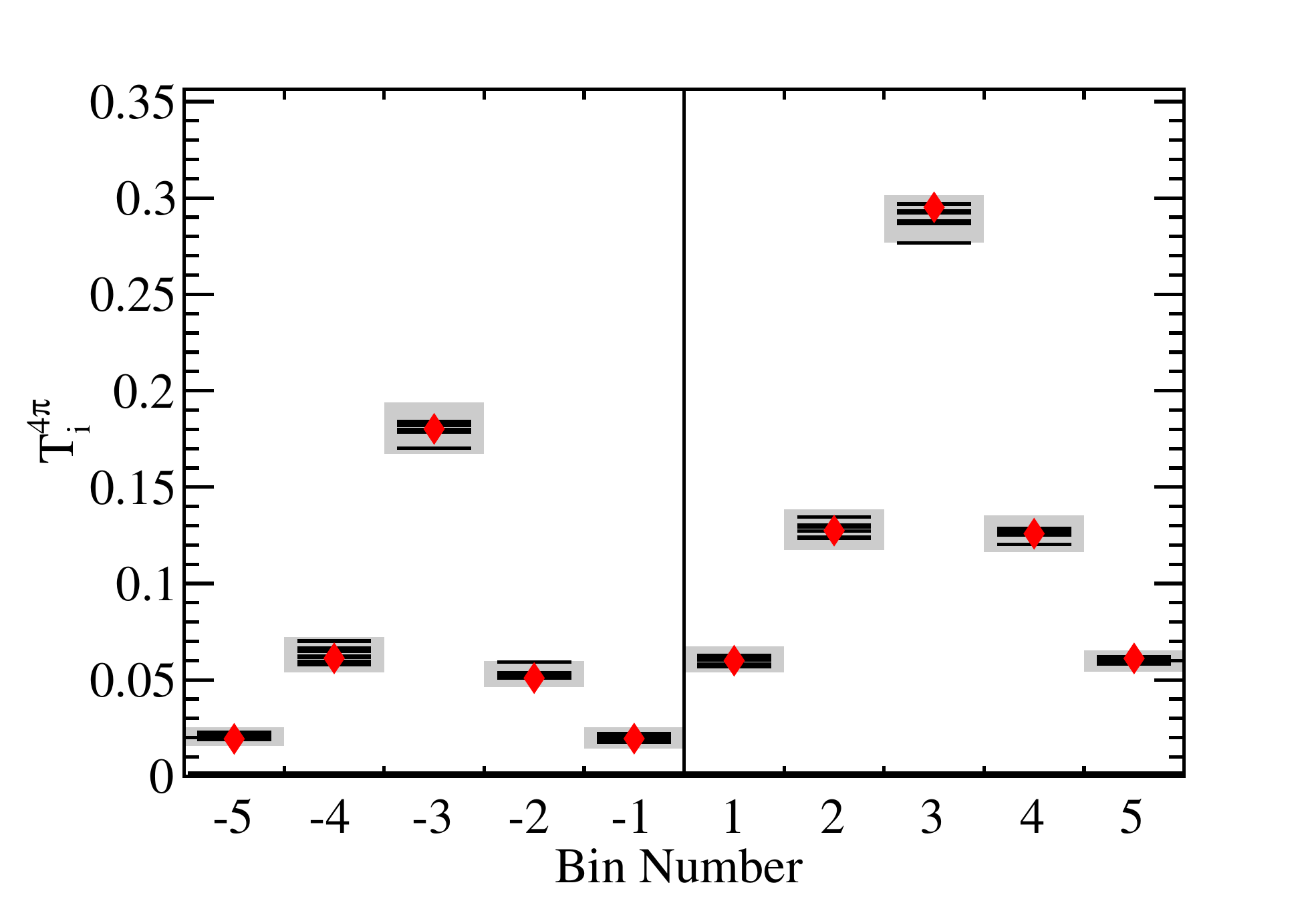}    
            
\caption{  The model predictions of the hadronic parameters for the (left) equal $\Deldelfourpi$ binning (centre) variable $\Deldelfourpi$ binning (right) alternative binning, with $\mathcal{N}=5$. (top row) The red diamonds (black dots) show the \ci and \si predictions calculated from the baseline model (alternative models). The grey shaded ellipses shows \ci and \si model prediction and uncertainty as described in the text. (bottom row) The (red diamonds) black horizontal lines show the \Fri and \Frbi predictions calculated from the baseline model (alternative models). The grey shaded band shows the \Fri and \Frbi model prediction and uncertainty as described in the text. \label{fig:hadPars1} }
\end{figure}

Using the integral expressions in \eqnsref{eqn:kkbar}{eqn:si} it is possible to calculate the hadronic parameters for a given amplitude model and binning scheme.
This is done using the baseline and alternative amplitude models given in Ref.~\cite{fourpimodel}.
Since the baseline-model is used to determine the \decay{\D}{\fourpi} binning schemes, using the hadronic parameters predicted with this model could result in a bias.
Therefore, the arithmetic-mean of the hadronic parameters from all alternative models is used as the model prediction, and the covariance of the results is used to determine a model-uncertainty.
To determine the statistical and systematic uncertainties, the hadronic parameters are calculated many times using the baseline model, each time varying the model parameters within their statistical and systematic uncertainties.
The covariance of the results is used to determine a combined statistical and systematic uncertainty, which is added to the model-uncertainty in quadrature to obtain the total uncertainty.
The model predictions for the equal / variable $\Deldelfourpi$ binning are shown in \figref{fig:hadPars1}.

\subsection{Alternate binning}

One drawback of the $\Deldelfourpi$ binning schemes is that the variation of $r^{4\pi}_{\mathbf{p}}$ across each bin is not considered, leading to $\ki \sim \kbi$, as seen in \figref{fig:hadPars1}. 
This means that the interference term in the \BmtoDKm decay rate, given in \eqnref{eqn:bmtodkbinned}, is relatively small in all phase space bins.
Ideally, one would choose to have $r^{4\pi}_{\mathbf{p}} \ll 1$ in half of the phase space bins, enhancing the interference in these regions (and therefore the sensitivity to $\gamma$). 
The $r^{4\pi}_{\mathbf{p}} \ll 1$ condition is satisfied in the \KSpipi final state, where many bins are dominated by DCS amplitudes.
Although the SCS \fourpi final state has no clear line of symmetry that divides favoured from suppressed phase space regions, the amplitude model can be used to define such a split. 
Any point \pspointf that that satisfies $r^{4\pi}_{\mathbf{p}} < 1$ is assigned a bin number $+i$, whereas any satisfying $r^{4\pi}_{\mathbf{p}} > 1$ is assigned a bin number $-i$. 
The $+i$ bin numbers are assigned using,
\begin{align}
 +i &:=  \forall \mathbf{p} : \bigg [ -\pi + \frac{2\pi}{\mathcal{N}} (i - 1) < + \Deldelfourpi < -\pi + \frac{2\pi}{\mathcal{N}} i \bigg ] \& \bigg [  r^{4\pi}_{\mathbf{p}} > 1  \bigg ], \label{eqn:kspipistylebins} 
\end{align}
which also uniquely defines the $-i$ bin numbers \ie
\begin{align}
  -i &:=  \forall \mathbf{p} : \bigg [ -\pi + \frac{2\pi}{\mathcal{N}} (i - 1) <  - \Deldelfourpi < -\pi + \frac{2\pi}{\mathcal{N}} i \bigg ] \& \bigg [  r^{4\pi}_{\mathbf{p}} < 1  \bigg ] .
\end{align}
The same hypervolumes from the equal $\Deldelfourpi$ binning schemes are used for the alternative binning schemes, but the bin number associated to each hypervolume is reassigned using \eqnref{eqn:kspipistylebins}.
The model predictions for the alternative binning with $\mathcal{N}=5$ is shown in \figref{fig:hadPars1}.

\subsection{Optimal binning}
\label{sec:optimalBinning}

To determine how sensitive a binning scheme is to a measurement of $\gamma$, the $Q_{\pm}$ values are defined~\cite{Bondar:2008hh},
\begin{align}
Q^2_{\pm} =
\frac{ 
\mathlarger{\sum_i} \left ( \frac{1}{\sqrt{N_{\Bpm}^{i}}} \frac{\mathrm{d}N_{\Bpm}^{i}}{\mathrm{d}x_\pm} \right )^2 
+ \left ( \frac{1}{\sqrt{N_{\Bpm}^{i}}} \frac{\mathrm{d}N_{\Bpm}^{i}}{\mathrm{d}y_\pm} \right )^2 
}
{ 
\mathlarger{\int_{D}}  \left [  \left ( \frac{1}{ \sqrt{\Gamma_{\Bpm}(\mathbf{p}) } } \frac{ \mathrm{d} \Gamma_{\Bpm}(\mathbf{p})  }{ \mathrm{d} x_\pm} \right )^2  
+ \left ( \frac{1}{ \sqrt{\Gamma_{\Bpm}(\mathbf{p}) } } \frac{ \mathrm{d} \Gamma_{\Bpm}(\mathbf{p})  }{ \mathrm{d} y_\pm} \right )^2  \right ] \mathrm{d} \mathbf{p} 
}, \label{eqn:qval}
\end{align}
where $N_{\Bpm}^{i}$ is the number of $\BtoDK, \decay{\D}{f}$ decays expected in bin $i$ (\eqnref{eqn:bmtodkunbinned} and \eqnref{eqn:bptodkunbinned}), and $\Gamma_{\Bpm}(\mathbf{p})$ gives the differential decay rate (\eqnref{eqn:bmtodkbinned} and \eqnref{eqn:bptodkbinned}).
The value of $Q_{\pm}$ gives the statistical sensitivity on the parameters $x_\pm$ and $y_\pm$ from a binned analysis of $\BtoDK, \decay{\D}{f}$ decays, divided by the statistical sensitivity from an analysis with infinitely many bins.
Substituting \eqnsref{eqn:bmtodkunbinned}{eqn:bptodkbinned} into \eqnref{eqn:qval} gives,
\begin{align}
Q^{2}_{\pm}=  1  -  \sum_i \frac{  \ki \kbi \left (  1 - (\ci)^2 - (\si)^2  \right )  }{ N_{\Bpm}^{i} } \bigg{/} \sum_i \ki
\end{align}
The $Q$ value, $Q^2 = \half(Q^{2}_{+} + Q^{2}_{-})$, is then used to rank the sensitivity of different binning schemes to $\gamma$.
The values $\delta_B = 140^\circ$, $\gamma = 70^\circ$ and $r_B = 0.1$ are used to determine $Q$. 
For the optimisation of the \KSpipi binning schemes in Ref.~\cite{KzPiPiCiSi}, a simplified $Q$ value was used where it was assumed $r_B = 0$.
Since the relative size of \ki and \kbi does not need to be optimised for \KSpipi (due to the division at $m_+^2 = m_-^2$), this assumption works well. 
For \fourpi decays, the simplified expression gives solutions where $\ki\sim\kbi$, so the full expression is used instead.

An iterative algorithm is used to take any hyper-binning scheme (i.e. a collection of hypervolumes, each with a bin number, that span the \decay{\D}{\fourpi} phase space) and reassign the bin numbers in order to maximise the model-prediction of $Q$. 
Each iteration of the algorithm involves looping over every hypervolume in the hyper-binning. 
For each hypervolume, every possible bin number ($-\mathcal{N}, ..., -1, +1, +\mathcal{N}$) is assigned, and $Q$ is recalculated; the bin number that gave the largest $Q$ is then kept. 
The algorithm keeps running until no hypervolumes change their bin number, typically taking around $20-50$ iterations. 

Since the number of free parameters being optimised is so large, it is unavoidable that the optimisation procedure will fall into a local maximum. 
The outcome is therefore dependent on the starting values (i.e. the bin numbers assigned to each hypervolume).
The starting bin numbers are therefore assigned using two methods: the equal \Deldelfourpi binning scheme (\eqnref{eqn:equalDelBins}); and the alternate binning scheme (\eqnref{eqn:kspipistylebins}).
The two sets of starting values give the `optimal binning' and `optimal-alternative binning', respectively. 
The set of hypervolumes used for the optimisation must have sufficient flexibility to describe the optimal binning.
For all optimal binning schemes, the hypervolumes are first taken from the equal $\Deldelfourpi$ binning scheme with $\mathcal{N} = 8$, then further divided so that,  for the sample sizes used in this paper, the probability of any single hypervolume being populated is less than $1/50$.

After running the $Q$ optimisation procedure it was found that occasionally the results had very small values of $\ki + \kbi$ for one or more bin pairs. 
For this reason a small change was made to the optimisation metric,
\begin{align}
Q^{'2} =  Q^{2} + \frac{1}{10}\sum_ {i = 1}^{\mathcal{N}}  \left\{\begin{matrix}  \ki + \kbi < t :&  \left [ \frac{ \ki + \kbi - t }{ t } \right ]^2 \\ \ki + \kbi > t : & 0 \end{matrix}\right. ,
\end{align}
where $t = \frac{2}{3\mathcal{N}} \sum_ {i = 1}^{\mathcal{N}} ( \ki + \kbi )$ is the lower threshold at which a constraint is applied to $\ki + \kbi$.

\begin{figure}[h]
     \centering

      \includegraphics[width=0.63\textwidth]{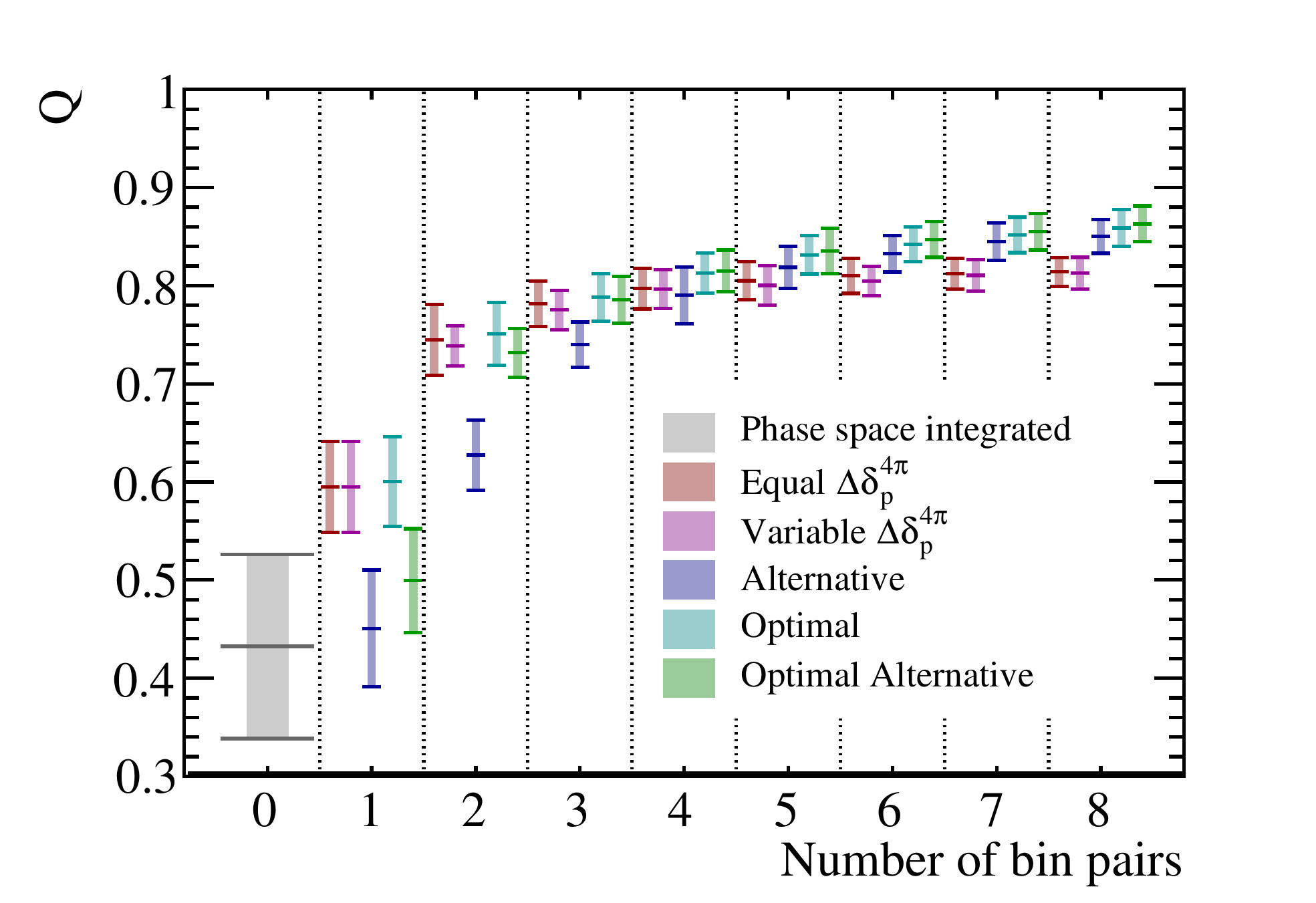}
      
\caption{ The model predictions of the $Q$ values found from all binning schemes considered in this paper. 
The uncertainties are found by varying the model predictions of the hadronic parameters within their uncertainties. \label{fig:qvals} }
      
\end{figure}

The $Q$ value for the optimal and optimal-alternative binning schemes is shown in \figref{fig:qvals} for $\mathcal{N} = 1 - 8$.
Also shown are the $Q$ values for the other binning schemes discussed in this paper.

\section{Event Selection}
\label{sec:selection}
The data set analysed consists of $e^+e^-$ collisions produced by the Cornell Electron Storage Ring (CESR) at $\sqrt{s}=3.77$~GeV corresponding to an integrated luminosity of  818~$\rm pb^{-1}$ and collected with the CLEO-c detector.  
The CLEO-c detector is described in detail elsewhere~\cite{CLEOREF1,CLEOREF2,CLEOREF3,CLEOREF4}.  
Monte Carlo (MC) simulated samples of signal decays are used to estimate selection efficiencies. 
Possible background contributions are determined from a generic $\Dz\Dzb$ simulated sample corresponding to
approximately fifteen times the integrated luminosity of the data set.  
The EVTGEN generator~\cite{EVTGEN} is used to simulate the decays.  
The detector response is modelled using the GEANT software package~\cite{GEANT}.

\Tabref{tab:listOfTags} lists all \D decay final states that are reconstructed in conjunction with a \decay{\D}{\fourpi} decay, referred to as double-tagged decays.
Underlined in \tabref{tab:listOfTags} are the \D decay final states that are also reconstructed alone, referred to as single-tagged decays. 
Unstable final state particles are reconstructed in the following decay modes: \decay{\piz}{\gamma\gamma}; \decay{\KS}{\pip\pim}; \decay{\omega}{\pip\pim\piz}; \decay{\eta}{\gamma\gamma}; \decay{\eta}{\pip\pim\piz}; and \decay{\eta'}{\eta(\gamma\gamma)\pip\pim}.
\begin{table}[h]
\centering
\begin{tabular}{ c  c }
Type & Final States \\
\hline
Flavoured & \Kenu  \\
Quasi-Flavoured & $\underline{\Kpi}$  \ \Kpipiz \ \Kpipipi \\
\CP even & $\underline{\KK} $  \ $\underline{\pipi}$ \ $\underline{\KSpizpiz}$ \ \KLpiz \ \KLomega \\
\CP odd  & $\underline{\KSpiz}$  \  $\underline{\KSomega}$ \ $\underline{\KSeta}$ \ $\underline{\KSetap}$ \\
Self-conjugate  & \KSpipi  \  \KLpipi \  $\underline{\pipipiz}$  \\
\end{tabular}
\caption{ List of all \D decay final states that are reconstructed in conjunction with a \decay{\D}{\fourpi} decay (double-tag modes). The underlined final states are also reconstructed alone (single-tag modes).  \label{tab:listOfTags}}
\end{table}

The selection procedure used for this paper is intended to be almost identical to that in Ref.~\cite{FPlusFourPi}.
The only change is to the selection criteria used to reject peaking background from \decay{\D}{\KSpipi} decays that are reconstructed as \decay{\D}{\fourpi}; henceforth referred to as \KSpipi background.
In Ref.~\cite{FPlusFourPi} any $\pip\pim$ pair with an invariant mass in the range $[0.470, 0.530] \gev$ is required to have a reconstructed vertex that is compatible with the $e^+e^-$ collision point.
In this paper, any $\pip\pim$ pair with an invariant mass in the range $[0.480, 0.505] \gev$ is rejected, regardless of its compatibility with the $e^+e^-$ collision point.
The \fourpi phase space bins defined in \secref{sec:binning} have the same region of phase space removed, so no corrections to the measured hadronic parameters are needed.
In addition to the tags in Ref.~\cite{FPlusFourPi}, this analysis also uses the flavour-tags \Kenu, and the quasi-flavour-tags \Kpi, \Kpipiz and \Kpipipi. 
These decays are selected following the same criteria as Ref.~\cite{KzPiPiCiSi}.

\begin{figure}[h]
      
      \centering
      \includegraphics[width=0.46\textwidth]{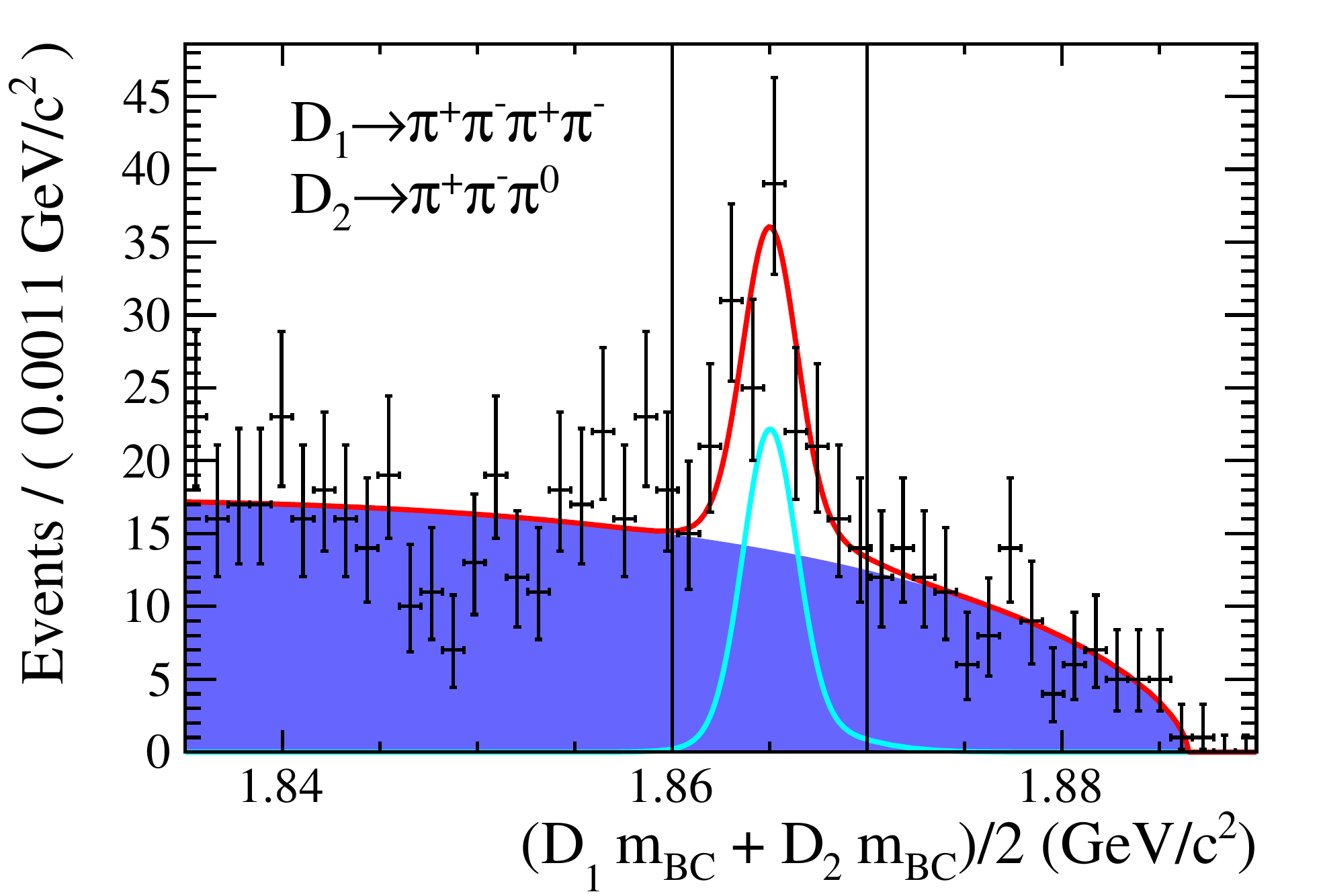}  
      \includegraphics[width=0.35\textwidth]{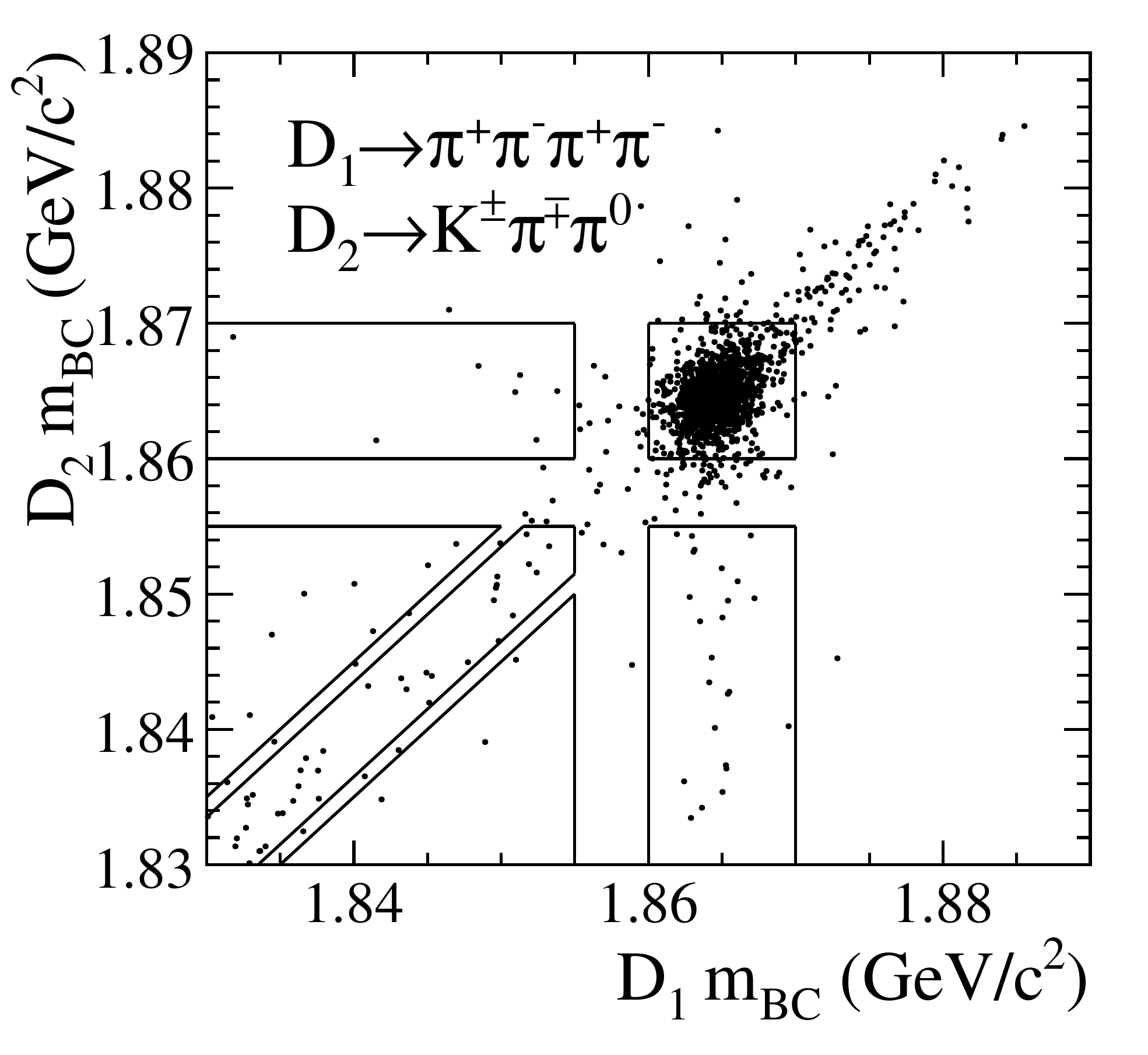}  \
      
\caption{  (left) The $m^{\mathrm{ave}}_{\mathrm{bc}}$ distribution of selected double-tagged \pipipiz candidates. Superimposed with a red line is the result of a unbinned maximum likelihood fit described in the text. The shaded purple area shows the background PDF, and the blue line shows the signal PDF. The vertical lines show the signal region. (right) Two dimensional $\D_1\,m_{\mathrm{bc}}$ vs. $\D_2\,m_{\mathrm{bc}}$ distribution of selected double tagged \Kpipiz candidates. The square box covering the range $1.86-1.87\gev$ shows the signal region, and the remaining boxes show the various sideband regions that are used to determine the combinatorial background contribution.\label{fig:fullyrecofit} }
\end{figure}

The final states that do not include a neutrino or a \KL are fully reconstructed using the beam-constrained candidate mass, $\mbc \equiv \sqrt{s /(4c^4) - \mathbf{p}_{D}^2/c^2}$, where $\mathbf{p}_{D}$ is the \D-candidate momentum, and $\Delta E \equiv E_D - \sqrt{s}/2$, where $E_D$ is the \D-candidate energy.
Requirements are first placed on the value of $\Delta E$, then \mbc is used as the discriminating variable to distinguish signal from non-peaking backgrounds. 
For double-tags that are dominated by background from continuum production of light quark-antiquark pairs (\pipi, \KK, \pipipiz and \fourpi), the signal yield is determined using an unbinned maximum likelihood fit to the average \mbc of the two \D decays, $\avembc \equiv \frac{1}{2} ( \D_1\,\mbc + \D_2\,\mbc )$.
The signal probability density function (PDF) is parameterised using the sum of a bifurcated Gaussian and a Gaussian, which have shape parameters fixed from a fit to samples of simulated signal decays\footnote{A bifurcated Gaussian has a different width below and above the mean.}. 
The background PDF is parameterised using an Argus function~\cite{ARGUS}. 
\Figref{fig:fullyrecofit} shows an example of this fit for double-tagged \pipipiz candidates - the signal yield is determined in the \avembc window $[1.86, 1.87]\gev$. 
For fully-reconstructed decays that are not continuum dominated, the double-tag yield is determined by counting events in signal and sideband regions of the two dimensional $\D_1\,m_{\mathrm{bc}}$ vs. $\D_2\,m_{\mathrm{bc}}$ plane, as indicated in \Figref{fig:fullyrecofit} for double-tagged \Kpipiz candidates. 

\begin{figure}[h]
      
      \centering
      \includegraphics[width=0.43\textwidth]{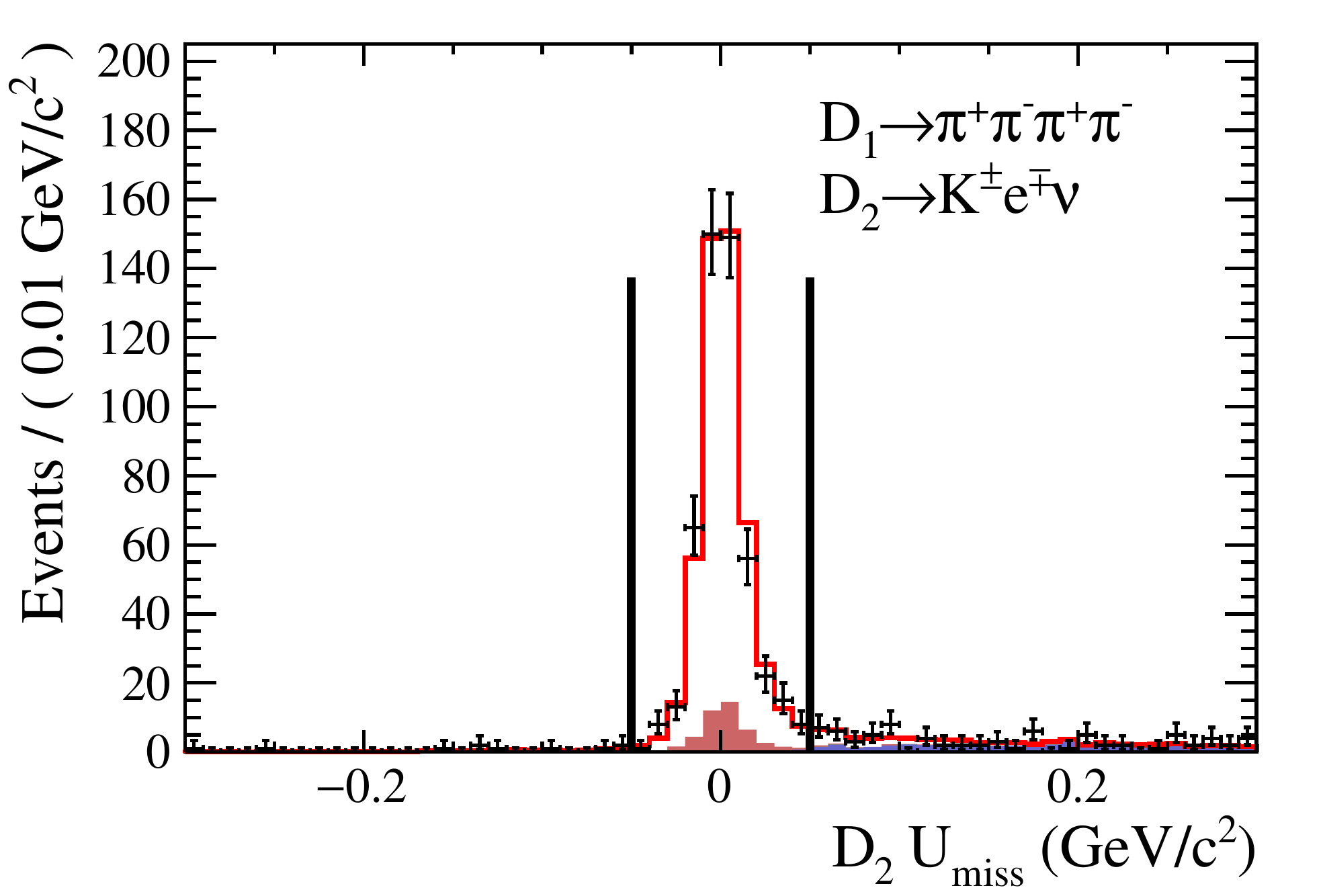}  
      \includegraphics[width=0.43\textwidth]{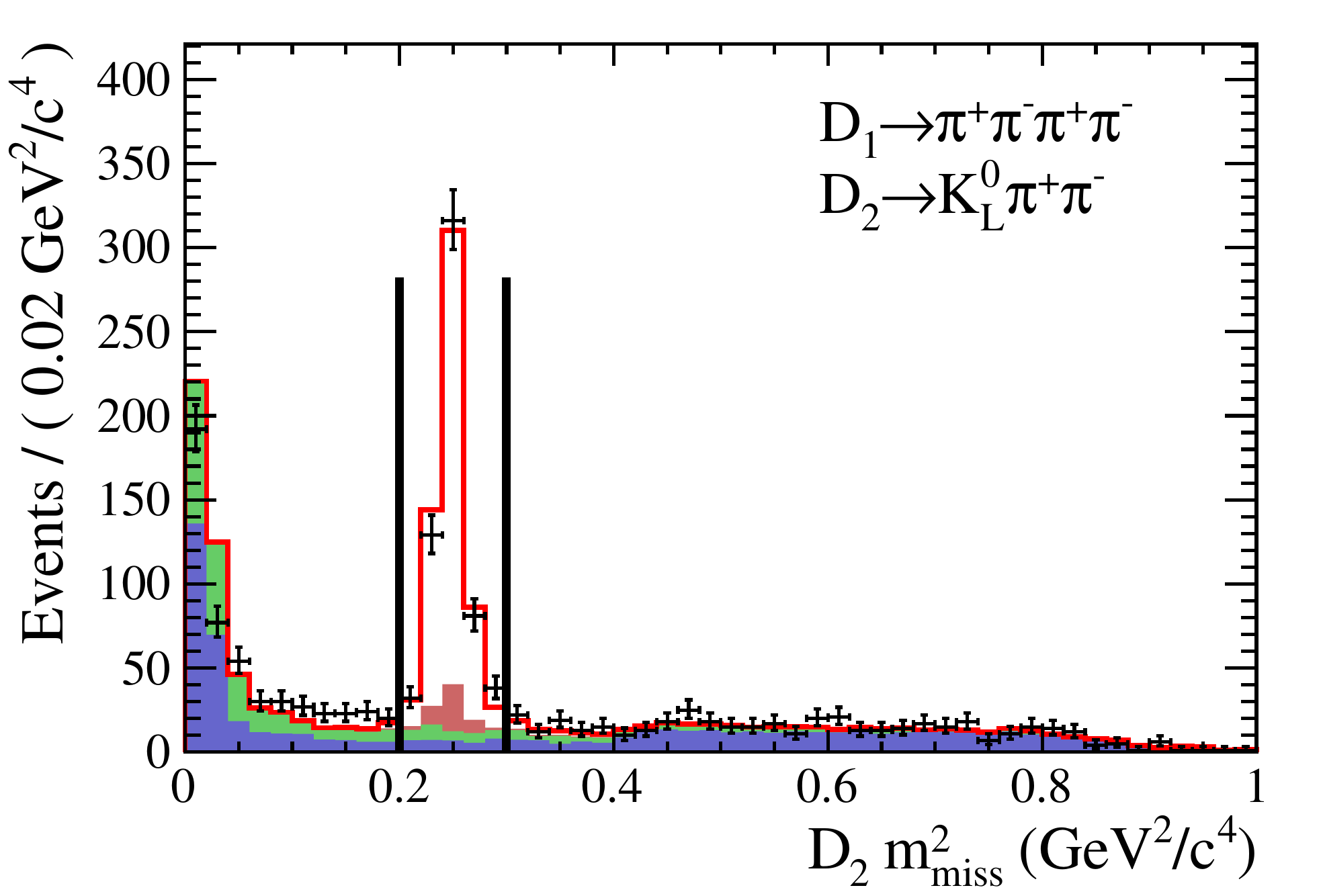}  
      
\caption{ Distribution of \umiss (\missm) for selected double tags containing a neutrino (\KL). Superimposed is the result of an unbinned maximum likelihood fit that is described in the text. 
The blue/green/red shaded area shows the distribution of combinatoric/continuum/peaking background respectively. The red line shows the total signal + background PDF. The black vertical lines indicate the events that fall within the signal region that are used for further analysis. \label{fig:partrecofit} }
\end{figure}

The final states containing a neutrino or a \KL cannot be fully-reconstructed; the energy and momentum, $p_{\mathrm{miss}}$ and $E_{\mathrm{miss}}$, of the missing particle is inferred by using knowledge of the initial $e^+e^-$ state and conservation of energy and momentum.
The missing-mass squared, $m^2_{\mathrm{miss}} \equiv E_{\mathrm{miss}}^2 /c^4 - \mathbf{p}_\mathrm{miss}^2 /c^2$, and the quantity $U_{\mathrm{miss}} \equiv E_{\mathrm{miss}} - |\mathbf{p}_\mathrm{miss}|c$, are used to discriminate signal from background for decays involving a \KL or a neutrino, respectively. 
The double-tag yields are determined using an unbinned maximum likelihood fit to the discriminating variable, where the signal and background PDFs are taken from histograms of simulated data samples.
\Figref{fig:partrecofit} shows an example of this fit for double-tagged \Kenu and \KLpipi candidates - the signal yields are determined within the signal windows indicated.

The dominant peaking background contribution to all double-tags is from \KSpipi background, which is estimated from the generic MC sample of \DDb events, and typically constitutes about $5-10\%$ of the selected events.
A data-driven estimate of this background is also calculated using the events that are rejected by the $\pip\pim$ mass cut - this shows good agreement with the estimates from generic MC. 
All decays involving a \KL decay have a peaking background from the equivalent decay with a \KS instead of a \KL - these are referred to as cross-feed backgrounds.
Using the simulated samples of \decay{\D}{\KS X} decays it is possible to find the ratio of \decay{\D}{\KS X} decays that are incorrectly reconstructed as \decay{\D}{\KL X} to those correctly reconstructed as \decay{\D}{\KS X}.
Since for every $\decay{\D}{\KL X}$ decay considered in this paper, the equivalent $\decay{\D}{\KS X}$ decay is also considered, this allows the background to be estimated using the measured $\decay{\D}{\KS X}$ yields. 
The decay \pipipiz has a peaking background from \KSpiz that is largely suppressed by requiring the $\pip\pim$ vertex to be consistent with with $e^+e^-$ collision point. 
Since the decay \KSpiz is also considered in this paper, the \KSpiz signal yield can be used (in the same manner as for the cross-feed backgrounds) to estimate the background contribution.
All remaining peaking backgrounds are either negligible, or considered in the systematics uncertainties in \secref{sec:systematics}.

\begin{figure}[h]
      
      \centering
      \includegraphics[width=0.43\textwidth]{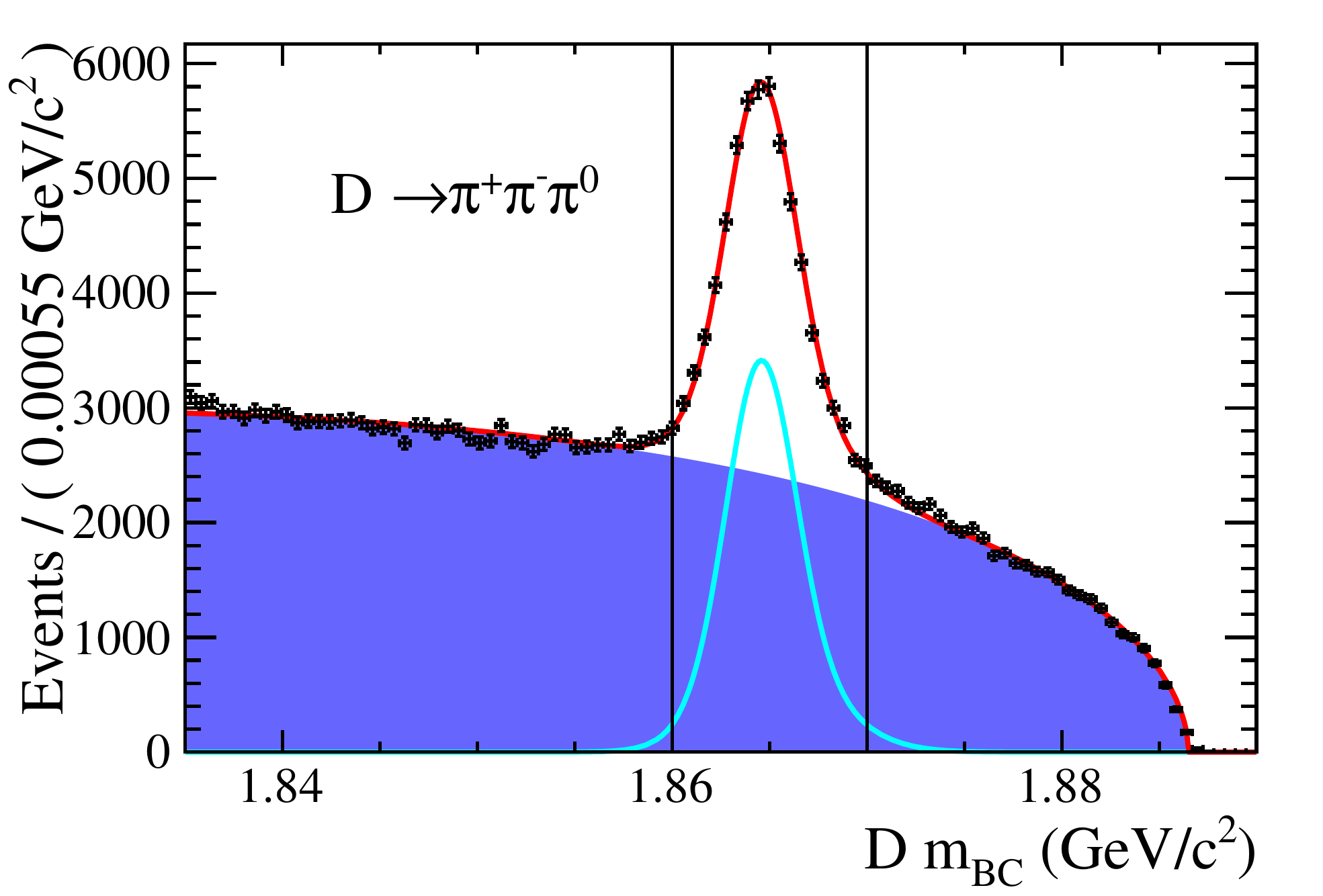}  
      
\caption{ The \mbc distribution of selected \pipipiz single tagged candidates. Superimposed with a red line is the result of a unbinned maximum likelihood fit described in the text. The shaded purple area shows the background PDF, and the blue line shows the signal PDF. The vertical lines show the signal region.   \label{fig:singletagfit} }
\end{figure}

Single-tagged candidates are selected using identical criteria to the corresponding double tags, with the exception of \pipi, \KK and \Kpi decays that have additional cuts to veto cosmic ray muon and radiative Bhabha events~\cite{FPlusPiPiPiz}.
The number of single-tags is estimated from a fit to the \mbc distribution. 
The signal and background PDFs are the same as those used in the fit to the $m^{\mathrm{ave}}_{\mathrm{bc}}$ distribution of continuum dominated double-tags. 
The signal shape parameters are fixed from a sample of simulated signal decays.
\Figref{fig:singletagfit} shows an example of this fit for single-tagged \pipipiz candidates - the signal yield is determined in the signal region indicated. 
Following Ref.~\cite{FPlusPiPiPiz}, a further uncertainty is assigned to each of the single-tag yields to account for any mismodelling of the signal PDF. 
For final states with no electromagnetically neutral final state particles (\KK, \pipi, \Kpi) the uncertainty assigned is 2.0\% of the measured signal yield. 
For final states where the neutrals are relatively hard (\KSpiz, $\KSeta(\gamma\gamma)$) or soft (all other modes), uncertainties of 2.5\% and 5.0\% are assigned, respectively.

In events where more than one single- or double-tagged candidate is reconstructed, an algorithm is used to select a single candidate based on information provided by the $\Delta E$ and \mbc variables. 
The particular choice of metric varies depending on the category of double-tag, and is optimised through simulation studies.

\begin{table}[h]
\centering
\footnotesize
\begin{tabular}{ l | l | l }
\hline
Decay Mode & $\pi^{+}\pi^{-}\pi^{+}\pi^{-}$ & All \\
\hline
$K^{+}K^{-}$ & $18.2\pm6.5$ & $11887.5\pm318.8$ \\
$\pi^{+}\pi^{-}$ & $3.0\pm8.3$ & $5599.5\pm170.8$ \\
$K^{0}_{S}\pi^{0}\pi^{0}$ & $18.2\pm5.6$ & $6989.7\pm374.1$ \\
$K^{0}_{L}\pi^{0}$ & $41.6\pm10.6$ & -- \\
$K^{0}_{L}\omega$ & $23.4\pm6.7$ & -- \\
$K^{0}_{S}\pi^{0}$ & $111.1\pm11.1$ & $19984.0\pm520.8$ \\
$K^{0}_{S}\omega$ & $47.4\pm7.3$ & $8033.6\pm413.3$ \\
$K^{0}_{S}\eta(\gamma\gamma)$ & $18.9\pm4.6$ & $2903.7\pm99.1$ \\
$K^{0}_{S}\eta(\pi^{+}\pi^{-}\pi^{0})$ & $6.7\pm2.7$ & $1283.2\pm80.3$ \\
$K^{0}_{S}\eta^{'}$ & $7.6\pm2.9$ & $1321.9\pm76.6$ \\
$K^{0}_{L}\pi^{+}\pi^{-}$ & $488.0\pm27.1$ & -- \\
$K^{0}_{S}\pi^{+}\pi^{-}$ & $237.4\pm16.6$ & -- \\
$\pi^{+}\pi^{-}\pi^{0}$ & $63.1\pm14.1$ & $30032.4\pm1553.9$ \\
$K^{\pm}e^{\mp}\nu$ & $484.5\pm22.1$ & -- \\
$K^{\pm}\pi^{\mp}$ & $595.6\pm24.7$ & $131613.0\pm2658.1$ \\
$K^{\pm}\pi^{\mp}\pi^{0}$ & $1243.4\pm36.5$ & -- \\
$K^{\pm}\pi^{\mp}\pi^{\pm}\pi^{\mp}$ & $923.7\pm41.2$ & -- \\
\end{tabular}
\caption{ Number of selected single- and double-tagged decays after background subtraction. \label{tab:totalYields}}
\end{table}

For double-tagged decays, the signal yields are evaluated in bins of \fourpi, \KSpipi and \KLpipi phase space. 
For these final states, the four-momenta of the \D daughters are determined with a constraint on the previously measured \Dz mass~\cite{PDG2014}, ensuring that all signal candidates fall within the kinematically allowed region of phase space.  
The \fourpi final state is binned using the schemes in \secref{sec:binning}. 
The \KSpipi and \KLpipi final states are binned according to the `Equal $\Delta\delta_D$ \babar 2008' scheme from Ref.~\cite{KzPiPiCiSi}, which is shown in \figref{fig:kspipiBinning}.
For non-continuum dominated decays, the binned yields are determined by counting the number of candidates in the signal region of the $\D_1\,\mbc$ vs. $\D_2\,\mbc$ plane - the background estimates are discussed in \secref{sec:fitter}.
For continuum-dominated final states a fit the \avembc distribution is performed in each phase space bin. 
In the case where 2 or more phase space bins have an identical decay rate (\eg \fourpi vs. \pipipiz has the same decay rate in bin $+i$ and $-i$) they are merged before determining the signal yield. 
The samples of flavour and quasi-flavour double-tags are split using the charge of the kaon before the binned yields are determined.
The phase space-integrated background subtracted event yields for all single- and double-tagged decays are given in \tabref{tab:totalYields}.

\section{Fit for \fourpi hadronic parameters}
\label{sec:fitter}
This section describes the fitting algorithm used to determine constraints on the \decay{\D}{\fourpi} hadronic parameters.
Following from \eqnref{eqn:masterFormula}, the expected number of $\psi(3770)\to\DDb\to\fsi\fsj$ signal decays is given by,
\begin{align}
N^{\mathrm{sig}}_{\fsi\fsj} = \notag \\
   N_{\DDb}\bigg\{ & \left (  1 + \frac{\ymix^2 - \xmix^2}{2} \right ) \left [ \ki \kbj +  \kbi \kj  
 - 2 \sqrt{  \ki \kbj \kbi \kj } \left ( \ci \cj + \si \sj \right ) \right ] \notag \\
 + &\left ( \phantom{ 1 + \ } \frac{\ymix^2 + \xmix^2}{2} \right ) \left [ \ki \kj +  \kbi \kbj  
 - 2 \sqrt{  \ki \kbj \kbi \kj } \left ( \ci \cj - \si \sj \right ) \right ] \bigg\} \label{eqn:expSig} 
\end{align}
where $N_{\DDb}$ is the total number of $\psi(3770)\to\DDb$ decays in the data sample. 
In the literature, different parameterisations of the hadronic parameters are used for different categories of final state, which sometimes differ from \ki, \kbi, \ci and \si parameterisation used to derive the formalism in this paper. 
The different parameterisations used are summarised in \tabref{tab:listFormalisms}, which are used as free parameters in the fit for the relevant final states.
The new parameters introduced are: the \CP-even fraction, $F_{+}^{f}$; the coherence factor, $R_{\D}^f$; the average strong phase difference, $\delta_{\D}^f$; and the ratio of \decay{\Dzb}{f} to \decay{\Dz}{f} amplitudes, $r_{\D}^f$.
The relationship between these and the \ki, \kbi, \ci and \si parameters is given in \tabref{tab:listFormalisms}.

\begin{table}[h]
\centering
\small
\begin{tabular}{ c  c  c  c  c   c }
Type & \ki & \kbi & \ci & \si \\
\hline
\Dz flavour tag   & $\mathrm{BF}(\decay{\Dz}{f})$ & 0 & 0 & 0  \\
\Dz quasi-flavour tag  &  $\mathrm{BF}(\decay{\Dz}{f})$ & $\mathrm{BF}(\decay{\Dz}{f}) (r_{D}^{f})^2$ & $R_{D}^{f} \cos \delta_{D}^{f}$ & $R_{D}^{f} \sin \delta_{D}^{f}$  \\
\CP tag  & $\mathrm{BF}(\decay{\Dz}{f})$ & $\mathrm{BF}(\decay{\Dz}{f})$ & $\eta_{\CP}$ & 0 \\
Self-conjugate tag  & $\mathrm{BF}(\decay{\Dz}{f})$ & $\mathrm{BF}(\decay{\Dz}{f})$ & $2F^{f}_{+}-1$ & 0 \\
\fourpi / $K^{0}_{S/L}\pi\pi$  & \ki & \kbi & \ci & \si \\
All \D decay final states  & 1 & 1 & \ymix & 0\\
\end{tabular}
\caption{ List of the different parameterisations used for the hadronic parameters of different categories of final state, and how they relate the \ki, \kbi, \ci and \si parameterisation used to derive the formalism in this paper.  \label{tab:listFormalisms}}
\end{table}

Substituting the various parameterisations in \tabref{tab:listFormalisms} into \eqnref{eqn:expSig}, it is clear that different categories of tag provide sensitivity to different hadronic parameters.
The flavour and quasi-flavour tags give sensitivity to \ki and \kbi; the \CP tags and \pipipiz tags give sensitivity to \ki, \kbi, and \ci; and the \KSpipi and \KLpipi tags give sensitivity to all hadronic parameters. 

The expected efficiency and background corrected yield is given by,

\begin{align}
N^{\mathrm{tot}}_{\fsi\fsj}  = N^{\mathrm{sig}}_{\fsi\fsj} \epsilon_{\fsi\fsj} + N^{\mathrm{bkg}}_{\fsi\fsj} \label{eqn:expSigBac}
\end{align}
where $\epsilon_{\fsi\fsj}$ is the reconstruction and selection efficiency for the decay in question, and $N^{\mathrm{bkg}}_{\fsi\fsj}$ is the expected number of background. 
The quantity $\epsilon_{\fsi\fsj}$ is determined from large samples of simulated signal decays, correcting for known discrepancies between data and simulation. 
Before efficiencies are calculated, the simulated samples containing \KSpipi and \fourpi decays are reweighted to their model expectations (using the \decay{\Dz}{\KSpipi} \babar model~\cite{BABAR2008} and the nominal \decay{\Dz}{\fourpi} model~\cite{fourpimodel}) including the effects of quantum correlations. 
The simulated sample of \KLpipi decays is also reweighted to the \KSpipi model with $\Deldelklpipi=-\Deldelkspipi$; this approximation holds in the scenario that only CF and DCS amplitudes contribute, and the two do not overlap in the Dalitz plot.
A systematic uncertainty is later assigned to account for any model dependence in the efficiency determination. 

The total background estimate is broken down into the following expression,
\begin{align}
N^{\mathrm{bkg}}_{\fsi\fsj} = N^{\KS\pi\pi}_{fg} \kappa^{\KS\pi\pi}_{\fsi\fsj} + N^{\mathrm{flat}}_{fg} \kappa^{\mathrm{flat}}_{\fsi\fsj}   + N^{\mathrm{sig}}_{\fsi h_j} \epsilon_{\fsi h_j} f^{h}_{g}  
\end{align}
where $N^{\KS\pi\pi}_{fg}$ and $N^{\mathrm{flat}}_{fg}$ are the total number of \KSpipi and combinatoric background in the $\DDb\to fg$ decay, respectively. 
The quantities $\kappa^{\KS\pi\pi}_{\fsi\fsj}$ and $\kappa^{\mathrm{flat}}_{\fsi\fsj}$ give the fraction of background that falls into the phase space bins $i$ and $j$.
The final term, $N^{\mathrm{sig}}_{\fsi h_j} \epsilon_{\fsi h_j} f^{h}_{g}$, gives the number of cross-feed background from the decay \decay{\DDb}{\fsi h_j}. 
The quantity $f^{h}_{g}$ gives the fraction of \decay{\DDb}{\fsi h_j} decays that are incorrectly reconstructed as \decay{\DDb}{\fsi\fsj}, to those correctly reconstructed.
The value of $N^{\KS\pi\pi}_{fg}$ is taken from generic MC, as was used for the determination of the of background subtracted yields in \tabref{tab:totalYields}.
The value of $\kappa^{\KS\pi\pi}_{\fsi\fsj}$ is found using a large sample of simulated \decay{\D}{\KSpipi} decays that are reconstructed as \decay{\D}{\fourpi}.
Before calculating $\kappa^{\KS\pi\pi}_{\fsi\fsj}$, the simulated sample is first reweighted to the model expectation, based on the phase space location of the generated \KSpipi decay, and including quantum correlations.
Since the \KSpipi model has been shown to give good agreement with model-independent measurements~\cite{KzPiPiCiSi}, any model dependent bias should be small, but this is considered as a systematic uncertainty later.
The value of $N^{\mathrm{flat}}_{fg}$ is determined from the sideband regions, as described in \secref{sec:selection}.
For continuum-dominated and single-tagged decays $N^{\mathrm{flat}}_{fg} = 0$, since the signal yields are determined from a fit to \avembc, so already have the combinatoric background component subtracted.
The value of $\kappa^{\mathrm{flat}}_{\fsi\fsj}$ is determined using simulated signal decays distributed according to the density of states (phase space).
Where possible, this assumption is checked using the sideband regions, which shows good agreement. 
Systematic uncertainties are assigned to cover any bias from this assumption.

The values of the \fourpi hadronic parameters are obtained by maximising the log-likelihood, $\log \mathcal{L}$.
The Poisson distribution, $P(k;\lambda) \equiv \lambda^k e^{-\lambda} / k!$, gives the probability of observing $k$ events when $\lambda$ are expected. 
For double-tagged decays that are not continuum-dominated the log-likelihood receives a term,
\begin{align}
\log \mathcal{L} \mathrel{{+}{=}} \log P( M_{\fsi\fsj} ; N^{\mathrm{tot}}_{\fsi\fsj} ),
\end{align}
where $M_{\fsi\fsj}$ is the number of events counted in the signal region of the decay $\DDb\to\fsi\fsj$.
For continuum-dominated double-tags and single-tags, the signal yield is obtained from a fit, which has an associated uncertainty $\sigma_{\fsi\fsj}$. 
Therefore, the log-likelihood receives a term,
\begin{align}
\log \mathcal{L} \mathrel{{+}{=}}  \log G( N^{\mathrm{tot}}_{\fsi\fsj}; M_{\fsi\fsj}, \sigma_{\fsi\fsj} ) \label{eqn:gauss}
\end{align}
where $G(k; \mu, \sigma)$ is a Gaussian distribution with mean $\mu$ and width $\sigma$.

External inputs are needed to constrain various parameters in the fit.
For the partially-reconstructed \CP final states \KLpiz and \KLomega it is not possible to obtain a single-tagged sample, which would provide the fitter with constraints on the product $N_{\DDb} \times \mathrm{BF}(\Dz\to f)$.
This constraint is important for normalising the respective double-tag yield, so an alternative method is followed for the \KLpiz and \KLomega final states. 
In order to constrain $N_{\DDb}$, the single-tagged $\Kpi$ yield is measured in conjunction with an external constraint on $\mathrm{BF}(\Dz\to\Kpi)$~\cite{PDG2014}. 
External constraints on $\mathrm{BF}(\Dz\to\KLpiz)$ and $\mathrm{BF}(\Dz\to\KLomega)$ then lead to the desired constraint on $N_{\DDb} \times \mathrm{BF}(\Dz\to f)$~\cite{PDG2014}.
For the quasi-flavour tags, external constraints are provided for the hadronic parameters $r_{D}^{f}$, $R_{D}^{f}$ and $\delta_{D}^{f}$, which are taken from Ref.~\cite{HFAGCKM2016} and Ref.~\cite{LHCbCLEOComb}. 
The self-conjugate final state \pipipiz is not a \CP eigenstate, so its \CP-even fraction, $F^{\pi\pi\piz}_{+}$, is constrained to its previously measured value~\cite{FPlusFourPi}.
The charm-mixing parameters are constrained to their world-average values~\cite{HFAGCKM2016}. 
The central values and uncertainties of the constraints are listed in \tabref{tab:externalConstraints}.
All constraints are applied by including a Gaussian constraint, similar to \eqnref{eqn:gauss}, in the $\log\mathcal{L}$; where available, correlations between the parameters are also included.

\begin{table}[h]
\centering
\begin{tabular}{ l  l  l }
Fit Parameter & Constraint & Source \\
\hline
$\mathrm{BF}(\Dz\to\Km\pip)$ & (3.93  $\pm$ 0.04)\%  & Ref.~\cite{PDG2014}\textsuperscript{*} \\
$\mathrm{BF}(\Dz\to\KL\piz)$ & (1.00  $\pm$ 0.07)\%  &  \\
$\mathrm{BF}(\Dz\to\KL\omega(3\pi))$ & (0.99 $\pm$ 0.05 $\pm$ 0.20)\% & \\
\hline
$r^{\Kmpi}$             & (5.90 $\pm$ 0.03)\% & Ref.~\cite{HFAGCKM2016}\textsuperscript{$\dagger$} \\
$\delta^{\Kmpi}$     & \phantom{(}3.41 $\pm$ 0.14 &  \\
\hline
$r^{\Kmpipiz}$        & (4.47 $\pm$ 0.12)\% & Ref.~\cite{LHCbCLEOComb} \\
$R^{\Kmpipiz}$       & \phantom{(}0.81 $\pm$ 0.06 &  \\
$\delta^{\Kmpipiz}$ & \phantom{(}3.46 $\pm$ 0.25 &  \\
$r^{\Km3\pi}$              & (5.49 $\pm$ 0.06)\% &  \\
$R^{\Km3\pi}$            & \phantom{(}0.43 $\pm$ 0.15 &  \\
$\delta^{\Km3\pi}$     & \phantom{(}2.23 $\pm$ 0.39 &  \\
\hline
\xmix                        & (0.322 $\pm$ 0.140)\% & Ref.~\cite{HFAGCKM2016}  \\
\ymix                        &  (0.688 $\pm$ 0.060)\% &  \\
\hline
$F^{\pi\pi\piz}_{+}$                        &  \phantom{(}0.973 $\pm$ 0.017 & Ref.~\cite{FPlusFourPi} \\

\vspace{-3mm}\\
\multicolumn{3}{p{9cm}}{\textsuperscript{*}\footnotesize{The constraint on $\mathrm{BF}(\Dz\to\KL\omega(3\pi))$ is taken from $\mathrm{BF}(\Dz\to\KS\omega(3\pi))$ with a systematic uncertainty of 20\%.}} \\
\multicolumn{3}{p{9cm}}{\textsuperscript{$\dagger$}\footnotesize{Ref.~\cite{HFAGCKM2016} uses the convention $\CP|\Dz\rangle=-|\Dzb\rangle$, so the transformation $\delta^{\Kmpi} \to \delta^{\Kmpi} + \pi$ is applied. }}
\end{tabular}
\caption{ List constraints used in the analysis. The right hand column gives the source of the constraint, along with any conventional adjustments that have to be made to use them in this analysis. \label{tab:externalConstraints} }
\end{table}

The hadronic parameters of the \KSpipi and \KLpipi final states are also constrained. 
The parameters \cikspipi, \sikspipi, \ciklpipi and \siklpipi are constrained using the covariance matrix for the \babar equal $\Delta\delta_D$ binning given in Ref.~\cite{KzPiPiCiSi}. 
An adjustment must be made to the constraints on \ciklpipi and \siklpipi, since a different convention is used in Ref.~\cite{KzPiPiCiSi} such that $\ciklpipi\to-\ciklpipi$ and $\siklpipi\to-\siklpipi$.
Constraints on the parameters \Friklpipi and \Frbiklpipi are taken directly from Ref.~\cite{BrisbaneThesis};
since it is the parameters \kiklpipi and \kbiklpipi that are used as free parameters in the fit, \Friklpipi and \Frbiklpipi are calculated dynamically so that the constraint can be applied.
The parameters \Frikspipi and \Frbikspipi are constrained from an average of \belle and \babar model predictions~\cite{BELLEModel,BABARModel}, as determined in Ref.~\cite{newCoherenceCLEO}. 
Since the amplitude models are fit to \D decay-time integrated samples of \decay{\Dstarp}{\Dz\pip} decays, small corrections must be made for \D-mixing using the expression~\cite{K3piLHCb},
\begin{align}
B_i^{\KS\pi\pi} \propto \Frikspipi - \sqrt{\Frikspipi \Frbikspipi} ( \cikspipi \ymix + \sikspipi \xmix  ) + \frac{\xmix^2+\ymix^2}{2}, \label{eqn:mixingcor}
\end{align}
where $B_i^{\KS\pi\pi}$ is the fraction of \decay{\Dstarp\to\Dz\pip,\Dz}{\KSpipi} decays in phase space bin $i$. 
Using external inputs from Refs.~\cite{KzPiPiCiSi,HFAGCKM2016}, the system of equations is solved to find \Frikspipi and \Frbikspipi.

In principle, the normalisation parameter $N_{\DDb}$ can be shared for every decay mode considered in the analysis, since the same $e^+e^-$ collision data are used.
In reality, however, this is not always desirable since the estimation of $N_{\DDb}$ relies on the absolute efficiencies (rather than the relative, bin-to-bin, efficiencies) determined from simulated samples.  
For the double-tagged samples of \KSpipi, \KLpipi, \Kpipiz, \Kpipipi, and \Kenu decays, almost all information comes from the relative bin-to-bin yields, so sharing a normalisation parameter provides little benefit while introducing a potential source of systematic uncertainty. 
Therefore, these final states each have their own normalisation parameter, $N_{\DDb}^f$, in the fit. 
On the other hand, the double- and single-tagged \Kpi samples share a normalisation constant, which allows the fitter to constrain $\mathrm{BF}(\decay{\Dz}{\fourpi})$, since $\mathrm{BF}(\decay{\Dz}{\Kmpi})$ has an external constraint (\tabref{tab:externalConstraints}). 
This normalisation constant is also shared with all single- and double-tagged \CP and \pipipiz final states.

The $\log \mathcal{L}$ expression is maximised numerically using the MINUIT software~\cite{MINUIT}. 
The maximisation procedure is repeated 5 times with different starting values to ensure the global maximum of $\log \mathcal{L}$ has been found (as opposed to a local maximum). 
Statistical uncertainties and correlations between fit parameters are provided by Minuit from evaluating the second derivatives of $\log \mathcal{L}$ with respect to the fit parameters.

The fitting procedure is tested using 400 simulated experiments that use the background and efficiency estimates from the fit to data.
The \decay{\D}{\fourpi} hadronic parameters used to generated the pseudo-experiments are taken from model predictions. 
The hadronic parameters of other final states are taken from their previously measured values, and randomly sampled from their associated uncertainties.
No statistically significant bias was found in the fit procedure.

The central values and statistical uncertainties of the \decay{\D}{\fourpi} hadronic parameters from the fit to data are given in \tabref{tab:results}, and the statistical correlations in \appref{app:correlations}.
In this paper only results using \fourpi binning schemes with $\mathcal{N}=5$ are presented, although the results for $\mathcal{N}=1-5$ can be found in the supplementary material.

\section{Systematics}
\label{sec:systematics}
The systematic uncertainties on the \fourpi hadronic parameters are broken down into several components, as listed in the systematic uncertainty breakdown in \tabref{tab:systematics}.
Each of these components will be discussed in the following.

\paragraph{Bin migration} Due to the finite detector resolution, it is possible for an event occurring in one phase space bin to be reconstructed in another; this bin-migration is relevant to the \fourpi, \KSpipi and \KLpipi final states. 
Since decays to these final states do not proceed by any narrow resonances, bin migration is not expected to significantly bias the result.
Using samples of simulated signal events (that are reweighted to their model expectations), a migration matrix is calculated, whose elements $M_{ik}$ give the probably of an event generated in bin $k$ to be reconstructed in bin $i$. 
For the fully-reconstructed final states \fourpi and \KSpipi, the diagonal elements of $M_{ik}$ are typically $\sim95\%$, whereas for the partially reconstructed \KLpipi final state they are $\sim85\%$.
The migration matrices are used in the calculation of the expected yields (\eqnref{eqn:expSigBac}) and the fit is rerun. 
The absolute difference between this result and the nominal result, which is obtained without correcting for bin-migration, is taken as a systematic uncertainty.

\paragraph{Multiple candidate selection} To check that the multiple candidate selection (MCS) procedure does not bias the result, an alternative MCS procedure is followed where one candidate is chosen at random (rather than based on a metric). 
The difference between the hadronic parameters determined using this selection and the nominal selection is taken as a systematic uncertainty.

\paragraph{Relative efficiencies} 
In the nominal fit, the relative efficiency between phase space bins is determined using simulated signal samples that are reweighted to their model expectation. 
To estimate an upper limit on the systematic uncertainty introduced by the model uncertainty, the efficiency estimates are redetermined with the simulated samples reweighted to phase space.
The absolute difference between the result using alternative efficiency estimates and the nominal result is taken as a systematic uncertainty.

\paragraph{Relative \KSpipi background distribution} To determine the relative distribution of \KSpipi background, a sample of simulated \decay{\D}{\KSpipi} decays, reconstructed as \decay{\D}{\fourpi}, is reweighted to its model prediction, including quantum correlations. 
In order to determine a systematic uncertainty, the quantum correlations are neglected (equivalent to setting $\ci=\si=0$ in \eqnref{eqn:expSig}) and the \KSpipi background distribution is recalculated. 
The absolute difference between this result and the nominal result is taken as a systematic uncertainty.

\paragraph{Absolute \KSpipi background yields} In the nominal fit, the total number of \KSpipi background events are estimated using the generic sample of simulated data.
Alternatively, this is determined using a data-driven technique. 
The relative event numbers in the \KS veto region and the signal region is determined from simulation for both \KSpipi background and \fourpi signal. 
These numbers are used to estimate the \KSpipi background contamination in the signal region based on the observed number of events in the \KS veto region.
The fit is rerun with the alternative \KSpipi background yields, and the absolute difference between this result and the nominal result is taken as a systematic uncertainty.

\paragraph{Relative flat background distribution}  The relative number of combinatorial background events across phase space bins is assumed to be distributed according to phase space. 
As an alternative method, the relative numbers are taken from the combinatorial background events in the generic MC sample. 
The absolute difference between this result and the nominal result is taken as a systematic uncertainty.

\paragraph{Absolute flat background yields} For fully-reconstructed non-continuum dominated double-tagged decays, the total number of combinatorial background is estimated from the number of events in five sideband regions of the two dimensional $\D_1\,m_{\mathrm{bc}}$ vs. $\D_2\,m_{\mathrm{bc}}$ plane (see \figref{fig:fullyrecofit}).
Each sideband region is associated with a particular background type, which is assumed to have the same density in the sideband and signal regions. 
Alternatively, the relative density of background between the sideband and signal regions is taken from generic MC. 
The alternative background estimates are used in the fit, and the difference between this result and the nominal result is taken as a systematic uncertainty.  

For partially-reconstructed double-tagged decays, the total number of combinatorial background is determined from a fit to the $m^2_{\mathrm{miss}}$ or $U_{\mathrm{miss}}$ distribution (see \figref{fig:partrecofit}). 
Alternatively, the combinatorial background yield is determined using a simpler sideband-subtraction approach. 
The alternative background estimates are used in the fit, and the difference between this result and the nominal result is taken as a systematic uncertainty.

\paragraph{Continuum dominated signal yields} For continuum-dominated double-tagged decays, the signal yield in each phase space bin is determined from a fit to the $m^{\mathrm{ave}}_{\mathrm{bc}}$ distribution. 
The fits are repeated with an alternative signal (sum of a Johnson function~\cite{JohnsonFunction} and a Gaussian) and background (second order polynomial) parameterisation, in a reduced $m^{\mathrm{ave}}_{\mathrm{bc}}$ range.
The alternative signal yields and uncertainties are used to determine the hadronic parameters, and the difference between this result and the nominal result is taken as a systematic uncertainty.

\paragraph{Non-resonant dilution} The final states \KSomega and $\KSeta(\pipipiz)$ have small contributions from non-resonant $\KS\pipipiz$ decays, which are estimated from generic MC. 
Since this background contributes to both the single-tagged and double-tagged modes, it can be accounted for by making a small adjustment to the \CP-even fraction of each final state, which would be identically zero (\CP-odd) in the case of no background. 
Since the \CP-content of this background is not known, it is conservatively assumed to be \CP-even. 
The fit is rerun with the updated \CP-even fractions, and the difference between this result and the nominal result is taken as a systematic uncertainty. 

\paragraph{Simulated sample statistics} In the nominal fit, the background and efficiency estimates all have an uncertainty due to limited statistics in simulated data samples.
The fit is rerun twenty times, each time randomly varying the efficiency and background estimates within their uncertainties.  
The covariance of the results obtained is used to determine a systematic uncertainty. 

A breakdown of the systematic uncertainties for the optimal alternative binning with $\mathcal{N}=5$ is given in \tabref{tab:systematics}.
The largest systematic uncertainty comes from imperfect knowledge of the combinatorial background.
The total systematic uncertainties for all binning schemes with $\mathcal{N}=5$ are given in \tabref{tab:results}, and the systematic correlations in \appref{app:correlations}. 
The equivalent information for the other binning schemes considered is provided in the supplementary material.
For all parameters the total uncertainty is statistically dominated.

\begin{table}[h]

\centering

 \resizebox{\textwidth}{!}{   
\begin{tabular}{ l | r  r  r  r  r  r  r  r  r  r }
   & $c^{4\pi}_{+1}$ [\%] & $c^{4\pi}_{+2}$ [\%] & $c^{4\pi}_{+3}$ [\%] & $c^{4\pi}_{+4}$ [\%] & $c^{4\pi}_{+5}$ [\%] & $s^{4\pi}_{+1}$ [\%] & $s^{4\pi}_{+2}$ [\%] & $s^{4\pi}_{+3}$ [\%] & $s^{4\pi}_{+4}$ [\%] & $s^{4\pi}_{+5}$ [\%]\\
\hline
Bin migration & 1.493 & 1.063 & 0.911 & 0.824 & 0.643 & 2.888 & 2.312 & 2.911 & 3.221 & 2.527\\
MCS & 4.753 & 1.858 & 0.438 & 0.734 & 0.058 & 5.211 & 3.659 & 1.914 & 2.294 & 11.313\\
Rel. Efficiency & 0.576 & 0.011 & 0.045 & 0.032 & 1.902 & 1.225 & 0.686 & 0.942 & 0.722 & 0.301\\
Abs. Flat Bkg. & 7.823 & 5.167 & 3.441 & 4.143 & 2.053 & 6.344 & 3.860 & 0.688 & 4.899 & 5.486\\
Rel. Flat Bkg. & 2.067 & 0.015 & 0.693 & 0.089 & 0.947 & 5.012 & 3.640 & 0.227 & 1.517 & 2.930\\
Cont. Dom. Fit & 2.058 & 1.146 & 0.347 & 0.791 & 2.220 & 0.079 & 0.075 & 0.005 & 0.005 & 0.278\\
Abs. \KSpipi Bkg. & 1.953 & 0.455 & 0.372 & 0.831 & 0.409 & 0.100 & 0.092 & 0.074 & 0.162 & 0.730\\
Rel. \KSpipi Bkg. & 0.628 & 0.193 & 0.716 & 0.454 & 0.058 & 0.388 & 0.061 & 0.144 & 0.049 & 0.279\\
Non Res. Dilution & 0.142 & 0.449 & 0.396 & 0.118 & 0.050 & 0.021 & 0.029 & 0.020 & 0.003 & 0.012\\
MC stats & 1.475 & 1.158 & 0.399 & 1.211 & 1.483 & 2.203 & 1.978 & 1.097 & 1.339 & 1.249\\
Total Sys. & 10.063 & 5.863 & 3.799 & 4.626 & 4.055 & 10.363 & 7.161 & 3.845 & 6.655 & 13.244\\
Total Stat. & 14.283 & 9.542 & 5.668 & 9.916 & 13.847 & 29.095 & 23.734 & 16.236 & 21.471 & 26.346\\
Total & 17.472 & 11.199 & 6.824 & 10.942 & 14.428 & 30.885 & 24.791 & 16.685 & 22.478 & 29.488\\
\end{tabular}

}

\vspace{1cm}

 \resizebox{\textwidth}{!}{   
\begin{tabular}{ l | r  r  r  r  r  r  r  r  r  r }
   & $T^{4\pi}_{+1}$ [\%] & $T^{4\pi}_{+2}$ [\%] & $T^{4\pi}_{+3}$ [\%] & $T^{4\pi}_{+4}$ [\%] & $T^{4\pi}_{+5}$ [\%] & $T^{4\pi}_{-1}$ [\%] & $T^{4\pi}_{-2}$ [\%] & $T^{4\pi}_{-3}$ [\%] & $T^{4\pi}_{-4}$ [\%] & $T^{4\pi}_{-5}$ [\%]\\
\hline
Bin migration & 0.049 & 0.011 & 0.091 & 0.089 & 0.027 & 0.041 & 0.101 & 0.115 & 0.059 & 0.060\\
MCS & 0.006 & 0.143 & 0.055 & 0.084 & 0.108 & 0.072 & 0.055 & 0.000 & 0.115 & 0.154\\
Rel. Efficiency & 0.153 & 0.260 & 0.107 & 0.020 & 0.110 & 0.076 & 0.091 & 0.128 & 0.011 & 0.051\\
Abs. Flat Bkg. & 0.099 & 0.149 & 0.427 & 0.406 & 0.135 & 0.469 & 0.276 & 0.052 & 0.337 & 0.186\\
Rel. Flat Bkg. & 0.041 & 0.033 & 0.100 & 0.084 & 0.062 & 0.129 & 0.059 & 0.066 & 0.054 & 0.029\\
Cont. Dom. Fit & 0.009 & 0.004 & 0.023 & 0.005 & 0.020 & 0.009 & 0.000 & 0.020 & 0.001 & 0.016\\
Abs. \KSpipi Bkg. & 0.002 & 0.008 & 0.005 & 0.005 & 0.004 & 0.019 & 0.005 & 0.002 & 0.002 & 0.005\\
Rel. \KSpipi Bkg. & 0.003 & 0.003 & 0.004 & 0.001 & 0.001 & 0.002 & 0.002 & 0.004 & 0.002 & 0.001\\
Non Res. Dilution & 0.000 & 0.001 & 0.001 & 0.000 & 0.000 & 0.000 & 0.000 & 0.001 & 0.000 & 0.000\\
MC stats & 0.080 & 0.104 & 0.133 & 0.146 & 0.071 & 0.055 & 0.076 & 0.161 & 0.112 & 0.067\\
Total Sys. & 0.209 & 0.350 & 0.483 & 0.457 & 0.228 & 0.503 & 0.327 & 0.251 & 0.382 & 0.265\\
Total Stat. & 0.517 & 0.568 & 0.743 & 0.605 & 0.463 & 0.391 & 0.438 & 0.699 & 0.506 & 0.385\\
Total & 0.558 & 0.667 & 0.886 & 0.758 & 0.516 & 0.637 & 0.546 & 0.743 & 0.634 & 0.467\\
\end{tabular}

}

\caption{ A breakdown of the systematic uncertainties for the optimal alternative binning scheme with $\mathcal{N} = 5$. \label{tab:systematics} }
\end{table}

\section{Results and consistency checks}
\label{sec:results}
\begin{figure}[h]
\centering

\begin{minipage}[c]{0.48\textwidth}
\centering
Equal \Deldelfourpi binning \\
\includegraphics[width=0.99\textwidth]{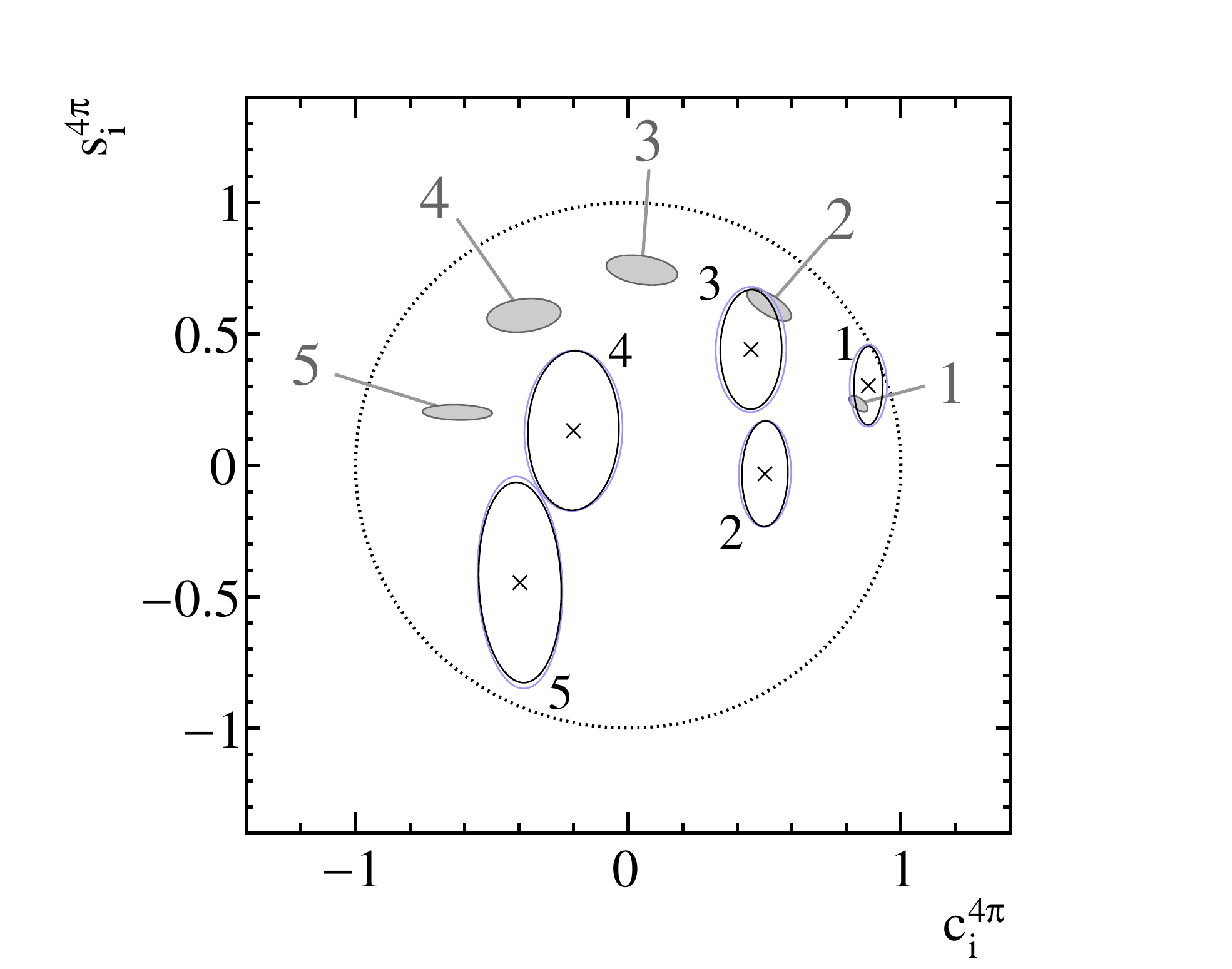} 
\end{minipage}
\begin{minipage}[c]{0.48\textwidth}
\centering
Variable \Deldelfourpi binning \\
\includegraphics[width=0.99\textwidth]{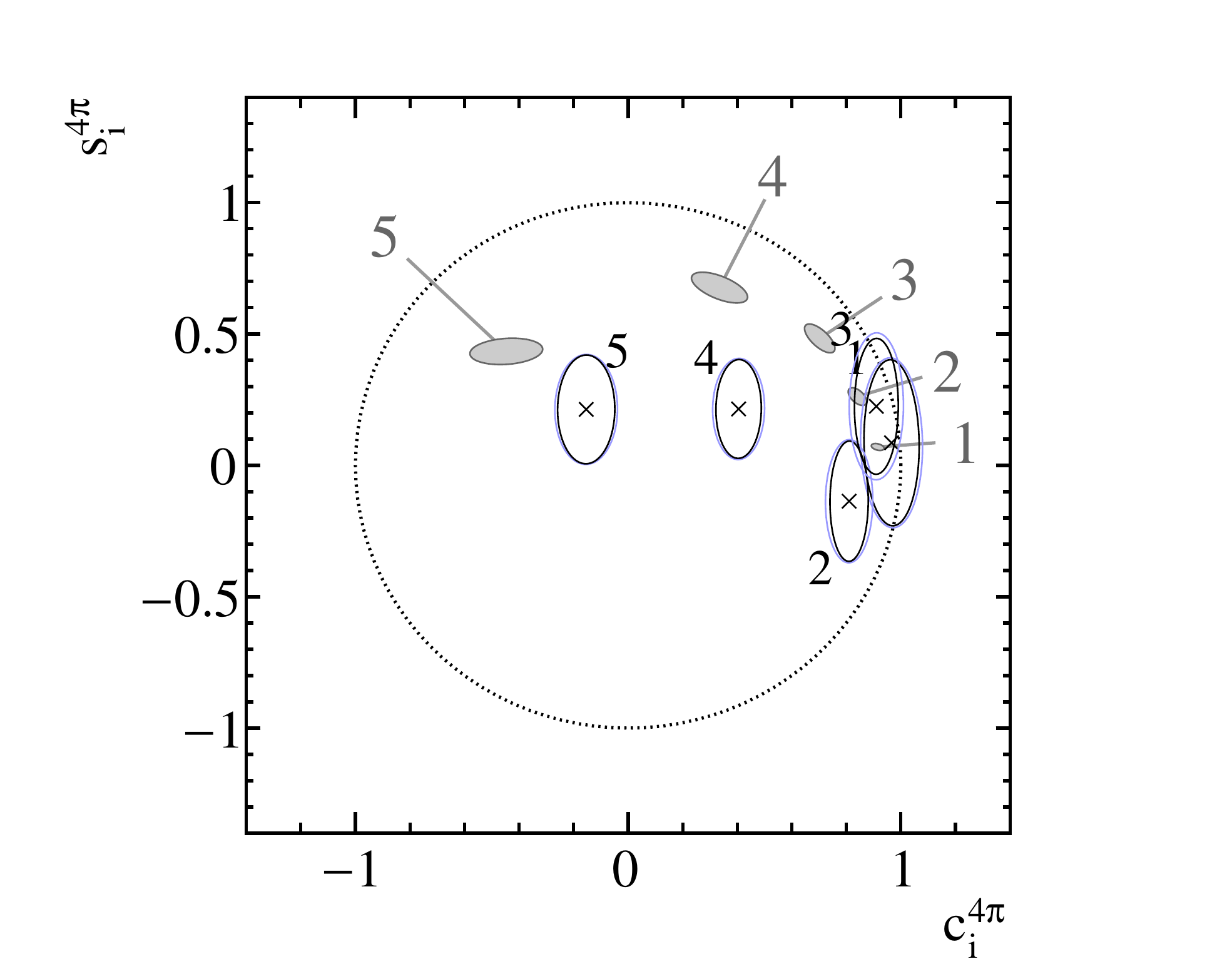} 
\end{minipage}

\begin{minipage}[c]{0.48\textwidth}
\centering
Alternative Binning \\
\includegraphics[width=0.99\textwidth]{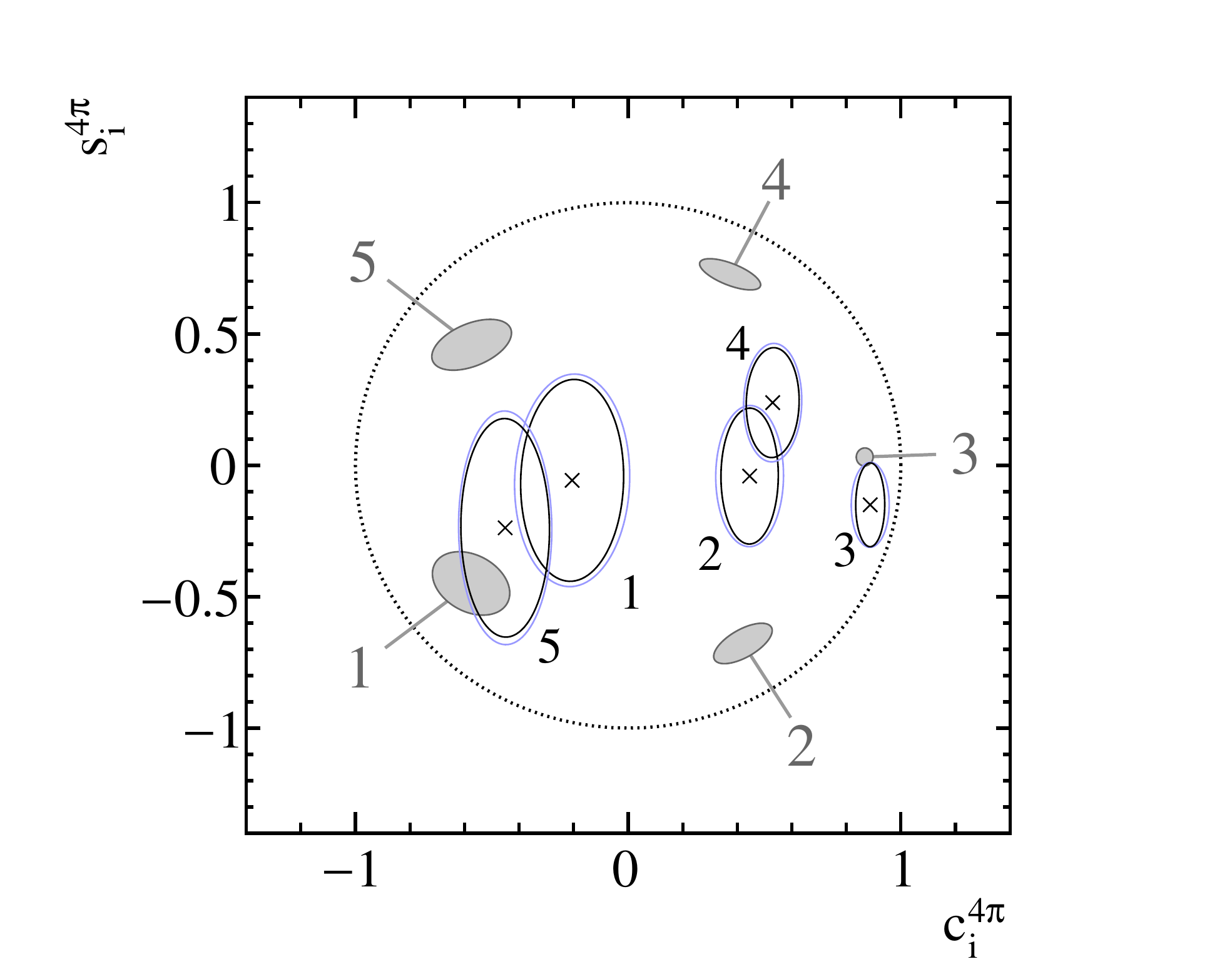} 
\end{minipage}
\begin{minipage}[c]{0.48\textwidth}
\centering
Optimal Binning \\
\includegraphics[width=0.99\textwidth]{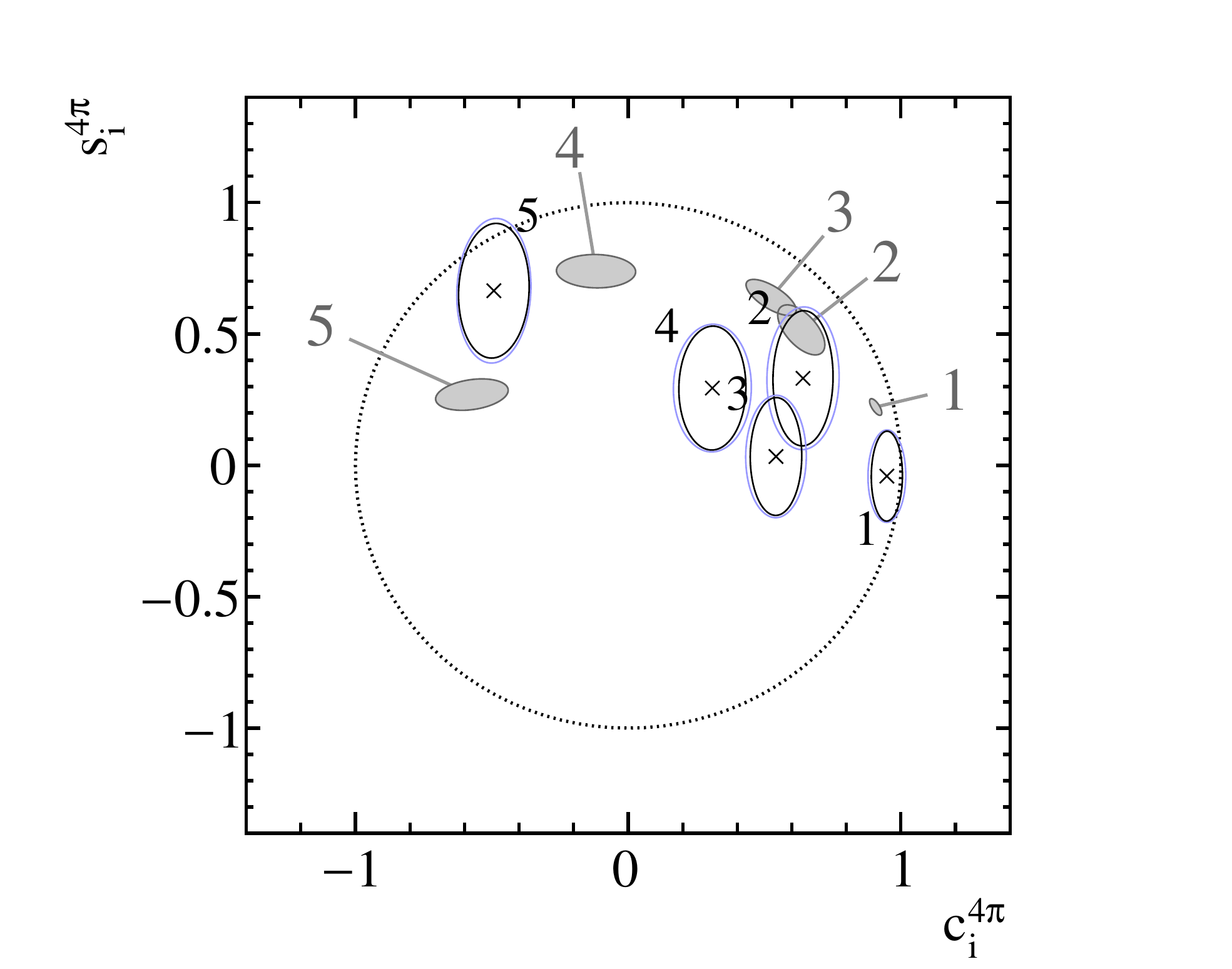} 
\end{minipage}

\begin{minipage}[c]{0.48\textwidth}
\centering
Optimal Alternative Binning \\
\includegraphics[width=0.99\textwidth]{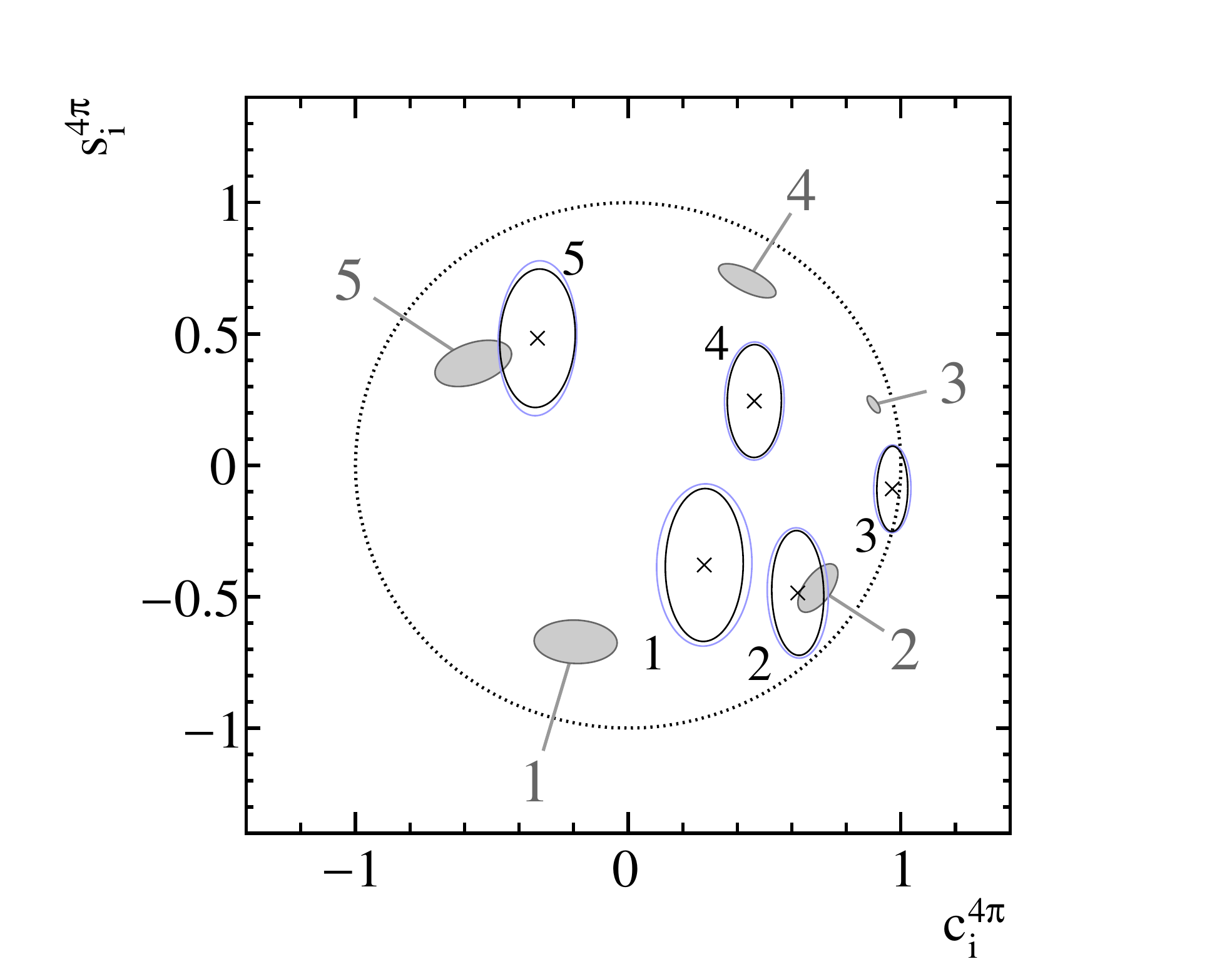} 
\end{minipage}

\caption{Each figure shows the hadronic parameters \cifourpi and \sifourpi measured using one of the \fourpi binning schemes discussed in \secref{sec:binning} where $\mathcal{N} = 5$. The grey shaded ellipses give the model predictions and uncertainties discussed in \secref{sec:binning}. The black (blue) ellipses show the measured values and statistical (statistical + systematic) uncertainties. In all cases the ellipse contains the $39.3\%$ confidence region, defined by the $\log \mathcal{L}_{\mathrm{max}} - \log \mathcal{L}= \half$ contour, where $\log \mathcal{L}_{\mathrm{max}}$ is the maximum value of $\log \mathcal{L}$.  \label{fig:cisiresults}    }
\end{figure}

\begin{figure}[h]
\centering

\begin{minipage}[c]{0.48\textwidth}
\centering
Equal \Deldelfourpi binning \\
\includegraphics[width=0.95\textwidth]{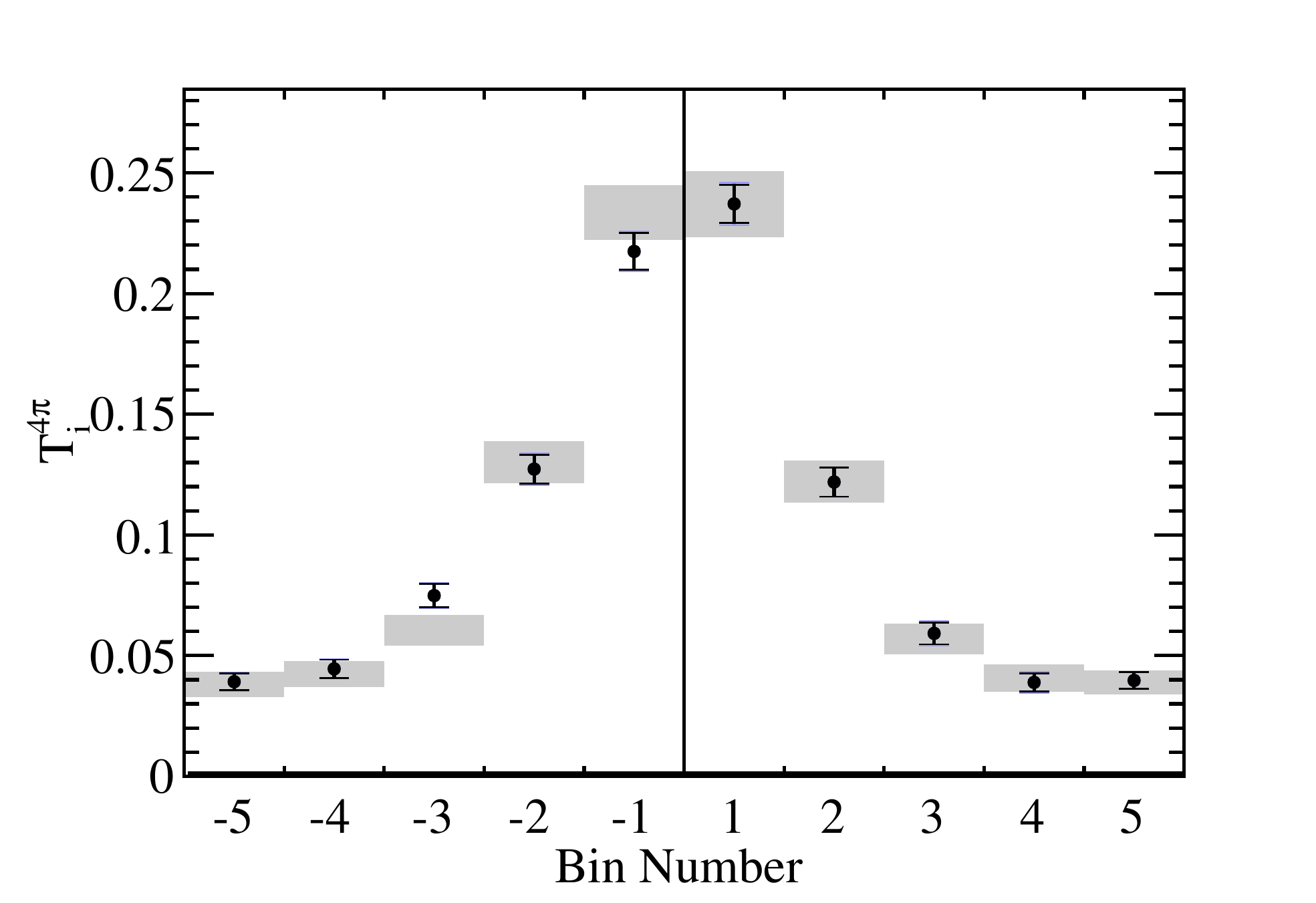} 
\end{minipage}
\begin{minipage}[c]{0.48\textwidth}
\centering
Variable \Deldelfourpi binning \\
\includegraphics[width=0.95\textwidth]{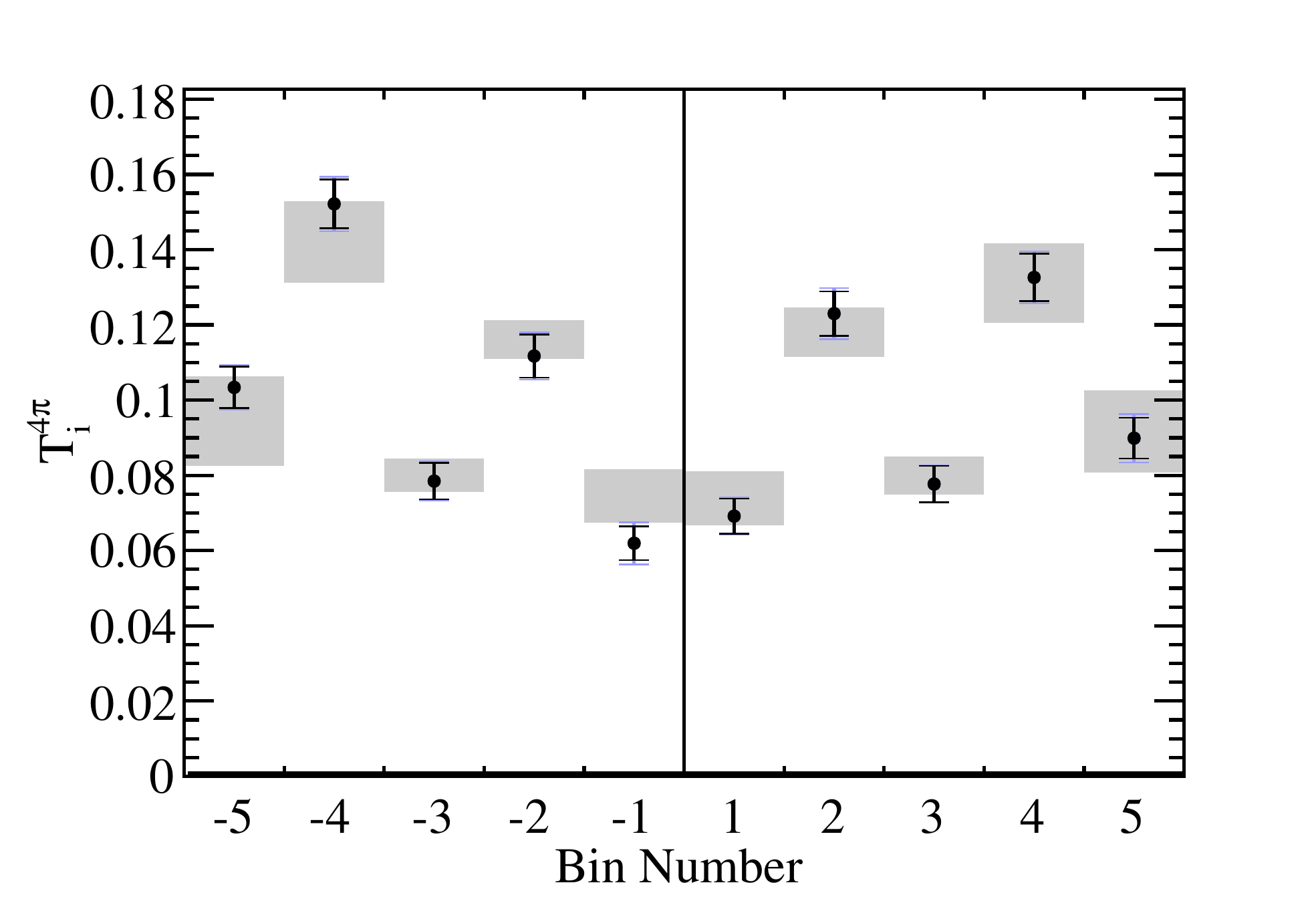} 
\end{minipage}

\begin{minipage}[c]{0.48\textwidth}
\centering
Alternative Binning \\
\includegraphics[width=0.95\textwidth]{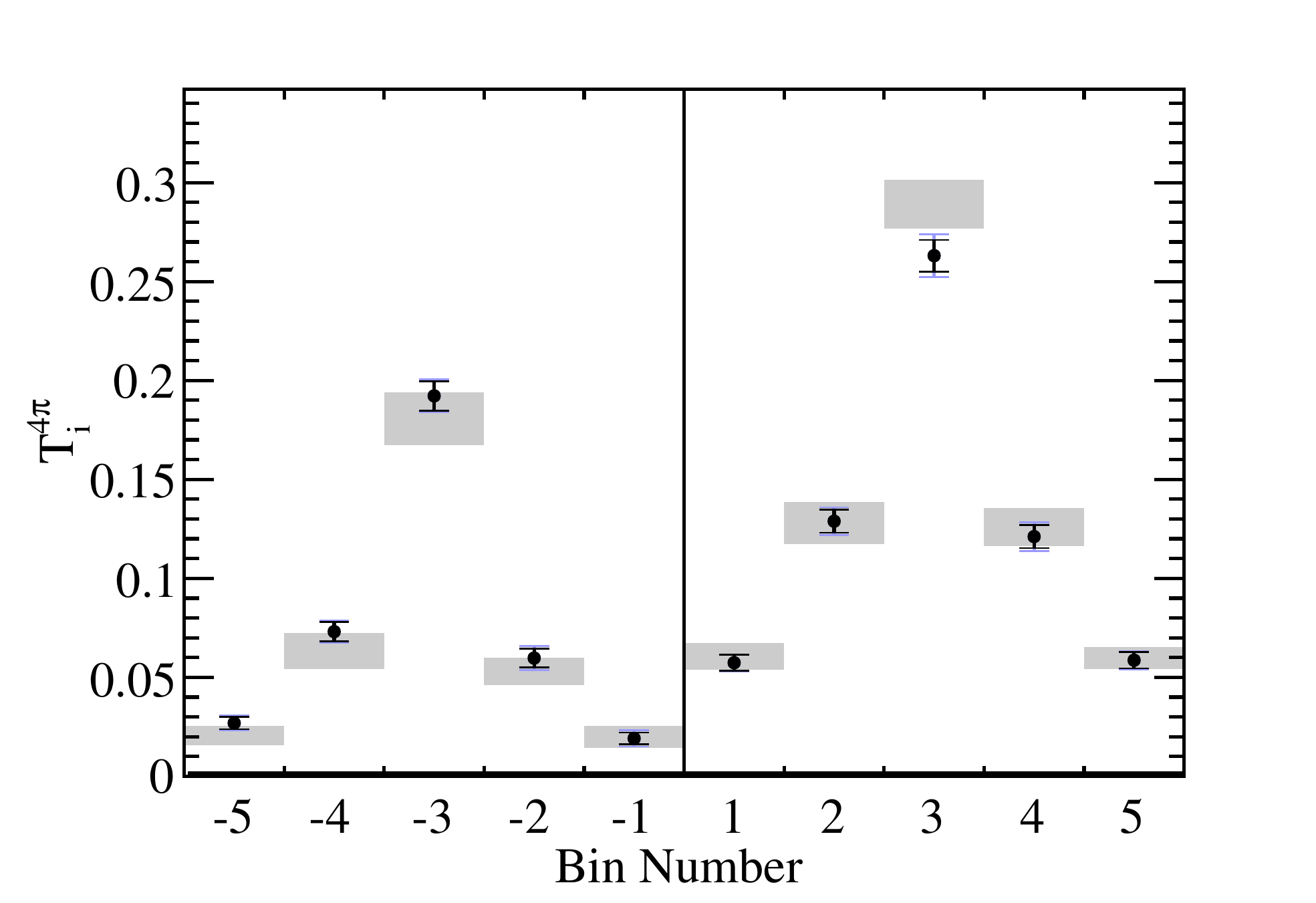} 
\end{minipage}
\begin{minipage}[c]{0.48\textwidth}
\centering
Optimal Binning \\
\includegraphics[width=0.95\textwidth]{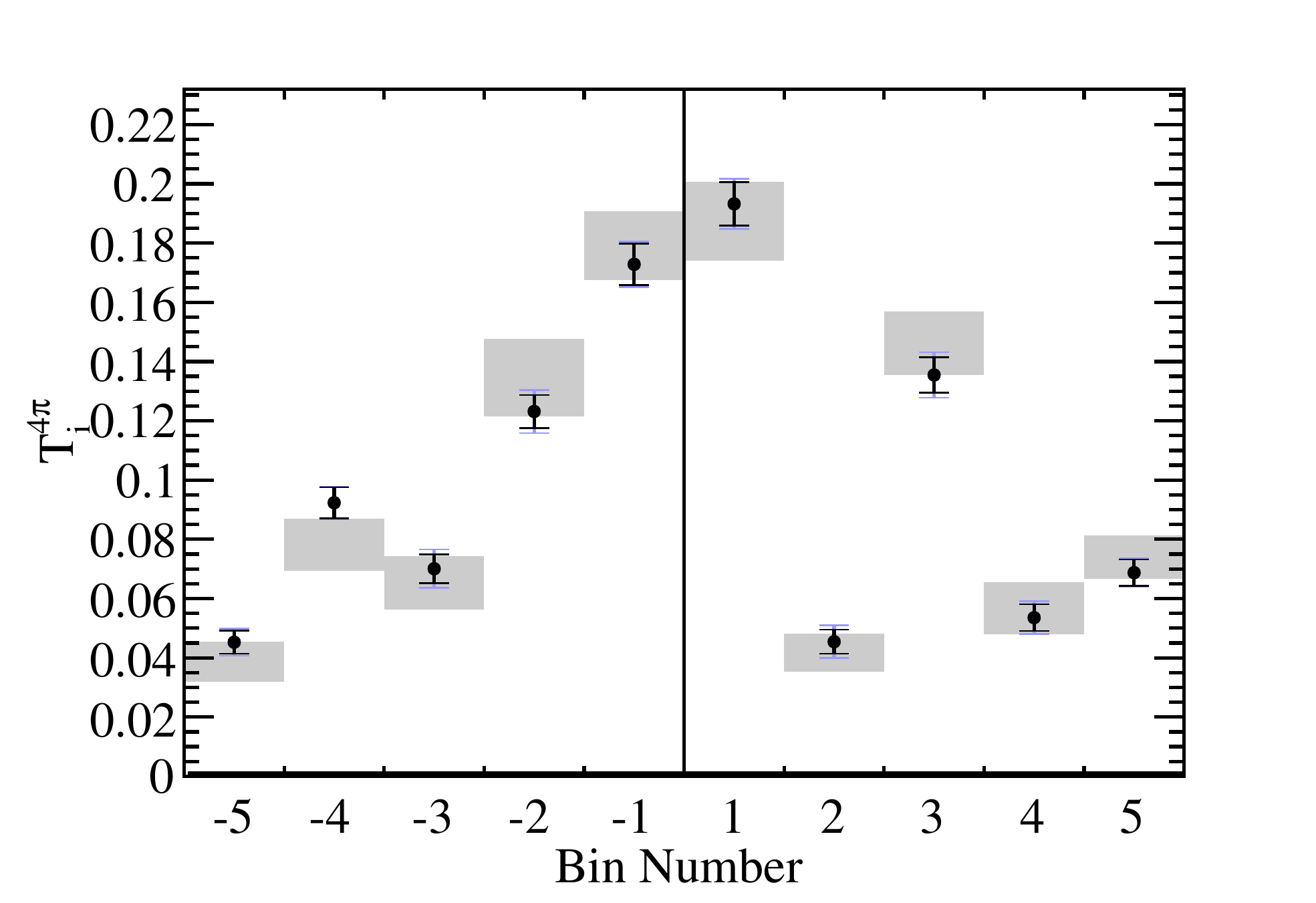} 
\end{minipage}

\begin{minipage}[c]{0.48\textwidth}
\centering
Optimal Alternative Binning \\
\includegraphics[width=0.95\textwidth]{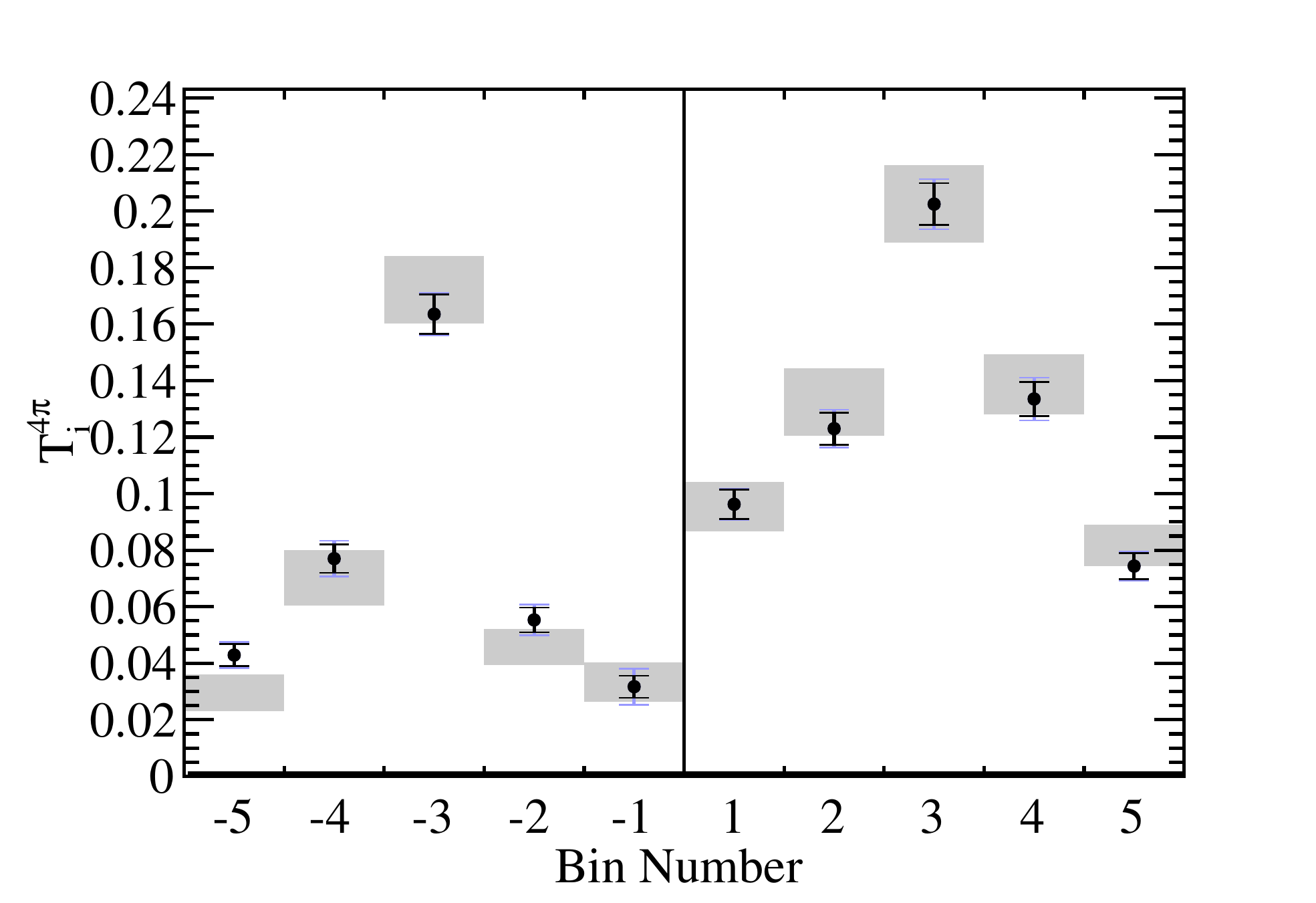} 
\end{minipage}

\caption{Each figure shows the hadronic parameters \Frifourpi and \Frbifourpi measured using one of the \fourpi binning schemes discussed in \secref{sec:binning} where $\mathcal{N} = 5$. The grey bands give the model predictions and uncertainties discussed in \secref{sec:binning}. The black (blue) points with errors give the measured values and statistical (statistical + systematic) uncertainties.   \label{fig:kikbiresults}   }
\end{figure}

The measurement of the \fourpi hadronic parameters with statistical and systematic uncertainties is given in \tabref{tab:results}, with correlations in \appref{app:correlations}.
The results are compared to the model predictions in \figref{fig:cisiresults} and \figref{fig:kikbiresults}.
The compatibility between the results and the model predictions is quantified by calculating the $\chi^2$ between to two, where all correlations are included.  
This is done independently for the \cifourpi/\sifourpi, and \Frifourpi/\Frifourpi parameters, and for the combination, with the results in \tabref{tab:modelDataChiSq}. 
The parameters \Frifourpi and \Frbifourpi show good agreement with the model predictions, which is expected since the model was determined from a fit to \Dz and \Dzb tagged data. 
The parameters \ci and \si are in slight tension with the model predictions, with p-values ranging from $0.03$ to $0.18$, but they clearly follow the same general trend in the \ci-\si plane. 
It is worth repeating here that any incompatibility with the model will not introduce additional systematic uncertainties to a measurement of $\gamma$, but will only increase the statistical uncertainty. 

\begin{table}
\resizebox{\textwidth}{!}{
\begin{tabular}{ l | c c | c c | c c }
Binning scheme & \multicolumn{2}{c|}{$\cifourpi$, $\sifourpi$} & \multicolumn{2}{c|}{$\Frifourpi$, $\Frbifourpi$} & \multicolumn{2}{c}{$\cifourpi$, $\sifourpi$, $\Frifourpi$, $\Frbifourpi$} \\
  & $\chi^2$ / ndof & (p-value) & $\chi^2$ / ndof & (p-value) & $\chi^2$ / ndof & (p-value) \\
\hline
Equal $\Deldelfourpi$ & 19.9 / 10& ( 0.03 ) & 7.4 / 9& ( 0.59 ) & 30.0 / 19& ( 0.05 ) \\ 
Variable $\Deldelfourpi$ & 13.9 / 10& ( 0.18 ) & 9.9 / 9& ( 0.36 ) & 27.9 / 19& ( 0.09 ) \\ 
Alternate & 16.6 / 10& ( 0.08 ) & 10.3 / 9& ( 0.33 ) & 27.0 / 19& ( 0.10 ) \\  
Optimal & 17.8 / 10& ( 0.06 ) & 9.9 / 9& ( 0.36 ) & 29.6 / 19& ( 0.06 ) \\ 
Optimal Alternate & 13.7 / 10& ( 0.19 ) & 17.2 / 9& ( 0.05 ) & 31.2 / 19& ( 0.04 ) \\
\end{tabular}
}
\caption{The compatibility of the measured \fourpi hadronic parameters with the model predictions for all binning schemes with $\mathcal{N} = 5$. \label{tab:modelDataChiSq} }
\end{table}

Using the measured \fourpi hadronic parameters, the \CP-even fraction of all phase space bins, $\tilde{{F}}^{4\pi}_{+}$, is calculated using the formula,
\begin{align}
\tilde{{F}}^{4\pi}_{+} = \frac{1}{2} + {\frac{1}{2} \sum_i{ \cifourpi \sqrt{\Frifourpi \Frbifourpi} } },
\end{align}
where the tilde indicates that a $\pip\pim$ mass window is excluded from the \decay{\D}{\fourpi} phase space i.e. ${F}^{4\pi}_{+}$ represents the \CP-even fraction for the entire phase space.
The values of $\tilde{{F}}^{4\pi}_{+}$ are presented in \tabref{tab:results}, and are consistent among binning schemes.
The nominal model is used to determine ${F}^{4\pi}_{+} - \tilde{{F}}^{4\pi}_{+} = -0.002 \pm 0.002$ which can be used as a correction factor to determine ${F}^{4\pi}_{+}$ from the values of $\tilde{{F}}^{4\pi}_{+}$ in \tabref{tab:results}.

The $Q$ value of each binning scheme is determined using \eqnref{eqn:qval}, and presented in \tabref{tab:results}; as expected, the optimal binning schemes give the largest $Q$ values.
The $Q$ value for a single phase space bin is calculated, using $\tilde{{F}}^{4\pi}_{+}$, to be $0.505$. 
Therefore, based on the relative $Q$ values, and using the optimal-alternative binning scheme with $\mathcal{N} = 5$, the increase in statistical power for a measurement of $\gamma$ is increased by $\sim2.2$ times with respect to the phase space integrated case.\footnote{ Note that it impossible to discuss the improvement in $\gamma$ sensitivity since an independent measurement of $\gamma$ is impossible in the phase space integrated regime. }

\begin{figure}[h]
  \centering
  
  \includegraphics[width=0.95\textwidth]{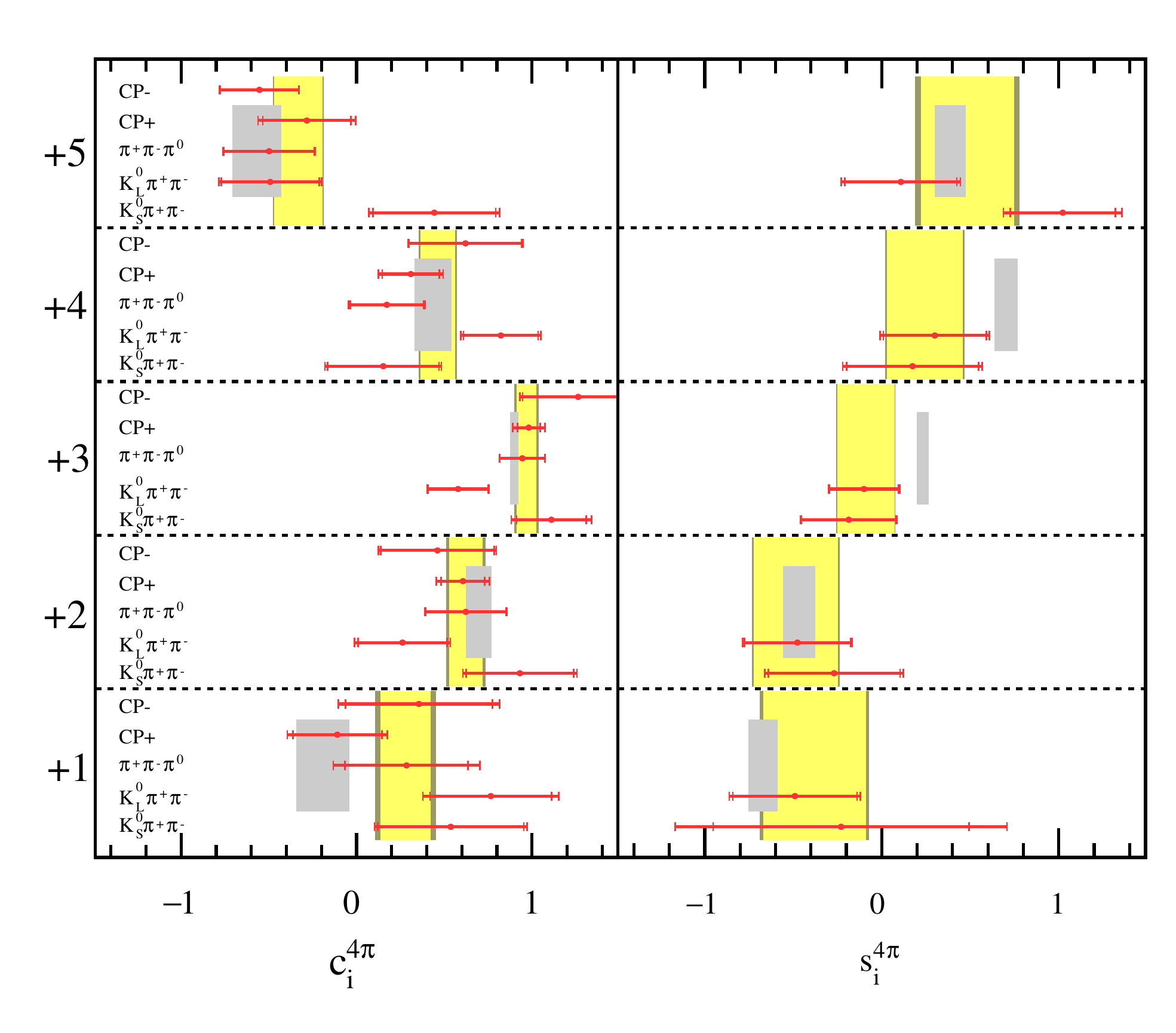}  
  
  \caption{ Constraints on the \cifourpi and \sifourpi parameters using the optimal alternative binning scheme with $\mathcal{N}=5$, determined using different subsets of tags. The grey bands show the model predictions and uncertainties. The red lines show the measured values and uncertainties when using a single subset of tags - the inner error bar shows the statistical uncertainty, and the outer error bar shows the combined statistical and systematic uncertainty. The yellow band shows the combined result using all subsets of tags - the lighter shade of yellow represents the statistical uncertainty, and the darker shade of yellow shows the combined statistical and systematic uncertainties.
 \label{fig:consistencyCheck}}
\end{figure}

\begin{figure}[h]
  \centering
  
  \includegraphics[width=0.95\textwidth]{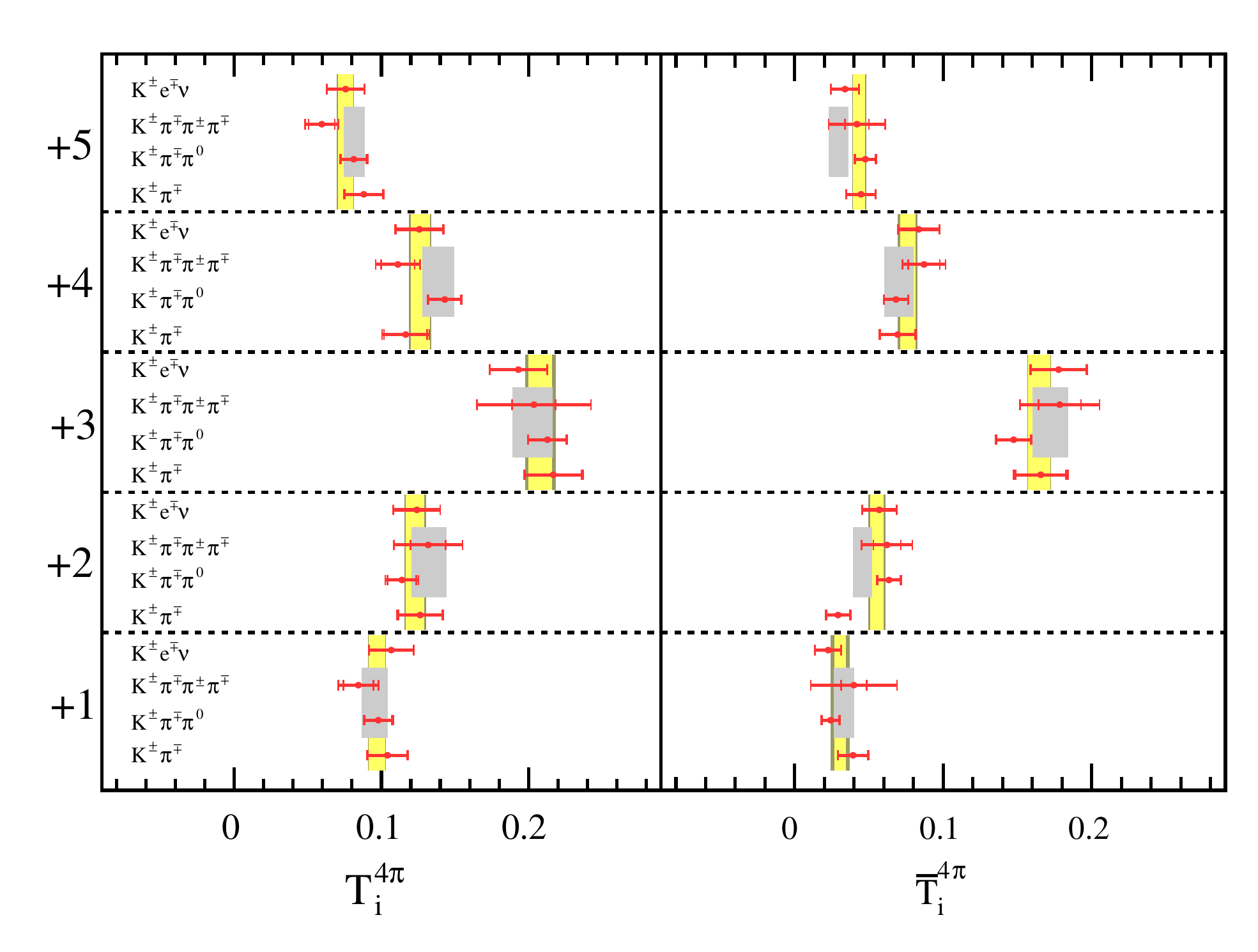}  
  
  \caption{ Constraints on the \Frifourpi and \Frbifourpi parameters (fraction of \Dz and \Dzb flavour tagged decays in each bin, repspectively) using the optimal alternative binning scheme with $\mathcal{N}=5$, determined using different subsets of tags. The grey bands show the model predictions and uncertainties. The red lines show the measured values and uncertainties when using a single subset of tags - the inner error bar shows the statistical uncertainty, and the outer error bar shows the combined statistical and systematic uncertainty. The yellow band shows the combined result using all subsets of tags - the lighter shade of yellow represents the statistical uncertainty, and the darker shade of yellow shows the combined statistical and systematic uncertainties.
 \label{fig:consistencyCheckK}}
\end{figure}

The consistency of the \cifourpi and \sifourpi constraints obtained using different categories of final state is shown in \figref{fig:consistencyCheck} and \figref{fig:consistencyCheckK} for the optimal alternative binning scheme with $\mathcal{N} = 5$.
For \figref{fig:consistencyCheck}, each fit to one of the five categories (\CPp, \CPm, \pipipiz, \KSpipi and \KLpipi) uses all flavour and quasi-flavour tags. 
The constraints obtained are consistent between all categories of final state.

\begin{figure}[h]
  \centering
  
  \includegraphics[width=0.95\textwidth]{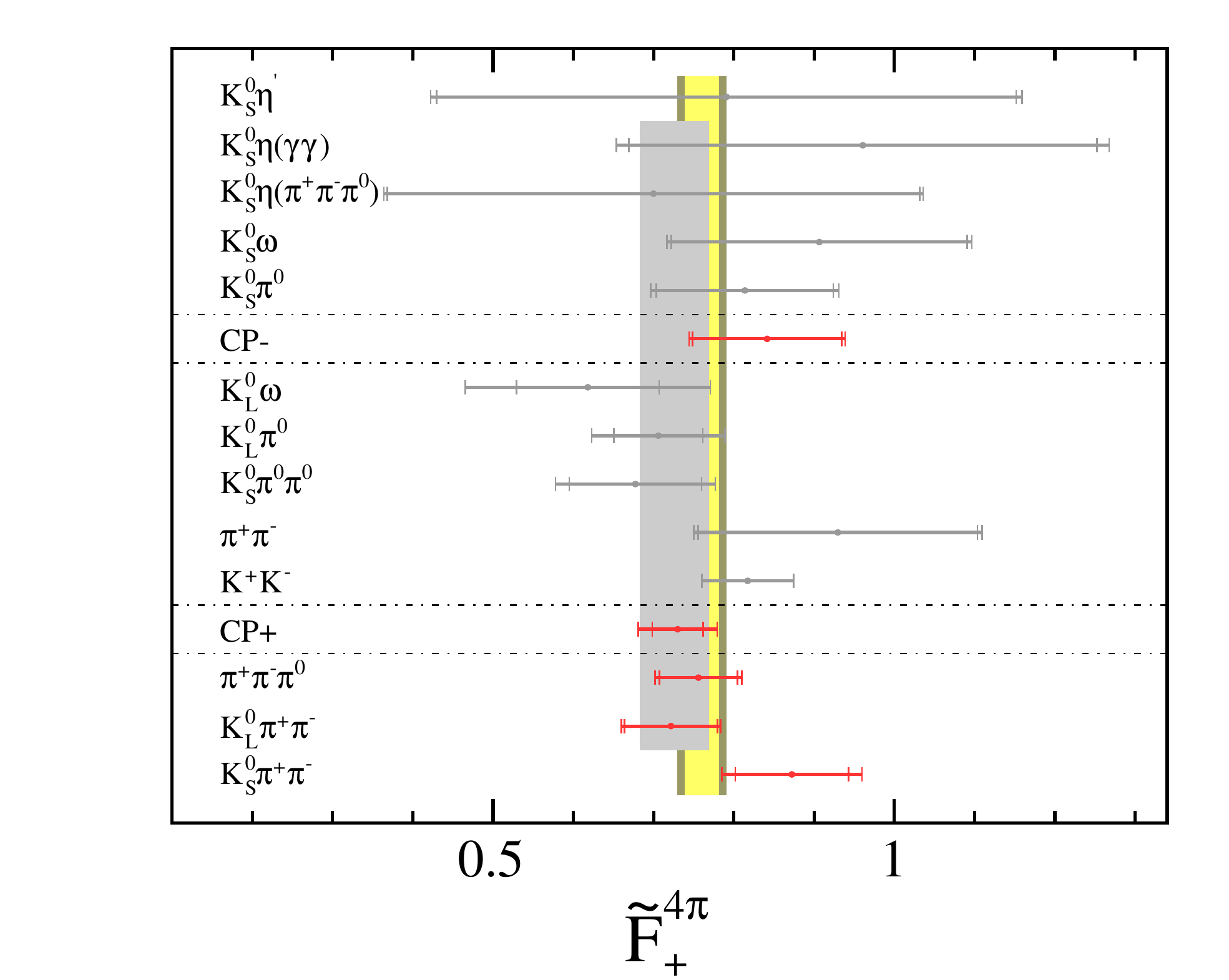}

  \caption{ The \CP-even fraction over all phase space bins, $\tilde{{F}}^{4\pi}_{+}$, determined using different subsets of tags. The grey bands show the model predictions and uncertainties. The red/grey lines show the measured values and uncertainties when using a single subset of tags - the inner error bar shows the statistical uncertainty, and the outer error bar shows the combined statistical and systematic uncertainty. The yellow band shows the combined result using all subsets of tags - the lighter shade of yellow represents the statistical uncertainty, and the darker shade of yellow shows the combined statistical and systematic uncertainties.
 \label{fig:fplusCheck}}
\end{figure}

The fit is also run using a single \fourpi phase space bin, which gives $\tilde{{F}}^{4\pi}_{+} = 0.760 \pm 0.021 \pm 0.021$.
The consistency of this result is checked between all final states in \figref{fig:fplusCheck}, following a similar method to the one used to obtain \figref{fig:consistencyCheck}.
Good consistency is observed. 

As a `default' binning scheme, we take the optimal-alternative binning with $\mathcal{N}=5$, as this has highest predicted $Q$ value. 
The default binning scheme also has the largest measured $Q$ value, although this information was not used to pick the default binning since it could bias the results. 
The value of $\tilde{{F}}^{4\pi}_{+}$ determined using the default model is $0.771 \pm 0.021 \pm 0.010$, which leads to ${{F}}^{4\pi}_{+} = 0.769 \pm 0.021 \pm 0.010 \pm 0.002$, where the final uncertainty is due to the \KS veto.
The value of ${{F}}^{4\pi}_{+}$ is an important input for determining the total \CP content of the neutral \D meson, which is related to the charm mixing parameter \ymix through \eqnref{eqn:citoy}~\cite{CPContOfDMeson}.

\begin{table}[h]
\centering
Equal \Deldelfourpi binning \\
\resizebox{0.95\columnwidth}{!}{
\begin{tabular}{ l | c  c  c  c } i & $c_i$ & $s_i$ & $T_i$ & $\overline{T}_i$ \\
\hline 
1 & \phantom{-}0.881 $\pm$ 0.053 $\pm$ 0.044 & \phantom{-}0.303 $\pm$ 0.149 $\pm$ 0.046 & 0.237 $\pm$ 0.008 $\pm$ 0.004 & 0.217 $\pm$ 0.008 $\pm$ 0.003\\
2 & \phantom{-}0.501 $\pm$ 0.084 $\pm$ 0.046 & -0.032 $\pm$ 0.201 $\pm$ 0.025 & 0.122 $\pm$ 0.006 $\pm$ 0.002 & 0.127 $\pm$ 0.006 $\pm$ 0.003\\
3 & \phantom{-}0.450 $\pm$ 0.113 $\pm$ 0.064 & \phantom{-}0.441 $\pm$ 0.228 $\pm$ 0.072 & 0.059 $\pm$ 0.004 $\pm$ 0.002 & 0.075 $\pm$ 0.005 $\pm$ 0.002\\
4 & -0.201 $\pm$ 0.167 $\pm$ 0.068 & \phantom{-}0.132 $\pm$ 0.304 $\pm$ 0.039 & 0.039 $\pm$ 0.004 $\pm$ 0.002 & 0.045 $\pm$ 0.004 $\pm$ 0.001\\
5 & -0.397 $\pm$ 0.152 $\pm$ 0.036 & -0.446 $\pm$ 0.381 $\pm$ 0.132 & 0.040 $\pm$ 0.004 $\pm$ 0.001 & 0.039 $\pm$ 0.004 $\pm$ 0.002\\
 $\tilde{F}_{+}^{4\pi}$ & \phantom{-}0.768 $\pm$ 0.021 $\pm$ 0.013 &  &  &  \\
 $Q$   & \phantom{-}0.733 $\pm$ 0.052 $\pm$ 0.035 &  &  &  \\
\end{tabular}

}
Variable \Deldelfourpi binning \\
\resizebox{0.95\columnwidth}{!}{
\begin{tabular}{ l | c  c  c  c } i & $c_i$ & $s_i$ & $T_i$ & $\overline{T}_i$ \\
\hline 
1 & \phantom{-}0.966 $\pm$ 0.101 $\pm$ 0.052 & \phantom{-}0.086 $\pm$ 0.316 $\pm$ 0.068 & 0.069 $\pm$ 0.005 $\pm$ 0.001 & 0.062 $\pm$ 0.004 $\pm$ 0.003\\
2 & \phantom{-}0.810 $\pm$ 0.070 $\pm$ 0.051 & -0.136 $\pm$ 0.229 $\pm$ 0.051 & 0.123 $\pm$ 0.006 $\pm$ 0.003 & 0.112 $\pm$ 0.006 $\pm$ 0.002\\
3 & \phantom{-}0.910 $\pm$ 0.080 $\pm$ 0.059 & \phantom{-}0.225 $\pm$ 0.259 $\pm$ 0.107 & 0.078 $\pm$ 0.005 $\pm$ 0.001 & 0.078 $\pm$ 0.005 $\pm$ 0.002\\
4 & \phantom{-}0.405 $\pm$ 0.083 $\pm$ 0.046 & \phantom{-}0.215 $\pm$ 0.188 $\pm$ 0.041 & 0.133 $\pm$ 0.006 $\pm$ 0.003 & 0.152 $\pm$ 0.006 $\pm$ 0.003\\
5 & -0.154 $\pm$ 0.105 $\pm$ 0.047 & \phantom{-}0.213 $\pm$ 0.207 $\pm$ 0.031 & 0.090 $\pm$ 0.005 $\pm$ 0.003 & 0.103 $\pm$ 0.006 $\pm$ 0.002\\
 $\tilde{F}_{+}^{4\pi}$ & \phantom{-}0.772 $\pm$ 0.021 $\pm$ 0.010 &  &  &  \\
 $Q$   & \phantom{-}0.698 $\pm$ 0.049 $\pm$ 0.020 &  &  &  \\
\end{tabular}

}
Alternative binning \\
\resizebox{0.95\columnwidth}{!}{
\begin{tabular}{ l | c  c  c  c } i & $c_i$ & $s_i$ & $T_i$ & $\overline{T}_i$ \\
\hline 
1 & -0.205 $\pm$ 0.189 $\pm$ 0.094 & -0.057 $\pm$ 0.384 $\pm$ 0.127 & 0.057 $\pm$ 0.004 $\pm$ 0.001 & 0.019 $\pm$ 0.003 $\pm$ 0.003\\
2 & \phantom{-}0.445 $\pm$ 0.105 $\pm$ 0.066 & -0.041 $\pm$ 0.259 $\pm$ 0.073 & 0.129 $\pm$ 0.006 $\pm$ 0.004 & 0.060 $\pm$ 0.005 $\pm$ 0.004\\
3 & \phantom{-}0.888 $\pm$ 0.053 $\pm$ 0.045 & -0.150 $\pm$ 0.159 $\pm$ 0.027 & 0.263 $\pm$ 0.008 $\pm$ 0.007 & 0.192 $\pm$ 0.007 $\pm$ 0.004\\
4 & \phantom{-}0.530 $\pm$ 0.097 $\pm$ 0.044 & \phantom{-}0.239 $\pm$ 0.209 $\pm$ 0.084 & 0.121 $\pm$ 0.006 $\pm$ 0.004 & 0.073 $\pm$ 0.005 $\pm$ 0.003\\
5 & -0.451 $\pm$ 0.162 $\pm$ 0.053 & -0.238 $\pm$ 0.416 $\pm$ 0.157 & 0.059 $\pm$ 0.004 $\pm$ 0.002 & 0.027 $\pm$ 0.003 $\pm$ 0.002\\
 $\tilde{F}_{+}^{4\pi}$ & \phantom{-}0.764 $\pm$ 0.022 $\pm$ 0.011 &  &  &  \\
 $Q$   & \phantom{-}0.702 $\pm$ 0.051 $\pm$ 0.027 &  &  &  \\
\end{tabular}

}
Optimal binning \\
\resizebox{0.95\columnwidth}{!}{
\begin{tabular}{ l | c  c  c  c } i & $c_i$ & $s_i$ & $T_i$ & $\overline{T}_i$ \\
\hline 
1 & \phantom{-}0.949 $\pm$ 0.057 $\pm$ 0.039 & -0.041 $\pm$ 0.171 $\pm$ 0.041 & 0.193 $\pm$ 0.007 $\pm$ 0.004 & 0.173 $\pm$ 0.007 $\pm$ 0.003\\
2 & \phantom{-}0.641 $\pm$ 0.110 $\pm$ 0.073 & \phantom{-}0.331 $\pm$ 0.257 $\pm$ 0.087 & 0.045 $\pm$ 0.004 $\pm$ 0.004 & 0.123 $\pm$ 0.006 $\pm$ 0.005\\
3 & \phantom{-}0.542 $\pm$ 0.094 $\pm$ 0.059 & \phantom{-}0.034 $\pm$ 0.224 $\pm$ 0.063 & 0.135 $\pm$ 0.006 $\pm$ 0.005 & 0.070 $\pm$ 0.005 $\pm$ 0.004\\
4 & \phantom{-}0.309 $\pm$ 0.123 $\pm$ 0.073 & \phantom{-}0.294 $\pm$ 0.236 $\pm$ 0.058 & 0.054 $\pm$ 0.005 $\pm$ 0.003 & 0.092 $\pm$ 0.005 $\pm$ 0.002\\
5 & -0.492 $\pm$ 0.130 $\pm$ 0.041 & \phantom{-}0.665 $\pm$ 0.256 $\pm$ 0.100 & 0.069 $\pm$ 0.004 $\pm$ 0.002 & 0.045 $\pm$ 0.004 $\pm$ 0.002\\
 $\tilde{F}_{+}^{4\pi}$ & \phantom{-}0.768 $\pm$ 0.021 $\pm$ 0.012 &  &  &  \\
 $Q$   & \phantom{-}0.757 $\pm$ 0.052 $\pm$ 0.026 &  &  &  \\
\end{tabular}

}
Optimal-alternative binning \\
\resizebox{0.95\columnwidth}{!}{
\begin{tabular}{ l | c  c  c  c } i & $c_i$ & $s_i$ & $T_i$ & $\overline{T}_i$ \\
\hline 
1 & \phantom{-}0.279 $\pm$ 0.143 $\pm$ 0.101 & -0.379 $\pm$ 0.291 $\pm$ 0.104 & 0.096 $\pm$ 0.005 $\pm$ 0.002 & 0.032 $\pm$ 0.004 $\pm$ 0.005\\
2 & \phantom{-}0.622 $\pm$ 0.095 $\pm$ 0.059 & -0.486 $\pm$ 0.237 $\pm$ 0.072 & 0.123 $\pm$ 0.006 $\pm$ 0.004 & 0.055 $\pm$ 0.004 $\pm$ 0.003\\
3 & \phantom{-}0.969 $\pm$ 0.057 $\pm$ 0.038 & -0.089 $\pm$ 0.162 $\pm$ 0.038 & 0.202 $\pm$ 0.007 $\pm$ 0.005 & 0.164 $\pm$ 0.007 $\pm$ 0.003\\
4 & \phantom{-}0.463 $\pm$ 0.099 $\pm$ 0.046 & \phantom{-}0.245 $\pm$ 0.215 $\pm$ 0.067 & 0.134 $\pm$ 0.006 $\pm$ 0.005 & 0.077 $\pm$ 0.005 $\pm$ 0.004\\
5 & -0.332 $\pm$ 0.138 $\pm$ 0.041 & \phantom{-}0.484 $\pm$ 0.263 $\pm$ 0.132 & 0.074 $\pm$ 0.005 $\pm$ 0.002 & 0.043 $\pm$ 0.004 $\pm$ 0.003\\
 $\tilde{F}_{+}^{4\pi}$ & \phantom{-}0.771 $\pm$ 0.021 $\pm$ 0.010 &  &  &  \\
 $Q$   & \phantom{-}0.760 $\pm$ 0.057 $\pm$ 0.017 &  &  &  \\
\end{tabular}

}
\caption{The hadronic parameters measured for each of the \fourpi binning schemes discussed in \secref{sec:binning} where $\mathcal{N} = 5$. The first uncertainty given is statistical, and the second systematic. Also given is the CP-even fraction, $\tilde{{F}}^{4\pi}_{+}$, and the $Q$ value, defined in \secref{sec:binning}; the uncertainties on these parameters are propagated from the statistical and systematic uncertainties on the hadronic parameters. \label{tab:results}   }
\end{table}

\clearpage

\clearpage

\section{Sensitivity studies}
\label{sec:gamma}

In this section the measured \fourpi hadronic parameters from \secref{sec:results} are used to simulate $\BtoDK,\decay{\D}{\fourpi}$ datasets, which in turn are used to estimate the sensitivity to $\gamma$. 
Three scenarios with different event yields are studied, based on measured and extrapolated $\BtoDK, \decay{\D}{\fourpi}$ event yields from \lhcb: 
``\lhcb run~I'', with event yields of $\sim1,500$ already recorded by \lhcb with $3\invfb$~\cite{cpObsTwoAndFourBody} of data; 
``\lhcb run II'', with plausible event yields of $\sim6,500$ at the end of the next \lhc data taking period with approximately twice the collision energy and an estimated $8\invfb$ of data; 
and ``\lhcb phase~1 upgrade'', with plausible event yields of $\sim100,000$ after phase 1 of the \lhcb upgrade. 
The increase in the heavy flavour cross section at higher collision energies is accounted for, along with the expected improvement in trigger efficiency at the \lhcb phase~1 upgrade~\cite{LHCbUpgradeCDR}. 
The extrapolations have of course large uncertainties.
The presence of background and systematic effects has been neglected in these studies, which is a reasonable assumption given previous measurements~\cite{cpObsTwoAndFourBody}.

Toy datasets of $\BtoDK,\decay{\D}{\fourpi}$ decays are generated using \eqnref{eqn:bmtodkbinned} and \eqnref{eqn:bptodkbinned} with $\delta_B = 140^\circ$, $\gamma = 70^\circ$ and $r_B = 0.1$. 
For each toy dataset, the central values of the \fourpi hadronic parameters are randomly sampled from the measured values and uncertainties.
When fitting the toy datasets, the parameters $\delta_B$, $\gamma$, $r_B$ and an overall normalisation parameter are allowed to float, whereas the \fourpi hadronic parameters are fixed to their measured values.
Therefore, the uncertainties obtained from the fit only account for the finite $\BtoDK,\decay{\D}{\fourpi}$ statistics, $\sigma_{\mathrm{stat}}$. 
The uncertainties on the parameters \cifourpi and \sifourpi are propagated to $\delta_B$, $\gamma$ and $r_B$ by repeating the fit 200 times, where for each fit \cifourpi and \sifourpi are randomly sampled from their associated covariance matrix. 
The covariance of the values obtained is used to assign an uncertainty, $\sigma_{\mathrm{had}}$. 
The parameters \kifourpi and \kbifourpi can be determined to an arbitrarily high precision at \lhcb using \decay{\Dstarp}{\Dz\pip} decays, so the uncertainties on these parameters are neglected.
As an alternative approach, the \cifourpi and \sifourpi parameters are Gaussian constrained in the fit, but this method was found to give a heavily biased estimate of $\gamma$, up to $70\%$ of the statistical uncertainty.  
The nominal fit method gives good coverage and small biases of less than $10\%$. 

The expected $\gamma$ uncertainties are presented in \tabref{tab:gamUncert} for several binning schemes. 
For each case the expected $\gamma$ uncertainty is median uncertainty determined from fits to 100 simulated datasets.
For each binning scheme type the uncertainty on $\gamma$ generally decreases with increasing numbers of bins - for illustration, the uncertainty on $\gamma$ is shown for the optimal-alternative binning scheme for $\mathcal{N} = 2-5$ in \tabref{tab:gamUncert}. 
The $\gamma$ uncertainties are also compared between different binning scheme types with $\mathcal{N} = 5$; all result in similar values of $\sigma_{\mathrm{stat}}(\gamma)$, although the values of $\sigma_{\mathrm{had}}(\gamma)$ are notably larger for the `variable \Deldelfourpi' and the `alternative' binning schemes. 
This is likely due to the measured central values of the \sifourpi parameters being consistent with zero for these schemes.
For the default binning (optimal alternative with $\mathcal{N} = 5$) the expected uncertainties are $(18 \oplus 13)^{\circ}$, $(10 \oplus 7)^{\circ}$ and $(2.5 \oplus 4.4)^{\circ}$ for the \lhcb ``Run~I'', ``Run~II'' and ``Phase~1 Upgrade'' scenarios, respectively, where the uncertainties are given in the form $\sigma_{\mathrm{stat}}\left(\gamma\right) \oplus \sigma_{\mathrm{had}}\left(\gamma\right)$. 

\begin{figure}[h]
     \centering

      \includegraphics[width=0.92\textwidth]{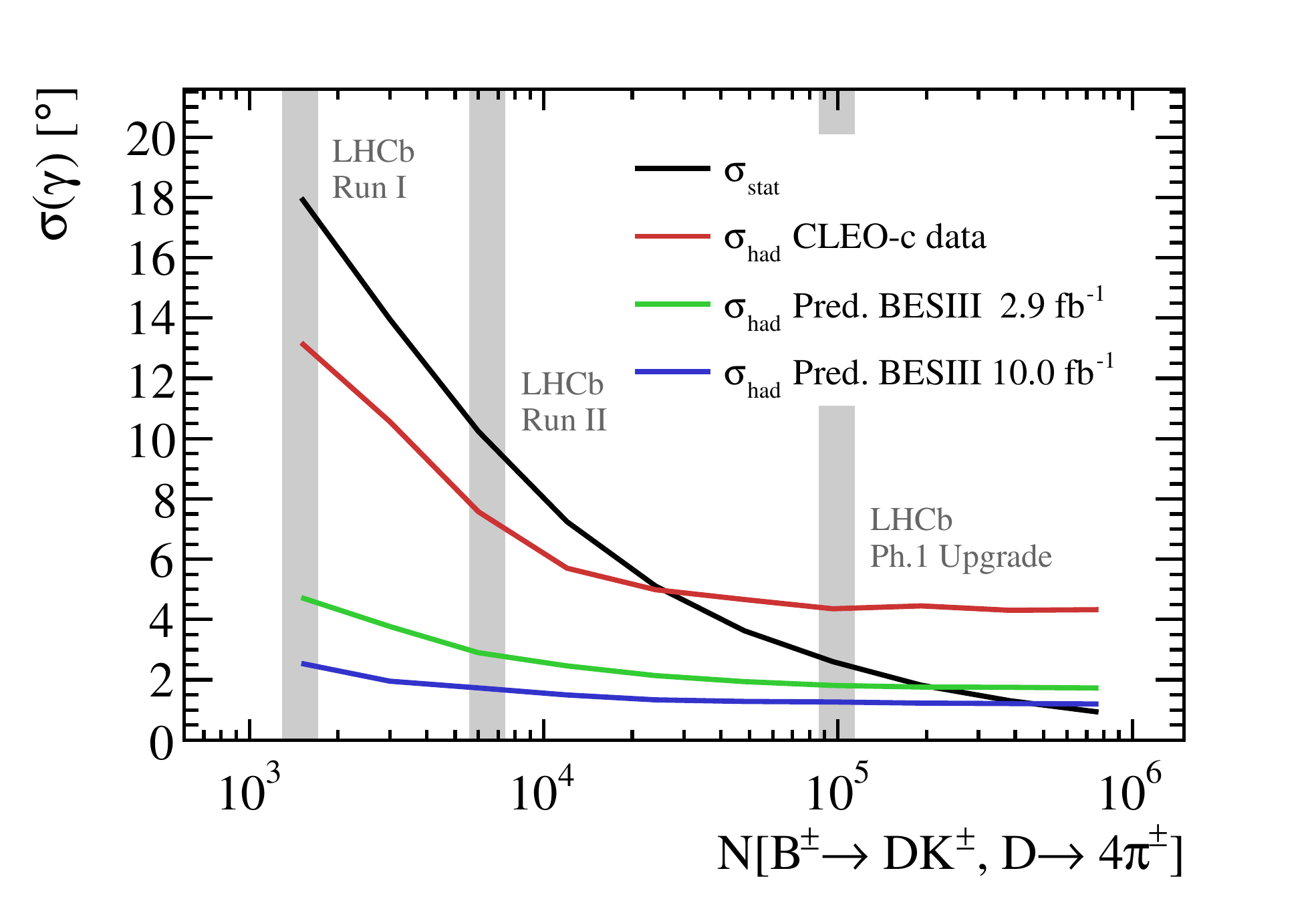}    
            
\caption{  Expected $\gamma$ uncertainties obtained using different numbers of $\BtoDK,\decay{\D}{\fourpi}$ decays and the default binning scheme. 
The black line shows the estimated uncertainty due to limited $\BtoDK,\decay{\D}{\fourpi}$ statistics. 
The red, green and blue lines shows the estimated uncertainty due to the measured/predicted constraints on the \fourpi hadronic parameters from \cleo-c data with 0.818\invfb, BES~III with 2.9\invfb, and BES~III with 10.0\invfb, respectively.
The grey bands highlight the event numbers that correspond to different \lhcb data taking periods.   \label{fig:gamSen} }
\end{figure}

Since $\sigma_{\mathrm{had}}\left(\gamma\right) \approx \sigma_{\mathrm{stat}}\left(\gamma\right)$ for the \lhcb ``Run~I'' and ``Run~II'' scenarios, and $\sigma_{\mathrm{had}}\left(\gamma\right) > \sigma_{\mathrm{stat}}\left(\gamma\right)$ for the ``Phase~1 Upgrade'' scenario, it is interesting to consider the impact that BESIII could have on reducing $\sigma_{\mathrm{had}}\left(\gamma\right)$.
Currently BESIII have collected 2.9\invfb of $\ep e^-$ collisions at the $\psi(3770)$ resonance, and a further $\sim7\invfb$ is planned for the future.
These datasets correspond to approximately 3.5 and 12 times the amount collected by \cleo-c, respectively. 
It is assumed that the uncertainties on the \fourpi hadronic parameters would be reduced by $1/\sqrt{3.5}$ and $1/\sqrt{12}$, respectively, compared to the constraints obtained in \secref{sec:results}.
The central values of the estimated BESIII measurements are different for each simulated dataset, and are randomly sampled from the constraints obtained in \secref{sec:results}.
\figref{fig:gamSen} shows, for the default binning scheme, the expected values of $\sigma_{\mathrm{had}}\left(\gamma\right)$ for different numbers of $\BtoDK,\decay{\D}{\fourpi}$ decays. 
This is shown for the hadronic parameter constraints measured in this paper, and the expected constraints for the two BESIII data taking periods. 
With $10.0\invfb$ of BESIII data, the expected $\gamma$ uncertainties become $(18 \oplus 3)^{\circ}$, $(10 \oplus 1.7)^{\circ}$ and $(2.5 \oplus 1.2)^{\circ}$ for the \lhcb ``Run~I'', ``Run~II'' and ``Phase~1 Upgrade'' scenarios, respectively.
It is also possible that BESIII could make further gains in sensitivity by using additional numbers of phase space bins. 
Improved constraints on the \fourpi hadronic parameters could be obtained using \D-mixing, as has been done for the $\Kpm\pimp\pipm\pimp$ final state in Ref.~\cite{K3piLHCb}; this would require further investigation.

\begin{table}
\centering
\begin{tabular}{ l l || C{2.5cm} | C{2.5cm} | C{2.5cm} } 
                          & &   \multicolumn{3}{c}{$\left( \sigma_{\mathrm{stat}}\left(\gamma\right) \oplus \sigma_{\mathrm{had}}\left(\gamma\right)  \right ) ~[^\circ]$ }         \\  \hline
               & & \lhcb & \lhcb & \lhcb   \\
 Binning scheme & $\mathcal{N}$ & Run~I & Run~II & Ph.1 Upgrade  \\
\hline
Optimal Alternative & 2&            20.4 $\oplus$ 27.0 &            16.2 $\oplus$ 20.5 & \phantom{0}4.6 $\oplus$ 15.6 \\
 & 3&             18.0 $\oplus$ 10.1 & \phantom{0}10.0 $\oplus$ \phantom{0}5.4 & \phantom{0}2.6 $\oplus$ \phantom{0}3.6 \\
 & 4&             18.2 $\oplus$ 15.9 &            10.5 $\oplus$ 10.6 & \phantom{0}2.9 $\oplus$ \phantom{0}6.5 \\
\hspace{0cm} {\it(default binning)} & 5&              18.0 $\oplus$ 13.2 & \phantom{0}9.7 $\oplus$ \phantom{0}7.4 & \phantom{0}2.5 $\oplus$ \phantom{0}4.4 \\
\hline
 Equal $\Deldelfourpi$ & 5 &         16.7 $\oplus$ 12.6 & \phantom{0}9.2 $\oplus$ \phantom{0}7.2 & \phantom{0}2.4 $\oplus$ \phantom{0}4.0 \\
 Variable $\Deldelfourpi$ & 5&           19.8 $\oplus$ 23.3 &            10.2 $\oplus$ 14.7 & \phantom{0}2.9 $\oplus$ 11.1 \\
Alternative & 5&        19.2 $\oplus$ 24.6 &            11.4 $\oplus$ 18.1 & \phantom{0}3.3 $\oplus$ 14.2 \\
Optimal  & 5&          17.3 $\oplus$ 13.9 & \phantom{0}10.0 $\oplus$ \phantom{0}7.9 & \phantom{0}2.6 $\oplus$ \phantom{0}5.0 \\
\end{tabular}
\caption{ Expected $\gamma$ sensitivity determined from simulated samples of $\BtoDK,\decay{\D}{\fourpi}$ decays for a variety of $\decay{\D}{\fourpi}$ binning schemes. 
Details of the simulation and fitting procedure can be found in the text.
The uncertainties are given for three different data taking periods of the \lhcb experiment, where the number of signal decays in each case it taken/extrapolated from existing measurements.
The uncertainty on $\gamma$ comes from two sources: the uncertainty due to to limited $\BtoDK,\decay{\D}{\fourpi}$ statistics, $\sigma_{\mathrm{stat}}\left(\gamma\right)$; and the uncertainty due to limited knowledge of the \fourpi hadronic parameters that are measured in this paper, $\sigma_{\mathrm{had}}\left(\gamma\right)$.
Both uncertainties are shown in the table, and are given in the format $\left( \sigma_{\mathrm{stat}}\left(\gamma\right) \oplus \sigma_{\mathrm{had}}\left(\gamma\right)  \right )$. 
All expected uncertainties are the median uncertainty from 100 simulated experiments. \label{tab:gamUncert} }
\end{table}

\section{Summary}
\label{sec:summary}
Using 818~$\rm pb^{-1}$ of $e^+e^-$ collision data collected by the CLEO-c detector, the hadronic parameters of the \decay{\D}{\fourpi} decay are measured in bins of phase space for the first time.
This allows the UT angle $\gamma$ to be determined using only \BtoDK decays where $\D$ decays to the \fourpi final state; previously only phase space integrated measurements have been possible~\cite{FPlusFourPi,cpObsTwoAndFourBody}, which need to be combined with other final states to obtain constraints on $\gamma$~\cite{cpObsTwoAndFourBody,LHCbGamCom}. 

The phase space of the \decay{\D}{\fourpi} decay is divided into bins based on the nominal amplitude model from Ref.~\cite{fourpimodel}. 
The equal and variable \Deldelfourpi binning schemes are based on an equal/variable division of \Deldelfourpi, whereas the alternate binning scheme also uses the relative magnitude of \decay{\Dz}{\fourpi} to \decay{\Dzb}{\fourpi} amplitudes. 
The optimal and optimal alternative binning schemes are defined to optimise the expected sensitivity to $\gamma$ in \BtoDK decays.
Although an amplitude model is used to inspire the binning schemes, the results are model-unbiased; any modelling deficiencies will only result in an increased statistical uncertainty on $\gamma$. 

Since amplitude models can be notoriously difficult to reproduce, it is useful to have a model-implementation independent method to represent a binning scheme. 
The phase space of the \decay{\D}{\fourpi} decay is five-dimensional, so using traditional techniques to divide the phase space into $N^{5}$ equally sized hypervolumes, where each is assigned a bin-number, would result in an unmanageable number of hypervolumes. 
An adaptive binning scheme is developed that uses an array of differently sized hypervolumes to drastically reduce the total number of hypervolumes needed, typically around $250,000$.

The measured values of the hadronic parameters are compared to the model-predictions, which show good agreement for the parameters \Frifourpi and \Frbifourpi, but a slight tension for \cifourpi and \sifourpi.
This could either be due to statistical fluctuations, which could be tested with larger datasets at BESIII, or a possible residual mismodelling of the phase motion across the \decay{\D}{\fourpi} phase space in Ref.~\cite{fourpimodel}.

The consistency of the results is checked using different subsets of final states, which give statistically compatible results. 
The \CP even fraction over all phase space bins, $\tilde{F}_{+}^{4\pi}$, is observed to be consistent between all binning schemes. 
Using the `default' binning scheme, ${F}_{+}^{4\pi}$ is determined as $0.769 \pm 0.021 \pm 0.010 \pm 0.002$ where the uncertainties are statistical, systematic, and from the \KS veto, respectively. 
This is the most precise determination ${F}_{+}^{4\pi}$ to date.

Using the \fourpi hadronic parameters measured in this paper, samples of $\BtoDK, \D\to\fourpi$ decays are simulated, then used to estimate the potential sensitivity to $\gamma$. 
It is shown that, using estimated sample sizes from \lhcb at the end of its current running period (``Run~II'') and the hadronic parameter constraints from this paper, constraints of $\sigma(\gamma) = (10 \oplus 7)^{\circ}$ could be obtained, potentially making \fourpi one of the most sensitive final states for a measurement of $\gamma$.
The first uncertainty is due to limited \BtoDK statistics, and the second is due to uncertainties on the \fourpi hadronic parameters.
It is shown that the latter uncertainty could be reduced to around $1.7^{\circ}$ by using current and future BESIII datasets.

\section{Acknowledgements}

This analysis was performed using CLEO-c data. The authors,
some of whom were members of CLEO, are grateful to the collaboration for the privilege
of using these data. We also thank Guy Wilkinson, Sneha Malde, Jim Libby, Tim Gershon, Philippe D'Argent, Kostas Petridis and Vincenzo Vagnoni for their valuable feedback on the paper draft.
We also acknowledge the support of the UK Science and Technology Facilities Council (STFC), and the European
Research Council's support under Framework 7 / ERC Grant Agreement number 307737.

\appendix

\clearpage
\section{Helicity variables}
\label{app:helicityVars}

In this paper the variables $\{  m_{+}, m_{-}, \cos\theta_{+}, \cos\theta_{-}, \phi  \}$ are used to parameterise a point in the \decay{\D}{\fourpi} phase space; their full definition is given in this appendix.
The variables $m_{+}$ and $m_{-}$ are defined,

\begin{align}
 m_{+}^{2} = (p_{\pi^{+}_{1}} + p_{\pi^{+}_{2}})^{2}, \\
 m_{-}^{2} = (p_{\pi^{-}_{1}} + p_{\pi^{-}_{2}})^{2}, 
\end{align}
where $p_{\pi^{+}_{1}}$ and $p_{\pi^{+}_{2}}$ ($p_{\pi^{-}_{1}}$ and $p_{\pi^{-}_{2}}$) are the four-vectors of the positively (negatively) charged pions in the final state.
The cosine of the two helicity angles, $\cos\theta_{+}$ and $\cos\theta_{-}$, are defined,

\begin{align}
\cos\theta_{+} = \frac{ \vec{ p}_{\pi^{+}_{1} } \cdot \vec{ p}_{D}  }{ |\vec{ p}_{\pi^{+}_{1} } || \vec{p}_{D}|  } \mathrm{ \ \ \ evaluated \ in \ the \ frame \ where \ \ \ } \vec{ p}_{\pi^{+}_{1} } + \vec{ p}_{\pi^{+}_{2} } = 0 \\
\cos\theta_{-} = \frac{ \vec{ p}_{\pi^{-}_{1} } \cdot \vec{ p}_{D}  }{ |\vec{ p}_{\pi^{-}_{1} } || \vec{p}_{D}|  } \mathrm{ \ \ \ evaluated \ in \ the \ frame \ where \ \ \ } \vec{ p}_{\pi^{-}_{1} } + \vec{ p}_{\pi^{-}_{2} } = 0
\end{align}
where $\vec{p}_{ \pi^{\pm}_{1,2} }$ is the three-vector associated to ${p}_{ \pi^{\pm}_{1,2} }$.
The angle between the $\pip\pip$ and $\pim\pim$ decay planes, $\phi$, is defined by,

\begin{align}
\sin \phi &= \left [ \frac{ (  \vec{ p}_{\pi^{+}_{1} } \times \vec{ p}_{\pi^{+}_{2} } )}{|  \vec{ p}_{\pi^{+}_{1} } \times \vec{ p}_{\pi^{+}_{2} } |} \times  \frac{(  \vec{ p}_{\pi^{-}_{1} } \times \vec{ p}_{\pi^{-}_{2} } ) }{  |  \vec{ p}_{\pi^{-}_{1} } \times \vec{ p}_{\pi^{-}_{2} } | } \right ]  \cdot \frac{ \vec{ p}_{\pi^{-}_{1} } + \vec{ p}_{\pi^{-}_{2} } }{ |\vec{ p}_{\pi^{-}_{1} } + \vec{ p}_{\pi^{-}_{2} }| } \mathrm{ \ \ \ evaluated \ in \ the \ }\vec{ p}_{D} = 0 \mathrm{ \ frame} \\
\cos \phi &= \left [ \frac{ (  \vec{ p}_{\pi^{+}_{1} } \times \vec{ p}_{\pi^{+}_{2} } )}{|  \vec{ p}_{\pi^{+}_{1} } \times \vec{ p}_{\pi^{+}_{2} } |} \ \cdot \ \frac{(  \vec{ p}_{\pi^{-}_{1} } \times \vec{ p}_{\pi^{-}_{2} } ) }{  |  \vec{ p}_{\pi^{-}_{1} } \times \vec{ p}_{\pi^{-}_{2} } | } \right ]  \hspace{2.3cm}\mathrm{ \ \ \ evaluated \ in \ the \ }\vec{ p}_{D} = 0 \mathrm{ \ frame}
\end{align}
Note that there is a conventional choice that can cause $\phi\to-\phi$, so it is important to copy these expressions exactly.

\clearpage

\clearpage
\section{Statistical and systematic correlations }
\label{app:correlations}
\begin{table}[h]
\centering
Equal \Deldelfourpi binning statistical correlations \\
\resizebox{1.05\columnwidth}{!}{
\begin{tabular}{ l | r  r  r  r  r  r  r  r  r  r  r  r  r  r  r  r  r  r  r  r }
   & $c^{4\pi}_{+1}$ & $c^{4\pi}_{+2}$ & $c^{4\pi}_{+3}$ & $c^{4\pi}_{+4}$ & $c^{4\pi}_{+5}$ & $s^{4\pi}_{+1}$ & $s^{4\pi}_{+2}$ & $s^{4\pi}_{+3}$ & $s^{4\pi}_{+4}$ & $s^{4\pi}_{+5}$ & $T^{4\pi}_{+1}$ & $T^{4\pi}_{+2}$ & $T^{4\pi}_{+3}$ & $T^{4\pi}_{+4}$ & $T^{4\pi}_{+5}$ & $T^{4\pi}_{-1}$ & $T^{4\pi}_{-2}$ & $T^{4\pi}_{-3}$ & $T^{4\pi}_{-4}$ & $T^{4\pi}_{-5}$\\
\hline
$c^{4\pi}_{+1}$ & 1.00 & 0.03 & 0.02 & 0.00 & 0.01 & 0.01 & 0.00 & 0.01 & 0.00 & 0.00 & -0.04 & 0.03 & 0.02 & 0.02 & 0.02 & -0.08 & 0.04 & 0.03 & 0.02 & 0.02\\
$c^{4\pi}_{+2}$ & 0.03 & 1.00 & 0.02 & 0.02 & 0.02 & -0.00 & 0.03 & 0.00 & 0.00 & -0.01 & 0.03 & -0.06 & 0.01 & 0.00 & 0.00 & 0.03 & -0.05 & 0.01 & 0.01 & 0.00\\
$c^{4\pi}_{+3}$ & 0.02 & 0.02 & 1.00 & 0.02 & 0.01 & 0.00 & 0.00 & 0.00 & -0.00 & 0.00 & 0.02 & 0.01 & -0.08 & 0.00 & 0.00 & 0.02 & 0.01 & -0.03 & 0.00 & 0.00\\
$c^{4\pi}_{+4}$ & 0.00 & 0.02 & 0.02 & 1.00 & 0.02 & -0.01 & 0.00 & -0.01 & 0.03 & -0.01 & 0.02 & 0.01 & 0.01 & -0.06 & 0.00 & 0.02 & 0.01 & 0.01 & -0.06 & 0.00\\
$c^{4\pi}_{+5}$ & 0.01 & 0.02 & 0.01 & 0.02 & 1.00 & -0.01 & -0.00 & -0.01 & -0.00 & -0.09 & 0.01 & 0.00 & 0.00 & -0.00 & -0.03 & 0.01 & 0.00 & 0.00 & -0.00 & -0.03\\
$s^{4\pi}_{+1}$ & 0.01 & -0.00 & 0.00 & -0.01 & -0.01 & 1.00 & -0.01 & 0.05 & 0.01 & 0.02 & -0.02 & 0.00 & -0.01 & 0.00 & 0.01 & 0.00 & 0.00 & 0.01 & 0.00 & 0.00\\
$s^{4\pi}_{+2}$ & 0.00 & 0.03 & 0.00 & 0.00 & -0.00 & -0.01 & 1.00 & -0.00 & 0.02 & 0.00 & -0.00 & 0.00 & -0.00 & -0.00 & 0.00 & 0.00 & -0.01 & 0.00 & 0.00 & -0.00\\
$s^{4\pi}_{+3}$ & 0.01 & 0.00 & 0.00 & -0.01 & -0.01 & 0.05 & -0.00 & 1.00 & -0.00 & 0.01 & 0.00 & 0.00 & -0.06 & 0.00 & 0.00 & 0.00 & 0.00 & 0.04 & 0.00 & 0.00\\
$s^{4\pi}_{+4}$ & 0.00 & 0.00 & -0.00 & 0.03 & -0.00 & 0.01 & 0.02 & -0.00 & 1.00 & -0.01 & -0.00 & 0.00 & -0.00 & 0.01 & 0.00 & 0.01 & 0.00 & 0.00 & -0.03 & -0.00\\
$s^{4\pi}_{+5}$ & 0.00 & -0.01 & 0.00 & -0.01 & -0.09 & 0.02 & 0.00 & 0.01 & -0.01 & 1.00 & -0.00 & -0.00 & -0.00 & 0.00 & -0.03 & -0.00 & -0.00 & 0.00 & 0.00 & 0.05\\
$T^{4\pi}_{+1}$ & -0.04 & 0.03 & 0.02 & 0.02 & 0.01 & -0.02 & -0.00 & 0.00 & -0.00 & -0.00 & 1.00 & -0.17 & -0.11 & -0.10 & -0.10 & -0.42 & -0.17 & -0.14 & -0.11 & -0.10\\
$T^{4\pi}_{+2}$ & 0.03 & -0.06 & 0.01 & 0.01 & 0.00 & 0.00 & 0.00 & 0.00 & 0.00 & -0.00 & -0.17 & 1.00 & -0.08 & -0.07 & -0.07 & -0.16 & -0.27 & -0.09 & -0.07 & -0.07\\
$T^{4\pi}_{+3}$ & 0.02 & 0.01 & -0.08 & 0.01 & 0.00 & -0.01 & -0.00 & -0.06 & -0.00 & -0.00 & -0.11 & -0.08 & 1.00 & -0.05 & -0.05 & -0.11 & -0.08 & -0.23 & -0.05 & -0.04\\
$T^{4\pi}_{+4}$ & 0.02 & 0.00 & 0.00 & -0.06 & -0.00 & 0.00 & -0.00 & 0.00 & 0.01 & 0.00 & -0.10 & -0.07 & -0.05 & 1.00 & -0.04 & -0.10 & -0.07 & -0.05 & -0.16 & -0.04\\
$T^{4\pi}_{+5}$ & 0.02 & 0.00 & 0.00 & 0.00 & -0.03 & 0.01 & 0.00 & 0.00 & 0.00 & -0.03 & -0.10 & -0.07 & -0.05 & -0.04 & 1.00 & -0.09 & -0.07 & -0.05 & -0.04 & -0.15\\
$T^{4\pi}_{-1}$ & -0.08 & 0.03 & 0.02 & 0.02 & 0.01 & 0.00 & 0.00 & 0.00 & 0.01 & -0.00 & -0.42 & -0.16 & -0.11 & -0.10 & -0.09 & 1.00 & -0.16 & -0.11 & -0.10 & -0.09\\
$T^{4\pi}_{-2}$ & 0.04 & -0.05 & 0.01 & 0.01 & 0.00 & 0.00 & -0.01 & 0.00 & 0.00 & -0.00 & -0.17 & -0.27 & -0.08 & -0.07 & -0.07 & -0.16 & 1.00 & -0.09 & -0.07 & -0.07\\
$T^{4\pi}_{-3}$ & 0.03 & 0.01 & -0.03 & 0.01 & 0.00 & 0.01 & 0.00 & 0.04 & 0.00 & 0.00 & -0.14 & -0.09 & -0.23 & -0.05 & -0.05 & -0.11 & -0.09 & 1.00 & -0.05 & -0.05\\
$T^{4\pi}_{-4}$ & 0.02 & 0.01 & 0.00 & -0.06 & -0.00 & 0.00 & 0.00 & 0.00 & -0.03 & 0.00 & -0.11 & -0.07 & -0.05 & -0.16 & -0.04 & -0.10 & -0.07 & -0.05 & 1.00 & -0.04\\
$T^{4\pi}_{-5}$ & 0.02 & 0.00 & 0.00 & 0.00 & -0.03 & 0.00 & -0.00 & 0.00 & -0.00 & 0.05 & -0.10 & -0.07 & -0.04 & -0.04 & -0.15 & -0.09 & -0.07 & -0.05 & -0.04 & 1.00\\
\end{tabular}

}

\vspace{0.7cm}

Equal \Deldelfourpi binning systematic correlations \\
\resizebox{1.05\columnwidth}{!}{
\begin{tabular}{ l | r  r  r  r  r  r  r  r  r  r  r  r  r  r  r  r  r  r  r  r }
   & $c^{4\pi}_{+1}$ & $c^{4\pi}_{+2}$ & $c^{4\pi}_{+3}$ & $c^{4\pi}_{+4}$ & $c^{4\pi}_{+5}$ & $s^{4\pi}_{+1}$ & $s^{4\pi}_{+2}$ & $s^{4\pi}_{+3}$ & $s^{4\pi}_{+4}$ & $s^{4\pi}_{+5}$ & $T^{4\pi}_{+1}$ & $T^{4\pi}_{+2}$ & $T^{4\pi}_{+3}$ & $T^{4\pi}_{+4}$ & $T^{4\pi}_{+5}$ & $T^{4\pi}_{-1}$ & $T^{4\pi}_{-2}$ & $T^{4\pi}_{-3}$ & $T^{4\pi}_{-4}$ & $T^{4\pi}_{-5}$\\
\hline
$c^{4\pi}_{+1}$ & 1.00 & -0.01 & 0.00 & 0.01 & -0.01 & -0.00 & -0.04 & -0.00 & -0.01 & 0.01 & 0.01 & -0.01 & 0.00 & 0.01 & -0.00 & -0.02 & 0.00 & 0.00 & -0.02 & 0.03\\
$c^{4\pi}_{+2}$ & -0.01 & 1.00 & 0.04 & 0.04 & 0.05 & 0.01 & 0.06 & 0.01 & 0.04 & 0.00 & -0.00 & 0.06 & 0.00 & -0.02 & -0.07 & -0.00 & -0.04 & 0.05 & 0.03 & -0.01\\
$c^{4\pi}_{+3}$ & 0.00 & 0.04 & 1.00 & 0.04 & 0.04 & 0.03 & -0.01 & 0.00 & -0.00 & -0.01 & -0.00 & 0.04 & -0.00 & 0.01 & -0.01 & -0.00 & -0.05 & 0.06 & -0.00 & -0.01\\
$c^{4\pi}_{+4}$ & 0.01 & 0.04 & 0.04 & 1.00 & 0.05 & 0.02 & 0.05 & 0.01 & -0.00 & 0.01 & 0.02 & -0.03 & -0.01 & -0.01 & 0.02 & 0.00 & -0.03 & 0.04 & 0.00 & -0.00\\
$c^{4\pi}_{+5}$ & -0.01 & 0.05 & 0.04 & 0.05 & 1.00 & 0.07 & -0.03 & 0.00 & -0.00 & -0.02 & 0.03 & -0.02 & -0.02 & -0.02 & 0.02 & 0.03 & -0.05 & 0.00 & -0.00 & -0.01\\
$s^{4\pi}_{+1}$ & -0.00 & 0.01 & 0.03 & 0.02 & 0.07 & 1.00 & 0.08 & -0.01 & -0.02 & -0.01 & -0.10 & 0.09 & 0.00 & -0.02 & 0.09 & 0.12 & -0.06 & 0.02 & -0.00 & -0.02\\
$s^{4\pi}_{+2}$ & -0.04 & 0.06 & -0.01 & 0.05 & -0.03 & 0.08 & 1.00 & 0.04 & 0.03 & -0.03 & -0.11 & -0.13 & -0.01 & -0.01 & 0.12 & 0.19 & 0.09 & 0.01 & 0.04 & -0.20\\
$s^{4\pi}_{+3}$ & -0.00 & 0.01 & 0.00 & 0.01 & 0.00 & -0.01 & 0.04 & 1.00 & 0.00 & -0.01 & 0.00 & -0.05 & -0.01 & -0.01 & 0.02 & 0.01 & 0.02 & -0.00 & 0.04 & -0.02\\
$s^{4\pi}_{+4}$ & -0.01 & 0.04 & -0.00 & -0.00 & -0.00 & -0.02 & 0.03 & 0.00 & 1.00 & -0.02 & -0.01 & -0.03 & 0.02 & -0.01 & -0.04 & 0.06 & -0.05 & 0.06 & -0.01 & -0.03\\
$s^{4\pi}_{+5}$ & 0.01 & 0.00 & -0.01 & 0.01 & -0.02 & -0.01 & -0.03 & -0.01 & -0.02 & 1.00 & 0.02 & 0.06 & 0.01 & -0.01 & -0.04 & -0.06 & 0.01 & -0.01 & -0.03 & 0.06\\
$T^{4\pi}_{+1}$ & 0.01 & -0.00 & -0.00 & 0.02 & 0.03 & -0.10 & -0.11 & 0.00 & -0.01 & 0.02 & 1.00 & 0.01 & -0.00 & 0.00 & -0.06 & -0.18 & -0.03 & -0.00 & -0.06 & 0.07\\
$T^{4\pi}_{+2}$ & -0.01 & 0.06 & 0.04 & -0.03 & -0.02 & 0.09 & -0.13 & -0.05 & -0.03 & 0.06 & 0.01 & 1.00 & 0.03 & -0.09 & -0.25 & -0.19 & -0.14 & 0.12 & -0.03 & 0.12\\
$T^{4\pi}_{+3}$ & 0.00 & 0.00 & -0.00 & -0.01 & -0.02 & 0.00 & -0.01 & -0.01 & 0.02 & 0.01 & -0.00 & 0.03 & 1.00 & -0.01 & -0.05 & 0.00 & 0.00 & -0.05 & -0.08 & 0.02\\
$T^{4\pi}_{+4}$ & 0.01 & -0.02 & 0.01 & -0.01 & -0.02 & -0.02 & -0.01 & -0.01 & -0.01 & -0.01 & 0.00 & -0.09 & -0.01 & 1.00 & 0.05 & -0.02 & -0.00 & 0.01 & -0.05 & -0.00\\
$T^{4\pi}_{+5}$ & -0.00 & -0.07 & -0.01 & 0.02 & 0.02 & 0.09 & 0.12 & 0.02 & -0.04 & -0.04 & -0.06 & -0.25 & -0.05 & 0.05 & 1.00 & 0.16 & 0.02 & -0.03 & 0.02 & -0.10\\
$T^{4\pi}_{-1}$ & -0.02 & -0.00 & -0.00 & 0.00 & 0.03 & 0.12 & 0.19 & 0.01 & 0.06 & -0.06 & -0.18 & -0.19 & 0.00 & -0.02 & 0.16 & 1.00 & 0.03 & -0.04 & 0.03 & -0.16\\
$T^{4\pi}_{-2}$ & 0.00 & -0.04 & -0.05 & -0.03 & -0.05 & -0.06 & 0.09 & 0.02 & -0.05 & 0.01 & -0.03 & -0.14 & 0.00 & -0.00 & 0.02 & 0.03 & 1.00 & -0.16 & 0.05 & -0.05\\
$T^{4\pi}_{-3}$ & 0.00 & 0.05 & 0.06 & 0.04 & 0.00 & 0.02 & 0.01 & -0.00 & 0.06 & -0.01 & -0.00 & 0.12 & -0.05 & 0.01 & -0.03 & -0.04 & -0.16 & 1.00 & 0.01 & 0.00\\
$T^{4\pi}_{-4}$ & -0.02 & 0.03 & -0.00 & 0.00 & -0.00 & -0.00 & 0.04 & 0.04 & -0.01 & -0.03 & -0.06 & -0.03 & -0.08 & -0.05 & 0.02 & 0.03 & 0.05 & 0.01 & 1.00 & -0.05\\
$T^{4\pi}_{-5}$ & 0.03 & -0.01 & -0.01 & -0.00 & -0.01 & -0.02 & -0.20 & -0.02 & -0.03 & 0.06 & 0.07 & 0.12 & 0.02 & -0.00 & -0.10 & -0.16 & -0.05 & 0.00 & -0.05 & 1.00\\
\end{tabular}

}

\caption{ The statistical and systematic correlations between the \fourpi hadronic parameters using the equal \Deldelfourpi binning scheme with $\mathcal{N} = 5$. \label{tab:correlations1}}
\end{table}


\begin{table}
\centering
Variable \Deldelfourpi binning statistical correlations \\
\resizebox{1.05\columnwidth}{!}{
\begin{tabular}{ l | r  r  r  r  r  r  r  r  r  r  r  r  r  r  r  r  r  r  r  r }
   & $c^{4\pi}_{+1}$ & $c^{4\pi}_{+2}$ & $c^{4\pi}_{+3}$ & $c^{4\pi}_{+4}$ & $c^{4\pi}_{+5}$ & $s^{4\pi}_{+1}$ & $s^{4\pi}_{+2}$ & $s^{4\pi}_{+3}$ & $s^{4\pi}_{+4}$ & $s^{4\pi}_{+5}$ & $T^{4\pi}_{+1}$ & $T^{4\pi}_{+2}$ & $T^{4\pi}_{+3}$ & $T^{4\pi}_{+4}$ & $T^{4\pi}_{+5}$ & $T^{4\pi}_{-1}$ & $T^{4\pi}_{-2}$ & $T^{4\pi}_{-3}$ & $T^{4\pi}_{-4}$ & $T^{4\pi}_{-5}$\\
\hline
$c^{4\pi}_{+1}$ & 1.00 & 0.03 & 0.02 & 0.01 & 0.00 & -0.05 & 0.02 & -0.00 & 0.01 & 0.01 & -0.08 & 0.02 & 0.02 & 0.02 & 0.02 & -0.14 & 0.02 & 0.02 & 0.03 & 0.02\\
$c^{4\pi}_{+2}$ & 0.03 & 1.00 & 0.02 & 0.02 & 0.01 & -0.01 & 0.01 & 0.00 & 0.01 & 0.01 & 0.01 & -0.05 & 0.02 & 0.02 & 0.01 & 0.01 & -0.07 & 0.02 & 0.02 & 0.02\\
$c^{4\pi}_{+3}$ & 0.02 & 0.02 & 1.00 & 0.02 & 0.01 & -0.01 & 0.01 & 0.01 & 0.01 & 0.01 & 0.01 & 0.02 & -0.07 & 0.02 & 0.01 & 0.01 & 0.02 & -0.08 & 0.02 & 0.02\\
$c^{4\pi}_{+4}$ & 0.01 & 0.02 & 0.02 & 1.00 & 0.04 & 0.00 & -0.00 & -0.00 & 0.02 & -0.00 & 0.02 & 0.03 & 0.02 & -0.09 & 0.01 & 0.02 & 0.03 & 0.02 & -0.06 & 0.02\\
$c^{4\pi}_{+5}$ & 0.00 & 0.01 & 0.01 & 0.04 & 1.00 & 0.01 & -0.01 & -0.00 & -0.00 & 0.00 & 0.01 & 0.01 & 0.01 & 0.01 & -0.04 & 0.01 & 0.01 & 0.01 & 0.01 & -0.04\\
$s^{4\pi}_{+1}$ & -0.05 & -0.01 & -0.01 & 0.00 & 0.01 & 1.00 & -0.16 & 0.03 & -0.09 & -0.08 & -0.01 & -0.00 & 0.00 & 0.00 & 0.00 & 0.01 & 0.00 & 0.00 & -0.01 & -0.00\\
$s^{4\pi}_{+2}$ & 0.02 & 0.01 & 0.01 & -0.00 & -0.01 & -0.16 & 1.00 & -0.08 & 0.09 & 0.12 & -0.00 & -0.00 & -0.00 & -0.01 & -0.00 & -0.00 & 0.01 & 0.00 & 0.01 & 0.00\\
$s^{4\pi}_{+3}$ & -0.00 & 0.00 & 0.01 & -0.00 & -0.00 & 0.03 & -0.08 & 1.00 & 0.01 & -0.01 & 0.00 & -0.00 & 0.01 & -0.00 & 0.00 & 0.00 & 0.00 & -0.01 & 0.00 & 0.00\\
$s^{4\pi}_{+4}$ & 0.01 & 0.01 & 0.01 & 0.02 & -0.00 & -0.09 & 0.09 & 0.01 & 1.00 & 0.08 & 0.00 & 0.00 & 0.00 & -0.01 & 0.00 & -0.00 & 0.00 & 0.00 & 0.00 & 0.01\\
$s^{4\pi}_{+5}$ & 0.01 & 0.01 & 0.01 & -0.00 & 0.00 & -0.08 & 0.12 & -0.01 & 0.08 & 1.00 & 0.00 & 0.00 & 0.00 & -0.00 & 0.01 & -0.00 & -0.00 & 0.00 & 0.00 & -0.01\\
$T^{4\pi}_{+1}$ & -0.08 & 0.01 & 0.01 & 0.02 & 0.01 & -0.01 & -0.00 & 0.00 & 0.00 & 0.00 & 1.00 & -0.08 & -0.06 & -0.09 & -0.08 & -0.22 & -0.08 & -0.07 & -0.10 & -0.08\\
$T^{4\pi}_{+2}$ & 0.02 & -0.05 & 0.02 & 0.03 & 0.01 & -0.00 & -0.00 & -0.00 & 0.00 & 0.00 & -0.08 & 1.00 & -0.09 & -0.12 & -0.10 & -0.08 & -0.27 & -0.09 & -0.13 & -0.11\\
$T^{4\pi}_{+3}$ & 0.02 & 0.02 & -0.07 & 0.02 & 0.01 & 0.00 & -0.00 & 0.01 & 0.00 & 0.00 & -0.06 & -0.09 & 1.00 & -0.09 & -0.08 & -0.06 & -0.08 & -0.22 & -0.10 & -0.09\\
$T^{4\pi}_{+4}$ & 0.02 & 0.02 & 0.02 & -0.09 & 0.01 & 0.00 & -0.01 & -0.00 & -0.01 & -0.00 & -0.09 & -0.12 & -0.09 & 1.00 & -0.11 & -0.09 & -0.12 & -0.10 & -0.30 & -0.12\\
$T^{4\pi}_{+5}$ & 0.02 & 0.01 & 0.01 & 0.01 & -0.04 & 0.00 & -0.00 & 0.00 & 0.00 & 0.01 & -0.08 & -0.10 & -0.08 & -0.11 & 1.00 & -0.07 & -0.10 & -0.08 & -0.12 & -0.23\\
$T^{4\pi}_{-1}$ & -0.14 & 0.01 & 0.01 & 0.02 & 0.01 & 0.01 & -0.00 & 0.00 & -0.00 & -0.00 & -0.22 & -0.08 & -0.06 & -0.09 & -0.07 & 1.00 & -0.08 & -0.06 & -0.09 & -0.08\\
$T^{4\pi}_{-2}$ & 0.02 & -0.07 & 0.02 & 0.03 & 0.01 & 0.00 & 0.01 & 0.00 & 0.00 & -0.00 & -0.08 & -0.27 & -0.08 & -0.12 & -0.10 & -0.08 & 1.00 & -0.08 & -0.12 & -0.10\\
$T^{4\pi}_{-3}$ & 0.02 & 0.02 & -0.08 & 0.02 & 0.01 & 0.00 & 0.00 & -0.01 & 0.00 & 0.00 & -0.07 & -0.09 & -0.22 & -0.10 & -0.08 & -0.06 & -0.08 & 1.00 & -0.10 & -0.08\\
$T^{4\pi}_{-4}$ & 0.03 & 0.02 & 0.02 & -0.06 & 0.01 & -0.01 & 0.01 & 0.00 & 0.00 & 0.00 & -0.10 & -0.13 & -0.10 & -0.30 & -0.12 & -0.09 & -0.12 & -0.10 & 1.00 & -0.12\\
$T^{4\pi}_{-5}$ & 0.02 & 0.02 & 0.02 & 0.02 & -0.04 & -0.00 & 0.00 & 0.00 & 0.01 & -0.01 & -0.08 & -0.11 & -0.09 & -0.12 & -0.23 & -0.08 & -0.10 & -0.08 & -0.12 & 1.00\\
\end{tabular}

}

\vspace{1cm}

Variable \Deldelfourpi binning systematic correlations \\
\resizebox{1.05\columnwidth}{!}{
\begin{tabular}{ l | r  r  r  r  r  r  r  r  r  r  r  r  r  r  r  r  r  r  r  r }
   & $c^{4\pi}_{+1}$ & $c^{4\pi}_{+2}$ & $c^{4\pi}_{+3}$ & $c^{4\pi}_{+4}$ & $c^{4\pi}_{+5}$ & $s^{4\pi}_{+1}$ & $s^{4\pi}_{+2}$ & $s^{4\pi}_{+3}$ & $s^{4\pi}_{+4}$ & $s^{4\pi}_{+5}$ & $T^{4\pi}_{+1}$ & $T^{4\pi}_{+2}$ & $T^{4\pi}_{+3}$ & $T^{4\pi}_{+4}$ & $T^{4\pi}_{+5}$ & $T^{4\pi}_{-1}$ & $T^{4\pi}_{-2}$ & $T^{4\pi}_{-3}$ & $T^{4\pi}_{-4}$ & $T^{4\pi}_{-5}$\\
\hline
$c^{4\pi}_{+1}$ & 1.00 & 0.01 & 0.01 & 0.01 & -0.02 & -0.01 & -0.02 & 0.00 & -0.00 & 0.03 & 0.01 & -0.00 & -0.00 & 0.00 & -0.01 & -0.01 & 0.01 & -0.01 & -0.02 & 0.06\\
$c^{4\pi}_{+2}$ & 0.01 & 1.00 & 0.00 & 0.01 & 0.01 & -0.02 & -0.00 & 0.00 & 0.02 & 0.01 & 0.01 & 0.02 & 0.01 & -0.01 & -0.02 & -0.01 & -0.01 & 0.02 & 0.00 & 0.02\\
$c^{4\pi}_{+3}$ & 0.01 & 0.00 & 1.00 & -0.01 & -0.01 & -0.00 & -0.01 & 0.00 & -0.00 & 0.01 & 0.00 & 0.00 & 0.03 & 0.01 & -0.00 & -0.01 & -0.01 & -0.01 & -0.02 & 0.04\\
$c^{4\pi}_{+4}$ & 0.01 & 0.01 & -0.01 & 1.00 & 0.04 & -0.00 & 0.02 & 0.00 & -0.01 & 0.00 & 0.03 & -0.00 & -0.01 & -0.02 & 0.00 & -0.00 & -0.02 & 0.06 & 0.01 & -0.02\\
$c^{4\pi}_{+5}$ & -0.02 & 0.01 & -0.01 & 0.04 & 1.00 & 0.05 & -0.03 & 0.00 & -0.01 & -0.02 & 0.04 & 0.01 & -0.01 & -0.01 & 0.01 & 0.00 & -0.04 & 0.02 & 0.01 & -0.02\\
$s^{4\pi}_{+1}$ & -0.01 & -0.02 & -0.00 & -0.00 & 0.05 & 1.00 & -0.05 & -0.02 & -0.05 & -0.03 & -0.11 & 0.04 & -0.00 & -0.01 & 0.04 & 0.05 & -0.07 & 0.01 & -0.02 & -0.01\\
$s^{4\pi}_{+2}$ & -0.02 & -0.00 & -0.01 & 0.02 & -0.03 & -0.05 & 1.00 & 0.00 & 0.03 & 0.02 & -0.09 & -0.06 & -0.01 & -0.00 & 0.02 & 0.04 & 0.09 & 0.02 & 0.04 & -0.12\\
$s^{4\pi}_{+3}$ & 0.00 & 0.00 & 0.00 & 0.00 & 0.00 & -0.02 & 0.00 & 1.00 & 0.00 & 0.01 & 0.01 & -0.02 & -0.02 & -0.01 & 0.00 & -0.01 & 0.02 & -0.02 & 0.03 & 0.00\\
$s^{4\pi}_{+4}$ & -0.00 & 0.02 & -0.00 & -0.01 & -0.01 & -0.05 & 0.03 & 0.00 & 1.00 & -0.02 & -0.00 & 0.02 & 0.07 & -0.04 & -0.03 & 0.02 & -0.05 & 0.09 & -0.00 & -0.02\\
$s^{4\pi}_{+5}$ & 0.03 & 0.01 & 0.01 & 0.00 & -0.02 & -0.03 & 0.02 & 0.01 & -0.02 & 1.00 & 0.03 & 0.03 & 0.07 & -0.01 & -0.01 & -0.05 & 0.04 & -0.07 & -0.04 & 0.09\\
$T^{4\pi}_{+1}$ & 0.01 & 0.01 & 0.00 & 0.03 & 0.04 & -0.11 & -0.09 & 0.01 & -0.00 & 0.03 & 1.00 & 0.04 & 0.01 & -0.03 & -0.03 & -0.10 & -0.05 & -0.02 & -0.05 & 0.12\\
$T^{4\pi}_{+2}$ & -0.00 & 0.02 & 0.00 & -0.00 & 0.01 & 0.04 & -0.06 & -0.02 & 0.02 & 0.03 & 0.04 & 1.00 & 0.04 & -0.10 & -0.05 & -0.03 & -0.06 & 0.06 & -0.02 & 0.05\\
$T^{4\pi}_{+3}$ & -0.00 & 0.01 & 0.03 & -0.01 & -0.01 & -0.00 & -0.01 & -0.02 & 0.07 & 0.07 & 0.01 & 0.04 & 1.00 & -0.01 & -0.01 & 0.01 & -0.01 & -0.06 & -0.17 & 0.03\\
$T^{4\pi}_{+4}$ & 0.00 & -0.01 & 0.01 & -0.02 & -0.01 & -0.01 & -0.00 & -0.01 & -0.04 & -0.01 & -0.03 & -0.10 & -0.01 & 1.00 & 0.04 & -0.04 & -0.04 & -0.02 & -0.11 & 0.01\\
$T^{4\pi}_{+5}$ & -0.01 & -0.02 & -0.00 & 0.00 & 0.01 & 0.04 & 0.02 & 0.00 & -0.03 & -0.01 & -0.03 & -0.05 & -0.01 & 0.04 & 1.00 & 0.02 & 0.01 & -0.02 & -0.01 & -0.04\\
$T^{4\pi}_{-1}$ & -0.01 & -0.01 & -0.01 & -0.00 & 0.00 & 0.05 & 0.04 & -0.01 & 0.02 & -0.05 & -0.10 & -0.03 & 0.01 & -0.04 & 0.02 & 1.00 & 0.03 & -0.02 & 0.03 & -0.10\\
$T^{4\pi}_{-2}$ & 0.01 & -0.01 & -0.01 & -0.02 & -0.04 & -0.07 & 0.09 & 0.02 & -0.05 & 0.04 & -0.05 & -0.06 & -0.01 & -0.04 & 0.01 & 0.03 & 1.00 & -0.20 & 0.07 & -0.08\\
$T^{4\pi}_{-3}$ & -0.01 & 0.02 & -0.01 & 0.06 & 0.02 & 0.01 & 0.02 & -0.02 & 0.09 & -0.07 & -0.02 & 0.06 & -0.06 & -0.02 & -0.02 & -0.02 & -0.20 & 1.00 & -0.01 & -0.04\\
$T^{4\pi}_{-4}$ & -0.02 & 0.00 & -0.02 & 0.01 & 0.01 & -0.02 & 0.04 & 0.03 & -0.00 & -0.04 & -0.05 & -0.02 & -0.17 & -0.11 & -0.01 & 0.03 & 0.07 & -0.01 & 1.00 & -0.08\\
$T^{4\pi}_{-5}$ & 0.06 & 0.02 & 0.04 & -0.02 & -0.02 & -0.01 & -0.12 & 0.00 & -0.02 & 0.09 & 0.12 & 0.05 & 0.03 & 0.01 & -0.04 & -0.10 & -0.08 & -0.04 & -0.08 & 1.00\\
\end{tabular}

}

\caption{ The statistical and systematic correlations between the \fourpi hadronic parameters using the Variable \Deldelfourpi binning scheme with $\mathcal{N} = 5$. \label{tab:correlations1}}
\end{table}

\begin{table}
\centering
Alternative binning statistical correlations \\
\resizebox{1.05\columnwidth}{!}{
\begin{tabular}{ l | r  r  r  r  r  r  r  r  r  r  r  r  r  r  r  r  r  r  r  r }
   & $c^{4\pi}_{+1}$ & $c^{4\pi}_{+2}$ & $c^{4\pi}_{+3}$ & $c^{4\pi}_{+4}$ & $c^{4\pi}_{+5}$ & $s^{4\pi}_{+1}$ & $s^{4\pi}_{+2}$ & $s^{4\pi}_{+3}$ & $s^{4\pi}_{+4}$ & $s^{4\pi}_{+5}$ & $T^{4\pi}_{+1}$ & $T^{4\pi}_{+2}$ & $T^{4\pi}_{+3}$ & $T^{4\pi}_{+4}$ & $T^{4\pi}_{+5}$ & $T^{4\pi}_{-1}$ & $T^{4\pi}_{-2}$ & $T^{4\pi}_{-3}$ & $T^{4\pi}_{-4}$ & $T^{4\pi}_{-5}$\\
\hline
$c^{4\pi}_{+1}$ & 1.00 & 0.02 & 0.01 & 0.02 & 0.02 & 0.04 & 0.00 & 0.01 & -0.00 & 0.01 & -0.04 & 0.01 & 0.02 & 0.01 & -0.00 & -0.05 & 0.00 & 0.01 & 0.00 & -0.00\\
$c^{4\pi}_{+2}$ & 0.02 & 1.00 & 0.03 & 0.03 & 0.02 & 0.00 & 0.01 & 0.01 & 0.00 & 0.00 & 0.01 & -0.01 & 0.04 & 0.02 & 0.01 & 0.00 & -0.15 & 0.03 & 0.01 & 0.00\\
$c^{4\pi}_{+3}$ & 0.01 & 0.03 & 1.00 & 0.03 & 0.00 & -0.01 & -0.00 & 0.01 & 0.01 & -0.01 & 0.02 & 0.04 & 0.00 & 0.04 & 0.02 & 0.01 & 0.02 & -0.13 & 0.03 & 0.01\\
$c^{4\pi}_{+4}$ & 0.02 & 0.03 & 0.03 & 1.00 & 0.02 & -0.00 & -0.01 & 0.00 & 0.04 & -0.01 & 0.01 & 0.01 & 0.03 & -0.00 & 0.00 & 0.00 & 0.01 & 0.02 & -0.11 & 0.00\\
$c^{4\pi}_{+5}$ & 0.02 & 0.02 & 0.00 & 0.02 & 1.00 & 0.01 & 0.00 & 0.01 & -0.00 & -0.02 & -0.00 & 0.00 & 0.01 & 0.00 & -0.05 & -0.00 & 0.00 & 0.01 & 0.00 & -0.01\\
$s^{4\pi}_{+1}$ & 0.04 & 0.00 & -0.01 & -0.00 & 0.01 & 1.00 & 0.01 & 0.00 & -0.00 & 0.12 & 0.02 & 0.00 & 0.00 & 0.00 & 0.00 & -0.04 & 0.00 & 0.00 & -0.00 & -0.00\\
$s^{4\pi}_{+2}$ & 0.00 & 0.01 & -0.00 & -0.01 & 0.00 & 0.01 & 1.00 & -0.02 & 0.04 & 0.02 & 0.00 & 0.01 & -0.00 & 0.00 & -0.00 & -0.00 & -0.01 & 0.00 & -0.00 & 0.00\\
$s^{4\pi}_{+3}$ & 0.01 & 0.01 & 0.01 & 0.00 & 0.01 & 0.00 & -0.02 & 1.00 & -0.01 & -0.05 & 0.00 & 0.00 & 0.00 & 0.00 & 0.00 & -0.00 & 0.00 & -0.01 & 0.00 & 0.00\\
$s^{4\pi}_{+4}$ & -0.00 & 0.00 & 0.01 & 0.04 & -0.00 & -0.00 & 0.04 & -0.01 & 1.00 & -0.00 & 0.00 & 0.00 & 0.00 & -0.00 & 0.00 & 0.00 & -0.00 & 0.00 & -0.01 & 0.00\\
$s^{4\pi}_{+5}$ & 0.01 & 0.00 & -0.01 & -0.01 & -0.02 & 0.12 & 0.02 & -0.05 & -0.00 & 1.00 & 0.00 & 0.00 & 0.00 & 0.00 & 0.01 & -0.01 & -0.00 & -0.00 & 0.00 & -0.02\\
$T^{4\pi}_{+1}$ & -0.04 & 0.01 & 0.02 & 0.01 & -0.00 & 0.02 & 0.00 & 0.00 & 0.00 & 0.00 & 1.00 & -0.08 & -0.13 & -0.08 & -0.05 & -0.15 & -0.06 & -0.10 & -0.06 & -0.04\\
$T^{4\pi}_{+2}$ & 0.01 & -0.01 & 0.04 & 0.01 & 0.00 & 0.00 & 0.01 & 0.00 & 0.00 & 0.00 & -0.08 & 1.00 & -0.18 & -0.12 & -0.08 & -0.05 & -0.24 & -0.15 & -0.09 & -0.06\\
$T^{4\pi}_{+3}$ & 0.02 & 0.04 & 0.00 & 0.03 & 0.01 & 0.00 & -0.00 & 0.00 & 0.00 & 0.00 & -0.13 & -0.18 & 1.00 & -0.18 & -0.13 & -0.08 & -0.13 & -0.41 & -0.13 & -0.09\\
$T^{4\pi}_{+4}$ & 0.01 & 0.02 & 0.04 & -0.00 & 0.00 & 0.00 & 0.00 & 0.00 & -0.00 & 0.00 & -0.08 & -0.12 & -0.18 & 1.00 & -0.08 & -0.05 & -0.08 & -0.15 & -0.24 & -0.06\\
$T^{4\pi}_{+5}$ & -0.00 & 0.01 & 0.02 & 0.00 & -0.05 & 0.00 & -0.00 & 0.00 & 0.00 & 0.01 & -0.05 & -0.08 & -0.13 & -0.08 & 1.00 & -0.03 & -0.06 & -0.11 & -0.06 & -0.14\\
$T^{4\pi}_{-1}$ & -0.05 & 0.00 & 0.01 & 0.00 & -0.00 & -0.04 & -0.00 & -0.00 & 0.00 & -0.01 & -0.15 & -0.05 & -0.08 & -0.05 & -0.03 & 1.00 & -0.04 & -0.07 & -0.04 & -0.02\\
$T^{4\pi}_{-2}$ & 0.00 & -0.15 & 0.02 & 0.01 & 0.00 & 0.00 & -0.01 & 0.00 & -0.00 & -0.00 & -0.06 & -0.24 & -0.13 & -0.08 & -0.06 & -0.04 & 1.00 & -0.11 & -0.06 & -0.04\\
$T^{4\pi}_{-3}$ & 0.01 & 0.03 & -0.13 & 0.02 & 0.01 & 0.00 & 0.00 & -0.01 & 0.00 & -0.00 & -0.10 & -0.15 & -0.41 & -0.15 & -0.11 & -0.07 & -0.11 & 1.00 & -0.11 & -0.07\\
$T^{4\pi}_{-4}$ & 0.00 & 0.01 & 0.03 & -0.11 & 0.00 & -0.00 & -0.00 & 0.00 & -0.01 & 0.00 & -0.06 & -0.09 & -0.13 & -0.24 & -0.06 & -0.04 & -0.06 & -0.11 & 1.00 & -0.04\\
$T^{4\pi}_{-5}$ & -0.00 & 0.00 & 0.01 & 0.00 & -0.01 & -0.00 & 0.00 & 0.00 & 0.00 & -0.02 & -0.04 & -0.06 & -0.09 & -0.06 & -0.14 & -0.02 & -0.04 & -0.07 & -0.04 & 1.00\\
\end{tabular}

}

\vspace{1cm}

Alternative binning systematic correlations \\
\resizebox{1.05\columnwidth}{!}{
\begin{tabular}{ l | r  r  r  r  r  r  r  r  r  r  r  r  r  r  r  r  r  r  r  r }
   & $c^{4\pi}_{+1}$ & $c^{4\pi}_{+2}$ & $c^{4\pi}_{+3}$ & $c^{4\pi}_{+4}$ & $c^{4\pi}_{+5}$ & $s^{4\pi}_{+1}$ & $s^{4\pi}_{+2}$ & $s^{4\pi}_{+3}$ & $s^{4\pi}_{+4}$ & $s^{4\pi}_{+5}$ & $T^{4\pi}_{+1}$ & $T^{4\pi}_{+2}$ & $T^{4\pi}_{+3}$ & $T^{4\pi}_{+4}$ & $T^{4\pi}_{+5}$ & $T^{4\pi}_{-1}$ & $T^{4\pi}_{-2}$ & $T^{4\pi}_{-3}$ & $T^{4\pi}_{-4}$ & $T^{4\pi}_{-5}$\\
\hline
$c^{4\pi}_{+1}$ & 1.00 & 0.01 & 0.00 & 0.03 & 0.01 & 0.01 & -0.00 & -0.02 & 0.00 & 0.01 & -0.02 & 0.00 & 0.01 & -0.01 & -0.01 & -0.00 & -0.01 & 0.02 & -0.02 & 0.00\\
$c^{4\pi}_{+2}$ & 0.01 & 1.00 & -0.00 & 0.03 & 0.04 & 0.01 & 0.02 & 0.01 & 0.01 & 0.00 & -0.01 & 0.01 & -0.00 & -0.01 & -0.03 & -0.01 & -0.03 & 0.05 & 0.02 & -0.01\\
$c^{4\pi}_{+3}$ & 0.00 & -0.00 & 1.00 & 0.00 & -0.01 & -0.00 & -0.01 & 0.02 & -0.01 & 0.01 & 0.01 & -0.00 & 0.01 & 0.01 & -0.00 & -0.01 & 0.00 & -0.01 & -0.02 & 0.01\\
$c^{4\pi}_{+4}$ & 0.03 & 0.03 & 0.00 & 1.00 & 0.03 & 0.01 & 0.03 & 0.04 & -0.00 & 0.01 & 0.00 & -0.02 & 0.00 & 0.01 & 0.01 & -0.01 & -0.02 & 0.04 & -0.01 & -0.00\\
$c^{4\pi}_{+5}$ & 0.01 & 0.04 & -0.01 & 0.03 & 1.00 & 0.03 & -0.01 & -0.01 & -0.00 & 0.00 & 0.02 & 0.00 & -0.00 & -0.00 & -0.00 & -0.01 & -0.02 & 0.02 & 0.01 & -0.00\\
$s^{4\pi}_{+1}$ & 0.01 & 0.01 & -0.00 & 0.01 & 0.03 & 1.00 & 0.01 & -0.02 & -0.01 & 0.01 & -0.03 & 0.01 & 0.00 & 0.00 & 0.01 & -0.00 & -0.01 & 0.00 & -0.01 & 0.00\\
$s^{4\pi}_{+2}$ & -0.00 & 0.02 & -0.01 & 0.03 & -0.01 & 0.01 & 1.00 & 0.02 & 0.01 & -0.00 & -0.03 & -0.03 & 0.00 & 0.00 & 0.03 & 0.02 & 0.03 & -0.00 & 0.01 & -0.03\\
$s^{4\pi}_{+3}$ & -0.02 & 0.01 & 0.02 & 0.04 & -0.01 & -0.02 & 0.02 & 1.00 & -0.01 & 0.00 & 0.04 & -0.06 & -0.01 & 0.00 & 0.02 & -0.01 & 0.02 & -0.01 & 0.06 & 0.02\\
$s^{4\pi}_{+4}$ & 0.00 & 0.01 & -0.01 & -0.00 & -0.00 & -0.01 & 0.01 & -0.01 & 1.00 & -0.01 & -0.01 & -0.00 & 0.01 & -0.01 & -0.01 & 0.00 & -0.01 & 0.02 & -0.01 & -0.00\\
$s^{4\pi}_{+5}$ & 0.01 & 0.00 & 0.01 & 0.01 & 0.00 & 0.01 & -0.00 & 0.00 & -0.01 & 1.00 & 0.00 & 0.02 & 0.01 & -0.01 & -0.02 & -0.01 & 0.00 & -0.02 & -0.01 & 0.02\\
$T^{4\pi}_{+1}$ & -0.02 & -0.01 & 0.01 & 0.00 & 0.02 & -0.03 & -0.03 & 0.04 & -0.01 & 0.00 & 1.00 & 0.01 & -0.00 & -0.01 & -0.02 & -0.03 & 0.00 & -0.03 & -0.01 & 0.02\\
$T^{4\pi}_{+2}$ & 0.00 & 0.01 & -0.00 & -0.02 & 0.00 & 0.01 & -0.03 & -0.06 & -0.00 & 0.02 & 0.01 & 1.00 & 0.01 & -0.05 & -0.05 & -0.01 & -0.01 & -0.00 & -0.01 & 0.02\\
$T^{4\pi}_{+3}$ & 0.01 & -0.00 & 0.01 & 0.00 & -0.00 & 0.00 & 0.00 & -0.01 & 0.01 & 0.01 & -0.00 & 0.01 & 1.00 & -0.01 & -0.01 & -0.00 & 0.00 & -0.04 & -0.04 & 0.01\\
$T^{4\pi}_{+4}$ & -0.01 & -0.01 & 0.01 & 0.01 & -0.00 & 0.00 & 0.00 & 0.00 & -0.01 & -0.01 & -0.01 & -0.05 & -0.01 & 1.00 & 0.03 & -0.01 & -0.01 & -0.00 & -0.04 & -0.01\\
$T^{4\pi}_{+5}$ & -0.01 & -0.03 & -0.00 & 0.01 & -0.00 & 0.01 & 0.03 & 0.02 & -0.01 & -0.02 & -0.02 & -0.05 & -0.01 & 0.03 & 1.00 & 0.02 & 0.02 & -0.04 & 0.01 & -0.02\\
$T^{4\pi}_{-1}$ & -0.00 & -0.01 & -0.01 & -0.01 & -0.01 & -0.00 & 0.02 & -0.01 & 0.00 & -0.01 & -0.03 & -0.01 & -0.00 & -0.01 & 0.02 & 1.00 & 0.01 & -0.01 & 0.02 & -0.02\\
$T^{4\pi}_{-2}$ & -0.01 & -0.03 & 0.00 & -0.02 & -0.02 & -0.01 & 0.03 & 0.02 & -0.01 & 0.00 & 0.00 & -0.01 & 0.00 & -0.01 & 0.02 & 0.01 & 1.00 & -0.07 & 0.02 & -0.01\\
$T^{4\pi}_{-3}$ & 0.02 & 0.05 & -0.01 & 0.04 & 0.02 & 0.00 & -0.00 & -0.01 & 0.02 & -0.02 & -0.03 & -0.00 & -0.04 & -0.00 & -0.04 & -0.01 & -0.07 & 1.00 & 0.01 & -0.01\\
$T^{4\pi}_{-4}$ & -0.02 & 0.02 & -0.02 & -0.01 & 0.01 & -0.01 & 0.01 & 0.06 & -0.01 & -0.01 & -0.01 & -0.01 & -0.04 & -0.04 & 0.01 & 0.02 & 0.02 & 0.01 & 1.00 & -0.01\\
$T^{4\pi}_{-5}$ & 0.00 & -0.01 & 0.01 & -0.00 & -0.00 & 0.00 & -0.03 & 0.02 & -0.00 & 0.02 & 0.02 & 0.02 & 0.01 & -0.01 & -0.02 & -0.02 & -0.01 & -0.01 & -0.01 & 1.00\\
\end{tabular}

}

\caption{ The statistical and systematic correlations between the \fourpi hadronic parameters using the Alternative binning scheme with $\mathcal{N} = 5$. \label{tab:correlations1}}
\end{table}

\begin{table}
\centering
Optimal binning statistical correlations \\
\resizebox{1.05\columnwidth}{!}{
\begin{tabular}{ l | r  r  r  r  r  r  r  r  r  r  r  r  r  r  r  r  r  r  r  r }
   & $c^{4\pi}_{+1}$ & $c^{4\pi}_{+2}$ & $c^{4\pi}_{+3}$ & $c^{4\pi}_{+4}$ & $c^{4\pi}_{+5}$ & $s^{4\pi}_{+1}$ & $s^{4\pi}_{+2}$ & $s^{4\pi}_{+3}$ & $s^{4\pi}_{+4}$ & $s^{4\pi}_{+5}$ & $T^{4\pi}_{+1}$ & $T^{4\pi}_{+2}$ & $T^{4\pi}_{+3}$ & $T^{4\pi}_{+4}$ & $T^{4\pi}_{+5}$ & $T^{4\pi}_{-1}$ & $T^{4\pi}_{-2}$ & $T^{4\pi}_{-3}$ & $T^{4\pi}_{-4}$ & $T^{4\pi}_{-5}$\\
\hline
$c^{4\pi}_{+1}$ & 1.00 & 0.02 & 0.02 & 0.02 & 0.01 & 0.01 & -0.01 & 0.00 & 0.01 & -0.00 & -0.05 & 0.02 & 0.04 & 0.02 & 0.02 & -0.10 & 0.04 & 0.02 & 0.03 & 0.02\\
$c^{4\pi}_{+2}$ & 0.02 & 1.00 & 0.02 & 0.01 & 0.02 & 0.00 & 0.03 & 0.00 & 0.00 & 0.00 & 0.02 & -0.21 & 0.02 & 0.01 & 0.01 & 0.02 & 0.04 & 0.01 & 0.01 & 0.01\\
$c^{4\pi}_{+3}$ & 0.02 & 0.02 & 1.00 & 0.02 & 0.02 & -0.00 & 0.00 & -0.00 & 0.00 & 0.00 & 0.03 & 0.01 & 0.01 & 0.01 & 0.01 & 0.03 & 0.02 & -0.15 & 0.01 & 0.00\\
$c^{4\pi}_{+4}$ & 0.02 & 0.01 & 0.02 & 1.00 & 0.02 & -0.00 & -0.00 & -0.00 & 0.02 & -0.01 & 0.03 & 0.01 & 0.02 & -0.13 & 0.01 & 0.03 & 0.02 & 0.01 & -0.04 & 0.00\\
$c^{4\pi}_{+5}$ & 0.01 & 0.02 & 0.02 & 0.02 & 1.00 & 0.01 & 0.01 & -0.01 & -0.00 & 0.06 & 0.01 & 0.00 & 0.00 & -0.00 & -0.05 & 0.01 & 0.01 & 0.00 & 0.00 & -0.01\\
$s^{4\pi}_{+1}$ & 0.01 & 0.00 & -0.00 & -0.00 & 0.01 & 1.00 & 0.03 & -0.03 & -0.01 & -0.02 & -0.02 & -0.00 & 0.00 & 0.00 & 0.00 & 0.01 & 0.00 & 0.00 & 0.00 & 0.00\\
$s^{4\pi}_{+2}$ & -0.01 & 0.03 & 0.00 & -0.00 & 0.01 & 0.03 & 1.00 & -0.06 & 0.01 & 0.04 & 0.00 & -0.03 & -0.00 & -0.00 & -0.00 & 0.00 & 0.02 & 0.00 & 0.00 & -0.00\\
$s^{4\pi}_{+3}$ & 0.00 & 0.00 & -0.00 & -0.00 & -0.01 & -0.03 & -0.06 & 1.00 & 0.02 & -0.02 & 0.00 & -0.00 & 0.02 & 0.00 & -0.00 & -0.00 & 0.00 & -0.03 & 0.00 & 0.00\\
$s^{4\pi}_{+4}$ & 0.01 & 0.00 & 0.00 & 0.02 & -0.00 & -0.01 & 0.01 & 0.02 & 1.00 & 0.01 & 0.00 & -0.00 & 0.00 & -0.00 & -0.00 & 0.00 & 0.01 & 0.00 & -0.02 & 0.01\\
$s^{4\pi}_{+5}$ & -0.00 & 0.00 & 0.00 & -0.01 & 0.06 & -0.02 & 0.04 & -0.02 & 0.01 & 1.00 & 0.01 & 0.00 & 0.00 & 0.00 & -0.01 & 0.01 & 0.00 & 0.00 & 0.00 & -0.03\\
$T^{4\pi}_{+1}$ & -0.05 & 0.02 & 0.03 & 0.03 & 0.01 & -0.02 & 0.00 & 0.00 & 0.00 & 0.01 & 1.00 & -0.09 & -0.16 & -0.11 & -0.12 & -0.36 & -0.15 & -0.11 & -0.13 & -0.10\\
$T^{4\pi}_{+2}$ & 0.02 & -0.21 & 0.01 & 0.01 & 0.00 & -0.00 & -0.03 & -0.00 & -0.00 & 0.00 & -0.09 & 1.00 & -0.07 & -0.04 & -0.05 & -0.08 & -0.22 & -0.05 & -0.06 & -0.05\\
$T^{4\pi}_{+3}$ & 0.04 & 0.02 & 0.01 & 0.02 & 0.00 & 0.00 & -0.00 & 0.02 & 0.00 & 0.00 & -0.16 & -0.07 & 1.00 & -0.08 & -0.09 & -0.15 & -0.12 & -0.24 & -0.11 & -0.08\\
$T^{4\pi}_{+4}$ & 0.02 & 0.01 & 0.01 & -0.13 & -0.00 & 0.00 & -0.00 & 0.00 & -0.00 & 0.00 & -0.11 & -0.04 & -0.08 & 1.00 & -0.06 & -0.10 & -0.08 & -0.06 & -0.22 & -0.05\\
$T^{4\pi}_{+5}$ & 0.02 & 0.01 & 0.01 & 0.01 & -0.05 & 0.00 & -0.00 & -0.00 & -0.00 & -0.01 & -0.12 & -0.05 & -0.09 & -0.06 & 1.00 & -0.11 & -0.09 & -0.07 & -0.08 & -0.16\\
$T^{4\pi}_{-1}$ & -0.10 & 0.02 & 0.03 & 0.03 & 0.01 & 0.01 & 0.00 & -0.00 & 0.00 & 0.01 & -0.36 & -0.08 & -0.15 & -0.10 & -0.11 & 1.00 & -0.14 & -0.10 & -0.12 & -0.09\\
$T^{4\pi}_{-2}$ & 0.04 & 0.04 & 0.02 & 0.02 & 0.01 & 0.00 & 0.02 & 0.00 & 0.01 & 0.00 & -0.15 & -0.22 & -0.12 & -0.08 & -0.09 & -0.14 & 1.00 & -0.09 & -0.10 & -0.07\\
$T^{4\pi}_{-3}$ & 0.02 & 0.01 & -0.15 & 0.01 & 0.00 & 0.00 & 0.00 & -0.03 & 0.00 & 0.00 & -0.11 & -0.05 & -0.24 & -0.06 & -0.07 & -0.10 & -0.09 & 1.00 & -0.08 & -0.05\\
$T^{4\pi}_{-4}$ & 0.03 & 0.01 & 0.01 & -0.04 & 0.00 & 0.00 & 0.00 & 0.00 & -0.02 & 0.00 & -0.13 & -0.06 & -0.11 & -0.22 & -0.08 & -0.12 & -0.10 & -0.08 & 1.00 & -0.06\\
$T^{4\pi}_{-5}$ & 0.02 & 0.01 & 0.00 & 0.00 & -0.01 & 0.00 & -0.00 & 0.00 & 0.01 & -0.03 & -0.10 & -0.05 & -0.08 & -0.05 & -0.16 & -0.09 & -0.07 & -0.05 & -0.06 & 1.00\\
\end{tabular}

}

\vspace{1cm}

Optimal binning systematic correlations \\
\resizebox{1.05\columnwidth}{!}{
\begin{tabular}{ l | r  r  r  r  r  r  r  r  r  r  r  r  r  r  r  r  r  r  r  r }
   & $c^{4\pi}_{+1}$ & $c^{4\pi}_{+2}$ & $c^{4\pi}_{+3}$ & $c^{4\pi}_{+4}$ & $c^{4\pi}_{+5}$ & $s^{4\pi}_{+1}$ & $s^{4\pi}_{+2}$ & $s^{4\pi}_{+3}$ & $s^{4\pi}_{+4}$ & $s^{4\pi}_{+5}$ & $T^{4\pi}_{+1}$ & $T^{4\pi}_{+2}$ & $T^{4\pi}_{+3}$ & $T^{4\pi}_{+4}$ & $T^{4\pi}_{+5}$ & $T^{4\pi}_{-1}$ & $T^{4\pi}_{-2}$ & $T^{4\pi}_{-3}$ & $T^{4\pi}_{-4}$ & $T^{4\pi}_{-5}$\\
\hline
$c^{4\pi}_{+1}$ & 1.00 & -0.00 & 0.00 & 0.01 & -0.01 & -0.02 & -0.02 & 0.00 & -0.01 & 0.00 & 0.01 & -0.00 & 0.00 & 0.01 & -0.01 & -0.02 & 0.00 & -0.00 & -0.03 & 0.02\\
$c^{4\pi}_{+2}$ & -0.00 & 1.00 & 0.02 & 0.02 & 0.01 & -0.02 & 0.00 & 0.01 & 0.01 & 0.00 & -0.00 & 0.00 & 0.00 & -0.00 & -0.02 & -0.00 & -0.01 & 0.01 & 0.03 & -0.01\\
$c^{4\pi}_{+3}$ & 0.00 & 0.02 & 1.00 & 0.03 & 0.03 & 0.02 & 0.00 & -0.00 & -0.00 & 0.00 & 0.01 & 0.01 & 0.01 & 0.01 & -0.01 & -0.01 & -0.02 & 0.02 & -0.02 & -0.01\\
$c^{4\pi}_{+4}$ & 0.01 & 0.02 & 0.03 & 1.00 & 0.03 & 0.01 & 0.01 & 0.01 & -0.01 & 0.00 & 0.01 & -0.00 & -0.01 & -0.01 & 0.01 & -0.00 & -0.01 & 0.01 & 0.01 & -0.01\\
$c^{4\pi}_{+5}$ & -0.01 & 0.01 & 0.03 & 0.03 & 1.00 & 0.07 & -0.01 & -0.00 & -0.01 & 0.00 & 0.04 & 0.00 & -0.00 & -0.01 & -0.00 & -0.01 & -0.02 & 0.00 & 0.01 & -0.00\\
$s^{4\pi}_{+1}$ & -0.02 & -0.02 & 0.02 & 0.01 & 0.07 & 1.00 & 0.04 & -0.05 & -0.04 & -0.00 & -0.11 & 0.03 & 0.01 & -0.03 & 0.09 & 0.15 & -0.04 & 0.01 & 0.01 & -0.03\\
$s^{4\pi}_{+2}$ & -0.02 & 0.00 & 0.00 & 0.01 & -0.01 & 0.04 & 1.00 & -0.00 & -0.00 & -0.00 & -0.04 & -0.01 & -0.00 & -0.01 & 0.03 & 0.05 & 0.01 & 0.00 & 0.02 & -0.04\\
$s^{4\pi}_{+3}$ & 0.00 & 0.01 & -0.00 & 0.01 & -0.00 & -0.05 & -0.00 & 1.00 & -0.00 & 0.00 & 0.02 & -0.01 & -0.01 & -0.00 & 0.01 & -0.03 & 0.01 & -0.01 & 0.03 & 0.00\\
$s^{4\pi}_{+4}$ & -0.01 & 0.01 & -0.00 & -0.01 & -0.01 & -0.04 & -0.00 & -0.00 & 1.00 & -0.01 & -0.00 & -0.00 & 0.02 & -0.01 & -0.03 & 0.02 & -0.01 & 0.01 & -0.00 & -0.01\\
$s^{4\pi}_{+5}$ & 0.00 & 0.00 & 0.00 & 0.00 & 0.00 & -0.00 & -0.00 & 0.00 & -0.01 & 1.00 & 0.01 & 0.01 & 0.02 & -0.00 & -0.02 & -0.03 & 0.01 & -0.01 & -0.04 & 0.01\\
$T^{4\pi}_{+1}$ & 0.01 & -0.00 & 0.01 & 0.01 & 0.04 & -0.11 & -0.04 & 0.02 & -0.00 & 0.01 & 1.00 & 0.01 & -0.00 & 0.00 & -0.05 & -0.14 & -0.02 & -0.00 & -0.06 & 0.05\\
$T^{4\pi}_{+2}$ & -0.00 & 0.00 & 0.01 & -0.00 & 0.00 & 0.03 & -0.01 & -0.01 & -0.00 & 0.01 & 0.01 & 1.00 & -0.00 & -0.01 & -0.04 & -0.04 & -0.01 & 0.02 & 0.00 & 0.02\\
$T^{4\pi}_{+3}$ & 0.00 & 0.00 & 0.01 & -0.01 & -0.00 & 0.01 & -0.00 & -0.01 & 0.02 & 0.02 & -0.00 & -0.00 & 1.00 & 0.00 & -0.02 & -0.01 & -0.01 & -0.01 & -0.09 & 0.01\\
$T^{4\pi}_{+4}$ & 0.01 & -0.00 & 0.01 & -0.01 & -0.01 & -0.03 & -0.01 & -0.00 & -0.01 & -0.00 & 0.00 & -0.01 & 0.00 & 1.00 & 0.02 & -0.03 & -0.01 & 0.00 & -0.07 & 0.01\\
$T^{4\pi}_{+5}$ & -0.01 & -0.02 & -0.01 & 0.01 & -0.00 & 0.09 & 0.03 & 0.01 & -0.03 & -0.02 & -0.05 & -0.04 & -0.02 & 0.02 & 1.00 & 0.09 & 0.00 & -0.01 & -0.01 & -0.04\\
$T^{4\pi}_{-1}$ & -0.02 & -0.00 & -0.01 & -0.00 & -0.01 & 0.15 & 0.05 & -0.03 & 0.02 & -0.03 & -0.14 & -0.04 & -0.01 & -0.03 & 0.09 & 1.00 & 0.02 & -0.02 & 0.05 & -0.10\\
$T^{4\pi}_{-2}$ & 0.00 & -0.01 & -0.02 & -0.01 & -0.02 & -0.04 & 0.01 & 0.01 & -0.01 & 0.01 & -0.02 & -0.01 & -0.01 & -0.01 & 0.00 & 0.02 & 1.00 & -0.03 & 0.04 & -0.02\\
$T^{4\pi}_{-3}$ & -0.00 & 0.01 & 0.02 & 0.01 & 0.00 & 0.01 & 0.00 & -0.01 & 0.01 & -0.01 & -0.00 & 0.02 & -0.01 & 0.00 & -0.01 & -0.02 & -0.03 & 1.00 & 0.02 & -0.00\\
$T^{4\pi}_{-4}$ & -0.03 & 0.03 & -0.02 & 0.01 & 0.01 & 0.01 & 0.02 & 0.03 & -0.00 & -0.04 & -0.06 & 0.00 & -0.09 & -0.07 & -0.01 & 0.05 & 0.04 & 0.02 & 1.00 & -0.05\\
$T^{4\pi}_{-5}$ & 0.02 & -0.01 & -0.01 & -0.01 & -0.00 & -0.03 & -0.04 & 0.00 & -0.01 & 0.01 & 0.05 & 0.02 & 0.01 & 0.01 & -0.04 & -0.10 & -0.02 & -0.00 & -0.05 & 1.00\\
\end{tabular}

}

\caption{ The statistical and systematic correlations between the \fourpi hadronic parameters using the Optimal binning scheme with $\mathcal{N} = 5$. \label{tab:correlations1}}
\end{table}

\begin{table}
\centering
Optimal alternative binning statistical correlations \\
\resizebox{1.05\columnwidth}{!}{
\begin{tabular}{ l | r  r  r  r  r  r  r  r  r  r  r  r  r  r  r  r  r  r  r  r }
   & $c^{4\pi}_{+1}$ & $c^{4\pi}_{+2}$ & $c^{4\pi}_{+3}$ & $c^{4\pi}_{+4}$ & $c^{4\pi}_{+5}$ & $s^{4\pi}_{+1}$ & $s^{4\pi}_{+2}$ & $s^{4\pi}_{+3}$ & $s^{4\pi}_{+4}$ & $s^{4\pi}_{+5}$ & $T^{4\pi}_{+1}$ & $T^{4\pi}_{+2}$ & $T^{4\pi}_{+3}$ & $T^{4\pi}_{+4}$ & $T^{4\pi}_{+5}$ & $T^{4\pi}_{-1}$ & $T^{4\pi}_{-2}$ & $T^{4\pi}_{-3}$ & $T^{4\pi}_{-4}$ & $T^{4\pi}_{-5}$\\
\hline
$c^{4\pi}_{+1}$ & 1.00 & 0.02 & 0.00 & 0.02 & 0.03 & 0.02 & 0.00 & -0.00 & 0.00 & -0.01 & -0.00 & 0.02 & 0.02 & 0.01 & 0.01 & -0.15 & 0.01 & 0.02 & 0.01 & 0.00\\
$c^{4\pi}_{+2}$ & 0.02 & 1.00 & 0.02 & 0.02 & 0.02 & -0.00 & -0.05 & 0.00 & -0.00 & 0.01 & 0.01 & 0.05 & 0.02 & 0.01 & 0.00 & 0.00 & -0.17 & 0.02 & 0.01 & 0.00\\
$c^{4\pi}_{+3}$ & 0.00 & 0.02 & 1.00 & 0.02 & 0.01 & -0.01 & 0.01 & 0.02 & -0.00 & 0.01 & 0.03 & 0.04 & -0.03 & 0.03 & 0.02 & 0.01 & 0.02 & -0.12 & 0.03 & 0.02\\
$c^{4\pi}_{+4}$ & 0.02 & 0.02 & 0.02 & 1.00 & 0.02 & -0.00 & 0.01 & -0.00 & 0.01 & 0.00 & 0.02 & 0.03 & 0.03 & -0.03 & 0.01 & 0.01 & 0.01 & 0.03 & -0.15 & 0.01\\
$c^{4\pi}_{+5}$ & 0.03 & 0.02 & 0.01 & 0.02 & 1.00 & 0.01 & 0.00 & -0.00 & -0.00 & 0.06 & 0.00 & 0.01 & 0.01 & 0.01 & -0.05 & 0.00 & 0.00 & 0.01 & 0.00 & -0.03\\
$s^{4\pi}_{+1}$ & 0.02 & -0.00 & -0.01 & -0.00 & 0.01 & 1.00 & -0.01 & 0.01 & -0.02 & -0.04 & 0.00 & -0.00 & -0.00 & 0.00 & 0.00 & 0.02 & 0.00 & -0.00 & -0.00 & -0.01\\
$s^{4\pi}_{+2}$ & 0.00 & -0.05 & 0.01 & 0.01 & 0.00 & -0.01 & 1.00 & -0.01 & 0.07 & -0.08 & -0.00 & -0.02 & -0.00 & 0.00 & 0.00 & 0.00 & 0.03 & 0.00 & -0.00 & -0.00\\
$s^{4\pi}_{+3}$ & -0.00 & 0.00 & 0.02 & -0.00 & -0.00 & 0.01 & -0.01 & 1.00 & 0.02 & -0.00 & 0.00 & 0.00 & -0.02 & 0.00 & -0.00 & -0.00 & -0.00 & 0.01 & 0.00 & 0.00\\
$s^{4\pi}_{+4}$ & 0.00 & -0.00 & -0.00 & 0.01 & -0.00 & -0.02 & 0.07 & 0.02 & 1.00 & -0.03 & 0.00 & 0.00 & 0.00 & 0.02 & 0.00 & 0.00 & -0.00 & 0.00 & -0.04 & 0.00\\
$s^{4\pi}_{+5}$ & -0.01 & 0.01 & 0.01 & 0.00 & 0.06 & -0.04 & -0.08 & -0.00 & -0.03 & 1.00 & 0.00 & 0.00 & 0.00 & 0.00 & 0.00 & 0.00 & 0.00 & 0.00 & 0.00 & -0.04\\
$T^{4\pi}_{+1}$ & -0.00 & 0.01 & 0.03 & 0.02 & 0.00 & 0.00 & -0.00 & 0.00 & 0.00 & 0.00 & 1.00 & -0.10 & -0.14 & -0.11 & -0.08 & -0.21 & -0.07 & -0.12 & -0.08 & -0.06\\
$T^{4\pi}_{+2}$ & 0.02 & 0.05 & 0.04 & 0.03 & 0.01 & -0.00 & -0.02 & 0.00 & 0.00 & 0.00 & -0.10 & 1.00 & -0.15 & -0.12 & -0.09 & -0.07 & -0.23 & -0.13 & -0.09 & -0.06\\
$T^{4\pi}_{+3}$ & 0.02 & 0.02 & -0.03 & 0.03 & 0.01 & -0.00 & -0.00 & -0.02 & 0.00 & 0.00 & -0.14 & -0.15 & 1.00 & -0.16 & -0.12 & -0.09 & -0.10 & -0.36 & -0.12 & -0.09\\
$T^{4\pi}_{+4}$ & 0.01 & 0.01 & 0.03 & -0.03 & 0.01 & 0.00 & 0.00 & 0.00 & 0.02 & 0.00 & -0.11 & -0.12 & -0.16 & 1.00 & -0.09 & -0.07 & -0.08 & -0.14 & -0.25 & -0.08\\
$T^{4\pi}_{+5}$ & 0.01 & 0.00 & 0.02 & 0.01 & -0.05 & 0.00 & 0.00 & -0.00 & 0.00 & 0.00 & -0.08 & -0.09 & -0.12 & -0.09 & 1.00 & -0.05 & -0.05 & -0.11 & -0.07 & -0.17\\
$T^{4\pi}_{-1}$ & -0.15 & 0.00 & 0.01 & 0.01 & 0.00 & 0.02 & 0.00 & -0.00 & 0.00 & 0.00 & -0.21 & -0.07 & -0.09 & -0.07 & -0.05 & 1.00 & -0.04 & -0.08 & -0.05 & -0.04\\
$T^{4\pi}_{-2}$ & 0.01 & -0.17 & 0.02 & 0.01 & 0.00 & 0.00 & 0.03 & -0.00 & -0.00 & 0.00 & -0.07 & -0.23 & -0.10 & -0.08 & -0.05 & -0.04 & 1.00 & -0.08 & -0.06 & -0.05\\
$T^{4\pi}_{-3}$ & 0.02 & 0.02 & -0.12 & 0.03 & 0.01 & -0.00 & 0.00 & 0.01 & 0.00 & 0.00 & -0.12 & -0.13 & -0.36 & -0.14 & -0.11 & -0.08 & -0.08 & 1.00 & -0.11 & -0.09\\
$T^{4\pi}_{-4}$ & 0.01 & 0.01 & 0.03 & -0.15 & 0.00 & -0.00 & -0.00 & 0.00 & -0.04 & 0.00 & -0.08 & -0.09 & -0.12 & -0.25 & -0.07 & -0.05 & -0.06 & -0.11 & 1.00 & -0.05\\
$T^{4\pi}_{-5}$ & 0.00 & 0.00 & 0.02 & 0.01 & -0.03 & -0.01 & -0.00 & 0.00 & 0.00 & -0.04 & -0.06 & -0.06 & -0.09 & -0.08 & -0.17 & -0.04 & -0.05 & -0.09 & -0.05 & 1.00\\
\end{tabular}

}

\vspace{1cm}

Optimal alternative binning systematic correlations \\
\resizebox{1.05\columnwidth}{!}{
\begin{tabular}{ l | r  r  r  r  r  r  r  r  r  r  r  r  r  r  r  r  r  r  r  r }
   & $c^{4\pi}_{+1}$ & $c^{4\pi}_{+2}$ & $c^{4\pi}_{+3}$ & $c^{4\pi}_{+4}$ & $c^{4\pi}_{+5}$ & $s^{4\pi}_{+1}$ & $s^{4\pi}_{+2}$ & $s^{4\pi}_{+3}$ & $s^{4\pi}_{+4}$ & $s^{4\pi}_{+5}$ & $T^{4\pi}_{+1}$ & $T^{4\pi}_{+2}$ & $T^{4\pi}_{+3}$ & $T^{4\pi}_{+4}$ & $T^{4\pi}_{+5}$ & $T^{4\pi}_{-1}$ & $T^{4\pi}_{-2}$ & $T^{4\pi}_{-3}$ & $T^{4\pi}_{-4}$ & $T^{4\pi}_{-5}$\\
\hline
$c^{4\pi}_{+1}$ & 1.00 & 0.00 & 0.00 & 0.03 & 0.02 & 0.01 & -0.01 & -0.00 & 0.00 & -0.00 & -0.00 & 0.00 & 0.00 & -0.01 & -0.01 & -0.00 & -0.01 & 0.03 & -0.01 & 0.00\\
$c^{4\pi}_{+2}$ & 0.00 & 1.00 & -0.00 & 0.01 & 0.01 & -0.01 & -0.01 & -0.00 & 0.02 & 0.00 & -0.01 & 0.03 & 0.00 & -0.01 & -0.04 & -0.00 & -0.02 & 0.06 & 0.01 & 0.01\\
$c^{4\pi}_{+3}$ & 0.00 & -0.00 & 1.00 & 0.00 & -0.01 & -0.00 & -0.01 & 0.01 & -0.01 & 0.00 & 0.02 & -0.00 & 0.01 & 0.01 & -0.00 & -0.01 & 0.00 & -0.02 & -0.01 & 0.01\\
$c^{4\pi}_{+4}$ & 0.03 & 0.01 & 0.00 & 1.00 & 0.05 & 0.00 & 0.01 & 0.02 & -0.01 & -0.00 & 0.01 & -0.01 & 0.00 & -0.01 & 0.00 & -0.01 & -0.02 & 0.06 & -0.01 & -0.01\\
$c^{4\pi}_{+5}$ & 0.02 & 0.01 & -0.01 & 0.05 & 1.00 & 0.03 & -0.02 & -0.01 & -0.00 & 0.00 & 0.04 & 0.02 & -0.00 & -0.01 & -0.01 & -0.01 & -0.02 & 0.02 & 0.01 & -0.00\\
$s^{4\pi}_{+1}$ & 0.01 & -0.01 & -0.00 & 0.00 & 0.03 & 1.00 & 0.00 & -0.01 & -0.01 & 0.00 & -0.03 & 0.01 & 0.00 & -0.00 & 0.02 & 0.00 & -0.01 & 0.01 & -0.00 & -0.00\\
$s^{4\pi}_{+2}$ & -0.01 & -0.01 & -0.01 & 0.01 & -0.02 & 0.00 & 1.00 & 0.02 & 0.00 & -0.01 & -0.03 & -0.03 & -0.00 & 0.01 & 0.03 & 0.01 & 0.03 & 0.00 & 0.01 & -0.04\\
$s^{4\pi}_{+3}$ & -0.00 & -0.00 & 0.01 & 0.02 & -0.01 & -0.01 & 0.02 & 1.00 & -0.01 & 0.01 & 0.02 & -0.04 & -0.01 & 0.00 & 0.02 & -0.00 & 0.02 & -0.02 & 0.02 & 0.01\\
$s^{4\pi}_{+4}$ & 0.00 & 0.02 & -0.01 & -0.01 & -0.00 & -0.01 & 0.00 & -0.01 & 1.00 & -0.00 & -0.02 & -0.00 & 0.02 & -0.00 & -0.02 & 0.00 & -0.01 & 0.05 & -0.01 & -0.01\\
$s^{4\pi}_{+5}$ & -0.00 & 0.00 & 0.00 & -0.00 & 0.00 & 0.00 & -0.01 & 0.01 & -0.00 & 1.00 & 0.01 & 0.01 & 0.01 & -0.01 & -0.01 & -0.00 & 0.00 & -0.02 & -0.01 & 0.01\\
$T^{4\pi}_{+1}$ & -0.00 & -0.01 & 0.02 & 0.01 & 0.04 & -0.03 & -0.03 & 0.02 & -0.02 & 0.01 & 1.00 & 0.01 & -0.00 & -0.01 & -0.02 & -0.02 & 0.01 & -0.07 & -0.01 & 0.04\\
$T^{4\pi}_{+2}$ & 0.00 & 0.03 & -0.00 & -0.01 & 0.02 & 0.01 & -0.03 & -0.04 & -0.00 & 0.01 & 0.01 & 1.00 & 0.01 & -0.06 & -0.06 & -0.01 & -0.02 & 0.03 & -0.00 & 0.02\\
$T^{4\pi}_{+3}$ & 0.00 & 0.00 & 0.01 & 0.00 & -0.00 & 0.00 & -0.00 & -0.01 & 0.02 & 0.01 & -0.00 & 0.01 & 1.00 & -0.01 & -0.01 & -0.00 & 0.00 & -0.06 & -0.05 & 0.01\\
$T^{4\pi}_{+4}$ & -0.01 & -0.01 & 0.01 & -0.01 & -0.01 & -0.00 & 0.01 & 0.00 & -0.00 & -0.01 & -0.01 & -0.06 & -0.01 & 1.00 & 0.04 & -0.01 & -0.00 & -0.02 & -0.04 & -0.01\\
$T^{4\pi}_{+5}$ & -0.01 & -0.04 & -0.00 & 0.00 & -0.01 & 0.02 & 0.03 & 0.02 & -0.02 & -0.01 & -0.02 & -0.06 & -0.01 & 0.04 & 1.00 & 0.01 & 0.02 & -0.06 & 0.00 & -0.03\\
$T^{4\pi}_{-1}$ & -0.00 & -0.00 & -0.01 & -0.01 & -0.01 & 0.00 & 0.01 & -0.00 & 0.00 & -0.00 & -0.02 & -0.01 & -0.00 & -0.01 & 0.01 & 1.00 & 0.01 & -0.01 & 0.01 & -0.01\\
$T^{4\pi}_{-2}$ & -0.01 & -0.02 & 0.00 & -0.02 & -0.02 & -0.01 & 0.03 & 0.02 & -0.01 & 0.00 & 0.01 & -0.02 & 0.00 & -0.00 & 0.02 & 0.01 & 1.00 & -0.10 & 0.02 & -0.01\\
$T^{4\pi}_{-3}$ & 0.03 & 0.06 & -0.02 & 0.06 & 0.02 & 0.01 & 0.00 & -0.02 & 0.05 & -0.02 & -0.07 & 0.03 & -0.06 & -0.02 & -0.06 & -0.01 & -0.10 & 1.00 & 0.00 & -0.02\\
$T^{4\pi}_{-4}$ & -0.01 & 0.01 & -0.01 & -0.01 & 0.01 & -0.00 & 0.01 & 0.02 & -0.01 & -0.01 & -0.01 & -0.00 & -0.05 & -0.04 & 0.00 & 0.01 & 0.02 & 0.00 & 1.00 & -0.02\\
$T^{4\pi}_{-5}$ & 0.00 & 0.01 & 0.01 & -0.01 & -0.00 & -0.00 & -0.04 & 0.01 & -0.01 & 0.01 & 0.04 & 0.02 & 0.01 & -0.01 & -0.03 & -0.01 & -0.01 & -0.02 & -0.02 & 1.00\\
\end{tabular}

}

\caption{ The statistical and systematic correlations between the \fourpi hadronic parameters using the Optimal alternative binning scheme with $\mathcal{N} = 5$. \label{tab:correlations1}}
\end{table}

\clearpage
\section{Supplementary Material}
\label{app:sup}


\subsection{List of files}
\label{app:files}

The supplementary material can be found at Ref.~\cite{SupGoogleDrive}.
The directory structure is organised so that each phase space binning scheme has its own directory.
Each of these directories has the same file structure inside, which is described in \tabref{tab:sup}.
Additionally there is a \root macro {\tt loadresults.C}, and a collection of \cpp functions in {\tt usehypbinning.cpp} that can be used to load the supplementary material files that are in \root format. 
All results are additionally given in text format for greater flexibility. 

\begin{table}[h]
\small
\begin{tabular}{l p{12cm}}
Filename & Description \\
\hline
\tt cisi.pdf & Figure of \cifourpi and \sifourpi measurements compared to the model predictions. \\ 
\tt kikbi.pdf & Figure of \Frifourpi and \Frbifourpi measurements compared to the model predictions. \\
\tt results.txt & The central values, statistical uncertainties, and systematic uncertainties for the measured hadronic parameters. \\
\tt statcor.txt & The statistical correlations between the measured hadronic parameters. \\
\tt systcor.txt & The systematic correlations between the measured hadronic parameters. \\
\tt stat.root & The central values, statistical uncertainties, and statistical correlations of the measured hadronic parameters in \root format. This can be loaded with the \root macro {\tt loadresults.C}. \\
\tt syst.root & The central values, systematic uncertainties, and systematic correlations of the measured hadronic parameters in \root format. This can be loaded with the \root macro {\tt loadresults.C}. \\
\tt statsyst.root & The central values, combined statistical and systematic uncertainties, and combined statistical and systematic correlations of the measured hadronic parameters in \root format. This can be loaded with the \root macro {\tt loadresults.C}. \\
\tt hypbinning.root & The hyper-binning scheme in \root format. Further description of how to use this file is described in \appref{app:hyperbinning}.  \\ 
\tt hypbinning.zip & A compressed directory containing the the hyper-binning scheme in a text file. Further description of how to use this file is described in \appref{app:hyperbinning}. \\
\tt benchmark.txt & The four-vectors associated to 100 phase space points, and their associated bin numbers. This can be used to check that the phase space binning has been correctly implemented.  \\
\tt modpred.txt & The central values and uncertainties of the hadronic parameter model predictions. \\
\tt modpredcor.txt & The correlations between the uncertainties of the hadronic parameter model predictions. \\
\tt modpred.root & The central values, uncertainties, and correlations of the hadronic parameter model predictions in \root format. This can be loaded with the \root macro {\tt loadresults.C}. \\
\tt modcompat.txt & The compatibility between the measured hadronic parameters and the model predictions. \\
\end{tabular}
\caption{ List of files in the supplementary material that are used to describe the measured hadronic parameters for a particular phase space binning scheme. \label{tab:sup} }
\end{table}


\subsection{Hyper-binning}
\label{app:hyperbinning}

For flexibility, the hyper-binning schemes are given in three different formats in the supplementary material, which will be discussed in this section. 
All binning schemes have been produced with a \Dz mass of $1864.84\mev$, and a $\pi^{\pm}$ mass of $139.57\mev$; this defines the boundaries of the $m_+$ and $m_-$ variables.

It is recommended to use the \root format ({\tt hypbinning.root}), which can be loaded using the {\sc HyperPlot} \cpp package located at,

{\
 \centering 
 
\url{http://samharnew.github.io/HyperPlot/index.html},

\ }

\noindent using the {\tt HyperHistogram} class. An example \cpp function is given in {\tt usehypbinning.cpp } that can be compiled with the {\sc HyperPlot} package to load any of the hyper-binning schemes.

The compressed directory {\tt hypbinning.zip} contains two text files; {\tt hypbinning.txt} and {\tt hypbinningwlinks.txt}.
Implementing the hyper-binning using the information in {\tt hypbinning.txt} is significantly easier than {\tt hypbinningwlinks.txt}, but the resulting code will be up to $10,000$ times slower (although this may still be fast enough for small event numbers).
Using the previously discussed \root format will automatically include this speed benefit.

The {\tt hypbinning.txt} file lists the low and high corner of each hypervolume in the binning scheme with its associated bin content. The bin content gives the phase space bin number $\in \{-\mathcal{N},...,-1, +1, ..., +\mathcal{N}$\}.
The coordinates are given in the order $\{  m_{+}'$, $m_{-}'$, $\cos\theta_{+}$, $\cos\theta_{-}$, $\phi  \}$; where invariant masses are given in units of \mev, and $\phi$ is given in radians.

\begin{figure}[h]
\centering
\includegraphics[width=0.7\textwidth]{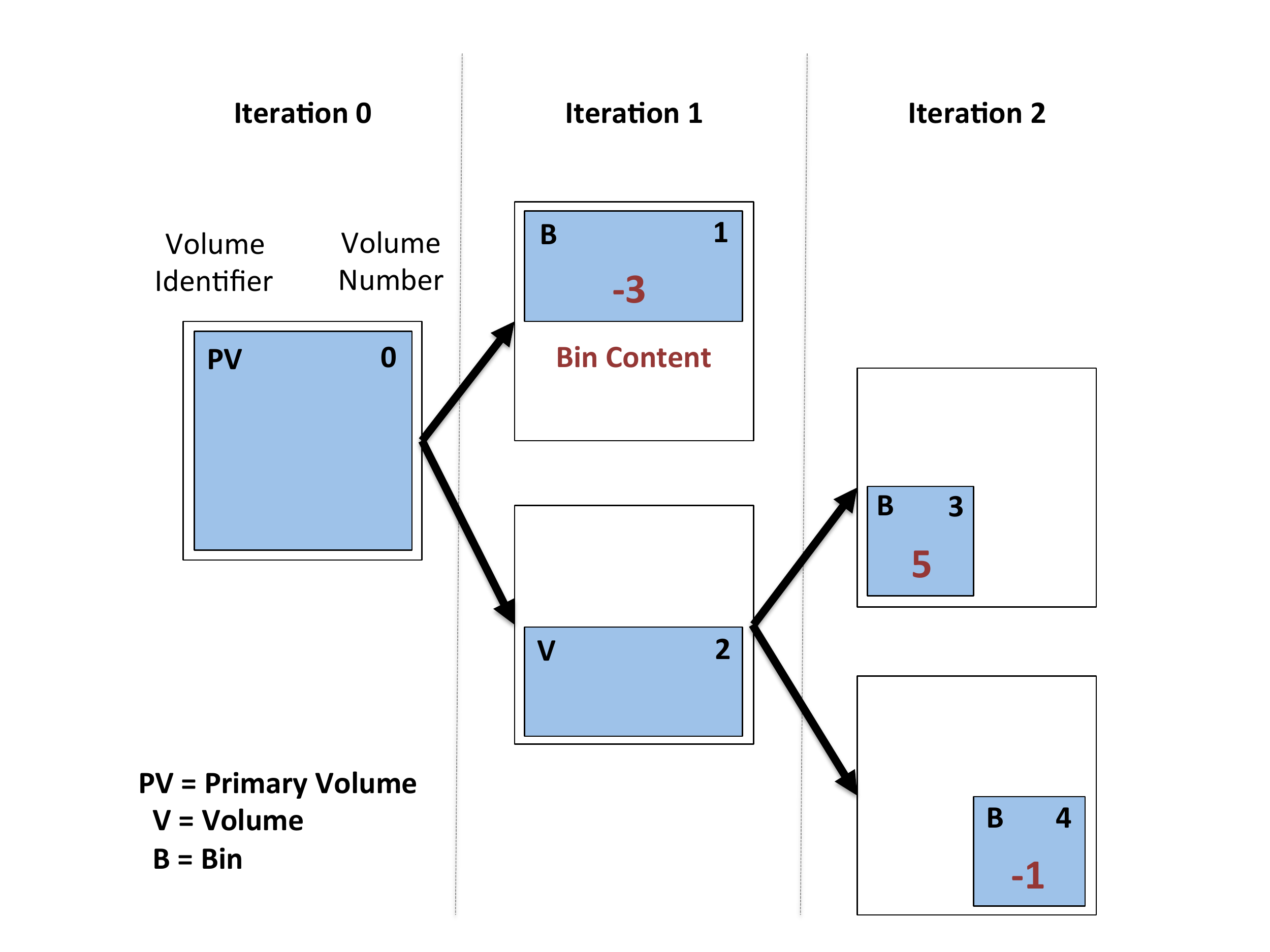}
\caption{ Simple example that demonstrates how the hyper-binning algorithm works. At iteration 0 there is a single primary volume (PV) with volume number 0. At iteration 1, the primary volume is split into two volumes with volume numbers 1 and 2. 
Volume number 1 is not split any further, so it is labelled as a `bin' (B) rather than a `volume' (V) - the content of this bin is -3. In iteration 2, volume number 2 is further divided into volume numbers 3 and 4; since this is the final iteration, these volumes are labelled as bins, which have bin contents of 5 and -1 respectively.   \label{fig:hypbinning} }
\end{figure}

To describe the format of the {\tt hypbinningwlinks.txt} file, it is useful to revisit how the binning algorithm works. 
At iteration 0, there is one hypervolume; at iteration 1, this gets split to give two hypervolumes; at iteration 2 each of these gets split to give 4 hypervolumes \etc
Rather than discard the hypervolumes from iteration 0 and iteration 1, these can be kept to speed up the binning process later.
The final set of hypervolumes that come out of the binning algorithm are known as `bins' (B). 
Other hypervolumes that were used during the binning algorithm (but were then further divided) are known simply as `volumes' (V). 
The first volume from iteration 0 is known as the primary volume (PV).
A simple example of a 2 dimensional binning scheme, iteration-by-iteration, is given in \figref{fig:hypbinning}, with bins and volumes labelled. 
Each volume and bin has a unique identifier called a `volume number'. 
Every volume has links to two volume numbers, whereas each bin has a bin content (which gives the phase space bin number).  
The simple binning scheme in \figref{fig:hypbinning} is described by the information in \figref{tab:hypbinningwlinks}, which has the same format as {\tt hypbinningwlinks.txt}. 
For comparison, the same binning scheme is described in the same format as {\tt hypbinning.txt} in \figref{tab:hypbinning}.

The general use case of a binning scheme is to find the bin (and its associated bin content), that an arbitrary phase space point, \pspointf, falls into. 
Using the information in {\tt hypbinning.txt} requires looping over every bin, and seeing which one contains \pspointf; on average this will take $\sim N/2$ operations, where $N$ is the number of bins. 
To use the information in {\tt hypbinningwlinks.txt}, one would first check if \pspointf is within the PV; if it is, then one would see which of the linked volumes \pspointf falls into \etc 
On average this will take $\sim \log_2 N$ operations.

\begin{figure}[h]
\centering

\begin{minipage}[c]{0.95\linewidth}

\tt
Vol \# \ Vol ID \ Low Corner \ \ \ \ High Corner \ \ \ \ Vol Links / Bin Cont  \\
------------------------------------------------------------------- \\
0 \ \ \ \ \ PV  \ \ \ \ \ ( 0.0, 0.0 ) \ \ ( 1.0, 1.0 ) \ \ \ \ 1 \ \ 2 \\
1 \ \ \ \ \ B \ \ \ \ \ \ ( 0.0, 0.5 ) \ \ ( 1.0, 1.0 ) \ \ \ -3 \\
2 \ \ \ \ \ V \ \ \ \ \ \ ( 0.0, 0.0 ) \ \ ( 1.0, 0.5 ) \ \ \ \ 3 \ \ 4 \\
3 \ \ \ \ \ B \ \ \ \ \ \ ( 0.0, 0.0 ) \ \ ( 0.5, 0.5 ) \ \ \ \ 5 \\
4 \ \ \ \ \ B \ \ \ \ \ \ ( 0.5, 0.0 ) \ \ ( 1.0, 0.5 ) \ \ \ -1 \\
\end{minipage}

\caption{ Representation of the hyper-binning in \figref{fig:hypbinning} using the same format as {\tt hypbinningwlinks.txt}. \label{tab:hypbinningwlinks}  }
\end{figure}

\begin{figure}[h]
\centering

\begin{minipage}[c]{0.60\linewidth}

\tt
Low Corner \ \ \ \ High Corner \ \ \ \ Bin Cont  \\
------------------------------------------- \\
( 0.0, 0.5 ) \ \ ( 1.0, 1.0 ) \ \ \ -3 \\
( 0.0, 0.0 ) \ \ ( 0.5, 0.5 ) \ \ \ \ 5 \\
( 0.5, 0.0 ) \ \ ( 1.0, 0.5 ) \ \ \ -1 \\
\end{minipage}

\caption{ Representation of the hyper-binning in \figref{fig:hypbinning} using the same format as {\tt hypbinning.txt}. \label{tab:hypbinning}  }
\end{figure}


\clearpage

\subsection{Binning schemes}
\label{app:binningSchemes}

As described in \secref{sec:binning}, the hyper-binning schemes are only defined in the region with corners 
$\{ m_{\mathrm{min}}, m_{\mathrm{min}}, 0, 0, 0 \}$ and 
$\{ m_{\mathrm{max}}, m_{\mathrm{max}}, +1, +1, +\pi \}$, 
which is $1/8$ of the entire phase space. 
The following algorithm can be used to determine the phase space bin of any given phase space point: 
\begin{itemize}
\item Calculate the variables $\{  m_{+}, m_{-}, \cos\theta_{+}, \cos\theta_{-}, \phi  \}$ using the formalism in \appref{app:helicityVars}.
\item Use the transformation in \eqnref{eqn:mprime} to determine $m_{+}'$ and $m_{-}'$.
\item Is $\cos\theta_{+} < 0$? If yes, $\cos\theta_{+} \to -\cos\theta_{+}$ and $\phi \to \phi - \pi$.
\item Is $\cos\theta_{-} < 0$? If yes, $\cos\theta_{-} \to -\cos\theta_{-}$ and $\phi \to \phi - \pi$.
\item Is $\phi < 0$? If no, $c_{\mathrm{flip}} = 1$. If yes, $\cos\theta_{+} \leftrightarrow \cos\theta_{-}$ and $m_{+}' \leftrightarrow m_{-}'$, $c_{\mathrm{flip}} = -1$.
\item After the above steps it is guaranteed that the transformed phase space point is in the region with corners $\{ m_{\mathrm{min}}, m_{\mathrm{min}}, 0, 0, 0 \}$ and $\{ m_{\mathrm{max}}, m_{\mathrm{max}}, +1, +1, +\pi \}$ (neglecting abitrary $2\pi$ rotations). 
\item Use the hyper-binning scheme to find the bin number, $i$, of the transformed point $\{  m_{+}', m_{-}', \cos\theta_{+}, \cos\theta_{-}, \phi  \}$ (see \secref{app:hyperbinning}).
\item The bin number of the original point is $c_{\mathrm{flip}} \times i$.
\end{itemize}

\clearpage



\bibliographystyle{JHEP}
\bibliography{bibliography}






\end{document}